%% file: ms.tex
\documentclass[11pt,letter]{article}
\pdfoutput=1

\usepackage[margin=1.5in]{geometry}
\usepackage{graphicx}
\usepackage{amsmath} 
\usepackage{amssymb}
\usepackage{amsfonts}
\usepackage{caption}
\usepackage{lscape}
\usepackage{rotating}
\usepackage[english]{babel}
\usepackage[T1]{fontenc}
\usepackage[utf8]{inputenc}
\usepackage[onehalfspacing]{setspace}
\usepackage{indentfirst}
\usepackage{bbm}
\usepackage{hyperref}
\usepackage{color,soul}
\usepackage[authoryear,sort&compress,round]{natbib} 
\usepackage{graphicx}

\usepackage{afterpage} % for landscaped floats, etc.
\usepackage[capposition=top,footfont=small,subfloatrowsep=none]{floatrow}
\usepackage[position=bottom]{subfig}

\usepackage{longtable}
\usepackage{booktabs}
\usepackage{multirow}
\usepackage{siunitx}
\usepackage{array}
\newcolumntype{C}[1]{>{\centering\arraybackslash\hspace{0pt}}p{#1}} 
\usepackage{makecell}

\usepackage{titlesec}

\usepackage{appendix}

\usepackage{chngcntr}

\title{Cash versus Kind:  \\ \Large{Benchmarking a Child Nutrition Program against \\  Unconditional Cash Transfers in Rwanda}%
\thanks{\protect\footnotesize We are grateful to DIV, Google.org, and USAID Rwanda for funding, and to USAID, CRS, GiveDirectly, CEGA/DIL, and IPA for their close collaboration.  We thank Leodimir Mfura and Marius Chabi for overseeing the fieldwork, and Richard Appell, Sarait Cardenas-Rodriguez, Chris Gray, Ali Hamza, and Bastien Koch for research assistance.  This project is covered by Rwanda National Ethics Committee IRB 143/RNEC/2017 and IPA IRB 13730, and the study is pre-registered with the AEA as trial AEARCTR-0002559. This study is made possible by the support of the American People through the United States Agency for International Development (USAID). The contents of this study are the sole responsibility of the authors and do not necessarily reflect the views of USAID or the United States Government.}
}

\author{ 
Craig McIntosh\thanks{\protect\footnotesize University of California, San Diego\href{mailto:ctmcintosh@ucsd.edu}{ctmcintosh@ucsd.edu}} \,  and
Andrew Zeitlin\thanks{\protect\footnotesize Georgetown University, \href{mailto:andrew.zeitlin@georgetown.edu}{andrew.zeitlin@georgetown.edu}}
}

\date{%
June 1, 2021 
}

\begin{document}

\maketitle

\begin{abstract}
\begin{singlespace}

We benchmark a multi-dimensional child nutrition intervention against an unconditional cash transfer of equal cost. Randomized variation in transfer amounts allows us to estimate impacts of cash transfers at expenditure levels equivalent to the in-kind program, as well as to estimate the return to increasing cash transfer values.  While neither the in-kind program nor a cost-equivalent transfer costing \$124 per household moves core child outcomes within a year, cash transfers create significantly greater consumption than the in-kind alternative.  A larger cash transfer costing \$517 substantially improves consumption and investment outcomes and drives modest improvements in dietary diversity and child growth. 

\bigskip

\noindent Keywords: Experimental Design, Cash Transfers, Malnutrition \\
\noindent JEL Codes: O12, C93, I15
\end{singlespace}
\vfill
\end{abstract}

\thispagestyle{empty} 

\clearpage 

\section{Introduction} 
\input{1-Gikuriro-Intro}

\section{Study design}
\label{s:design}
\input{2-Gikuriro-StudyDesign}

\section{Comparative Impact Analysis}
\label{s:mainresults}
\input{3-Gikuriro-Analysis}

\section{Heterogeneous comparative effects and tradeoffs across sub-populations}
\label{s:heterogeneity}
\input{4-Gikuriro-Heterogeneity}

\section{Program rigidity, choice, and beneficiary welfare}
\label{s:choice}
\input{5-GIkuriro-Choice}

\section{Conclusion}
\label{s:conclusion}
\input{6-Gikuriro-Conclusion}

%--------------------------------------------------------------------------%
%--  Bibliography  --%
\cleardoublepage 
\singlespacing
\bibliographystyle{aer}
\bibliography{Gikuriro}

---------%
\cleardoublepage

\newgeometry{margin=1in,bottom=1in,top=1in}

%-- BODY TABLES AND FIGURES AT END ---%
\section*{Tables and Figures}
\input{7-Gikuriro-Tables}

\clearpage 
\input{8-Gikuriro-Figures}

%%%%%%%%%%%%%%%%%%%%%%%%%%%%%%%%%%%%%%%%%%%%%%%%%%%%%%%%%%%%%%%%%%%%%%%%%%%%%%%%%%%
%  Appendices  
%%%%%%%%%%%%%%%%%%%%%%%%%%%%%%%%%%%%%%%%%%%%%%%%%%%%%%%%%%%%%%%%%%%%%%%%%%%%%%%%%%%
%--  Alphabetic indexing for appendix sections, with tables and figures in the appendix numbered within sections --% 
\cleardoublepage
\appendix

%  Appendix numbering style: A.1
\renewcommand\thesection{\Alph{section}}
\renewcommand\thesubsection{\Alph{section}.\arabic{subsection}}

%  Section and sub-section header formats
\titleformat{\section}{\centering\normalfont\Large}{Appendix \Alph{section}.}{1em}{}
\titleformat{\subsection}{\normalfont\large}{\Alph{section}.\arabic{subsection}.}{1em}{}

%%  And adding Appendix counter to table and figure numbers
\renewcommand\thetable{\Alph{section}.\arabic{table}}
\renewcommand\thefigure{\Alph{section}.\arabic{figure}}
\renewcommand{\thepage}{\Alph{section}.\arabic{page}}
\counterwithin{figure}{section}
\counterwithin{table}{section}
\counterwithin{page}{section}

%----------------------------------------------------------------------------------%
\cleardoublepage

\thispagestyle{empty}
\addtocounter{page}{-1} 

\hspace{0pt}
\vfill 
\begin{center}
\Large{\textbf{For Online Publication}}
\end{center}    
\vfill 

\clearpage 

\setcounter{page}{0}
\setcounter{table}{0} 
\setcounter{figure}{0}

\thispagestyle{empty}
\hspace{0pt}
\vfill 
\section{Supplementary Tables and Figures}
\setcounter{page}{0}
\vfill

\input{9-Gikuriro-Appendix-Tables}

%----------------------------------------------------------------------------------%
%--  ONLINE APPENDIX --% 
%----------------------------------------------------------------------------------%

\clearpage

\setcounter{page}{0}
\setcounter{table}{0} 
\setcounter{figure}{0}

\thispagestyle{empty}
\hspace{0pt}
\vfill
\section{Additional Supplementary Materials}
\vfill

\input{99-Gikuriro-Online_Appendix}

%----------------------------------------------------------------------------------%

\end{document}

%% file: 1-Gikuriro-Intro.tex
%!TEX root = ./Gikuriro.tex

Should governments and aid agencies provide assistance in kind, or would beneficiaries be better off if they just received the money?  There are several reasons to compare impacts of in-kind programs to those generated when its cost is distributed directly as a cash transfer.  From a practical perspective, cash transfers generate well-documented  benefits for households \citep{aizer2016long, haushofer2016short}, and their costs are falling across the developing world due to the penetration of mobile money \citep{suri2017mobile}.  Cash programs are flexible and can be rescaled in terms of targeting or expense, which allows them to be paired at cost-equivalent levels against more complex interventions.\footnote{In medical research, new interventions are typically benchmarked against the best current treatment, referred to as the `standard of care'.  Perhaps the best-validated anti-poverty tool in the development literature is BRAC's Targeting the Ultra-Poor program \citep{banerjee2015multifaceted}; however given that this program is expensive (\$3,700 per household), lengthy in duration (a year of intensive intervention), and targeted at very specific households (the ultra-poor), it represents a somewhat inflexible benchmark for the broad range of potential development interventions.}   By contrast, comparisons across programs in different contexts and at different costs rely on strong assumptions about external validity \citep{vivalt2015much} and linearity (both in cost-effectiveness ratios and the preferences of policymakers across individuals).  More fundamentally, decisions made over the use of unconditional transfers may reveal information about the preferences of the beneficiary households that is important in interpreting the welfare effects of more complex programs \citep{FinHen20jep}.  While missing markets, externalities, or other failures can justify the provision of in-kind programs, benchmarking the impact of such programs against cost-equivalent cash transfers forces policymakers to be explicit about the circumstances that would merit a more multidimensional, overhead-heavy approach.  For these reasons, recent years have seen increasing calls for head-to-head studies that use cash as a benchmark for more complex forms of international development assistance.\footnote{In particular \citet{blattman2014show} argue for the use of cash transfers as the `index funds' of international development, providing a reference rate of return that could be used to hold donors accountable.}

In this study we provide such a benchmark, comparing a nutritional and maternal health intervention to a cash transfer in Kayonza and Nyabihu districts of Rwanda.  
This is an ideal setting and intervention for such a comparison. Nutritional deprivation is a pressing issue in both districts, which feature under-five stunting levels of 34.8 and and 44.9 percent, respectively \citep{dhs16rwanda}.  The in-kind program takes a multi-dimensional approach widely used by USAID and other donors to combat malnutrition, one that has been broadly advocated in the public health literature \citep{ruel2013nutrition} and found to have positive impacts in neighboring Burundi \citep{leroy2016tubaramure} and Kenya \citep{Null18lancetgh}. The complex and overhead-intensive nature of such a holistic, bundled intervention makes a stark contrast with the pared-down nature of the cash-transfer alternative.  And yet there is reason to believe that cash transfers can improve child growth outcomes \citep{aguero2006impact,seidenfeld2014impact, BAIRD2019169}, and that gains achieved in early childhood might have lasting effects through the accrual of human capital \citep{currie2011human,HenSpru20qje}.

To form this comparison, we conduct a cluster-randomized trial across 248 Rwandan villages.\footnote{The village-level study design was motivated by the clustered nature of the Gikuriro intervention. In addition, this helps to allay concerns about the potential for spillovers of cash transfers.}   We enroll poor households containing children or pregnant women, with an emphasis on the 1,000-day window of opportunity from pregnancy until a child's second birthday \citep{currie2011human}. In the in-kind arm, eligible beneficiaries are offered Catholic Relief Service's \emph{Gikuriro} program.\footnote{Gikuriro means `well-growing child' in Kinyarwanda.}  Gikuriro aims to improve child nutrition through superior information, direct transfer of productive assets, and improvements in household diet and sanitation.  It consists of four components targeted directly at beneficiary households: a Village Nutritional School, Farmer Field Learning Schools, Savings and Internal Lending Communities, and a Water, Sanitation, and Hygiene  intervention, as well as a Behavior Change Communication intervention implemented at the village level.  

Gikuriro is compared to a control arm as well as to a study arm that received unconditional cash transfers via mobile money, implemented by the U.S. non-profit GiveDirectly.  Such transfers have been found to improve consumption and/or dietary diversity in multiple contexts across Sub-Saharan Africa \citep{haushofer2016short, aker2016payment}.  At the village level, transfer amounts were randomized in a range around the expected cost of Gikuriro, with a subset of villages receiving a much larger transfer.  Regression adjustment then allows us to calculate an exact cost-equivalent comparison of cash versus in-kind programming across \textit{modalities}, while the larger cash transfer allows us to compare benefits across the \textit{cost} of the intervention.  Within these GiveDirectly villages, we implemented a household-level experiment, whereby beneficiary households were randomized to receive transfers in a lump-sum or as a 12-month flow, or to have their choice between these options. The ability to understand the role of cash modality preferences gives us a window into the extent to which self- and other-control problems drive the use of unconditional cash.

Thirteen months after baseline, we measure impacts on five primary outcomes:  household consumption, household dietary diversity, child and maternal anemia, child growth, and household non-land net wealth.   These outcomes are documented as core drivers of improved long-run outcomes for children \citep{maluccio2009impact, hoddinott2013economic}, as well as being plausibly impacted by both intervention types.  In addition, we examine a set of secondary outcomes including borrowing and savings, fertility, health knowledge and sanitation practices, diseases and mortality, household assets, and the quality of housing stock.\footnote{All primary and secondary outcomes were registered prior to receipt of endline data on the American Economic Association RCT Registry, with ID AEARCTR-0002559.}  Prior to randomization the survey firm classified all households in study villages as `eligible' (identifiable using administrative data sources as containing underweight children, or households in the bottom two income categories with children 5 years old or younger or with pregnant or lactating women), or `ineligible' (everyone else).  We can therefore measure impacts both on the mutually agreed-upon intended target population as well as on the study villages as a whole, even in the presence of potential differences in actual targeting across implementers.  Using the eligible sample we can estimate experimental intention-to-treat effects, and using the full sample (population weighted) we estimate total causal effects on the average household in study villages.  

This study design, like that of the follow-on study that benchmarks cash against a youth employment program \citep{mcintosh2020hugukadukore}, allows us to assess tradeoffs  across \emph{outcomes}, across \emph{expenditure per beneficiary}, and across \emph{sub-populations}.\footnote{The successor study utilizes the same benchmarking methodology (comparing an in-kind program to randomized unconditional cash transfers) to answer applied questions from the labor literature: namely, how best to help underemployed youth gain access to productive work.}

Across outcomes, our results provide a nuanced view of the relative impact of a highly-tailored child malnutrition program and cost-equivalent cash transfers. Impacts of both cash and in-kind programs are multidimensional.  At a cost of \$142, Gikuriro was successful at delivering gains in savings, a domain that was the target of one core program component. 
It did not lead to improvements in consumption, dietary diversity, wealth, child anthropometrics, or anemia within the thirteen-month period of the study. A cost-equivalent cash transfer has significantly larger effects on consumption, allows households to pay down debt, and generates increased investment in productive and consumption assets.  
Here, then, is a first type of tradeoff that this design can reveal:  should policymakers prefer an increase in savings versus the consumption gains, debt reductions, and physical asset accumulation that households choose to engage in when programming is shifted from in-kind to cash?

Across expenditure levels, our results show that a much larger cash transfer costing \$567 led to across-the-board improvements in consumption-based welfare measures, a substantial improvement in dietary diversity, a drop in child mortality, and modest improvements of about 0.1 standard deviation in the anthropometric indicators of height-for-age, weight-for-age, and mid-upper arm circumference (all significant at 10 percent or above).  The large cash transfer delivers benefits  even on outcomes specifically targeted by the other program. While it is unsurprising that very large amounts of money show up in consumption and productive assets, the  improvements in diet and particularly child anthropometrics over such a short period of time are impressive.  Further, while it may be unsurprising that the impact of cash transfers scales with the amount spent in the way found here, the same may not be true of other types of development intervention that would quickly hit diminishing marginal returns once certain core objectives were achieved.  For a given target population, policymakers must therefore decide how to trade off between relatively intensive interventions that can only be provided to a small subset, or lower-cost interventions that can reach a larger proportion of this target group.

Across subpopulations, our analysis speaks to tradeoffs in three distinct ways.  First, policymakers face decisions about the extent to which they value outcomes among a narrowly targeted population, or the broader society in which they reside.  Our study design allows us to calculate not only benefits per dollar cost among the program-eligible population, but also to compare the total causal effect of each intervention on the full population of study villages in a cost-equivalent manner and at comparable levels of treatment saturation.  Despite 80 and 84 percent of the eligible households in treatment villages receiving Gikuriro and GiveDirectly, respectively (for the villages as a whole the treatment rates were 19 and 18 percent), neither treatment resulted in sufficiently widespread benefits as to be detectable at cost-equivalent levels in the general population, with the exception of an improvement in health knowledge and vaccination rates in Gikuriro villages and vaccination rates in GD large villages.  Second, on the other hand, policymakers need not place equal welfare weights on all subgroups within the targeted beneficiary  population, and the relative effects of cash versus in-kind transfers may differ across these groups.  We test for heterogeneity in relative returns along observed subgroups of a priori interest---using consumption and child-growth levels at baseline---and find limited evidence of such heterogeneity in this instance.  Third, since policymakers might also be able to fine-target intervention modalities based on beneficiary characteristics, we estimate the extent of heterogeneity in \emph{relative} treatment response by application of a generalized random forest \citep{WagAth18jasa,AthTibWag19annsat}. We find that while there is substantial heterogeneity in impacts of cash versus in-kind treatments for any one outcome dimension at a time, estimated subpopulation treatment effects are only weakly correlated across outcome dimensions. This suggests that fine targeting can exploit heterogeneity with tailored treatments to improve outcomes for policymakers with narrow objective functions, but that---at least in this case---such personalized treatments deliver only modest gains when policymaker objectives are broadly defined.

Finally, we build a sub-experiment into the cash arm of the study to shed light on the role that choice and self-control play in the comparative impact of cash transfers.  Both the conventional wisdom and the behavioral literature have been concerned with the idea that liquid resources with no strings attached will be wasted or spent in ways that the households later would regret \citep{ashraf2006tying,lai97qje}.  We speak to this issue using an incentivized choice experiment where households were asked to decide between receiving all their money in a lump-sum versus getting it in a flow over twelve months.   While demand for the more rigid monthly cash transfers is substantial, we find no evidence of `sophisticated' choice reflecting essential heterogeneity, and the benefits of flow or in-kind transfers are not larger for households that report self- or other-control problems at baseline.  The striking absence of behavioral heterogeneity suggests that the higher-cost and more rigid programs do not deliver better outcomes even for those with behavioral challenges.

This paper develops and demonstrates a methodology---combining randomization in cash transfer amounts with regression-based interpolation of cash impacts---that allows for exact cost-equivalent comparisons, while recognizing ex-ante uncertainty about the realized costs of both cash and in-kind programming. 
Such cost-equivalent comparisons are a special case of cost-benefit comparisons (which we also provide), where the cost-equivalent approach compares benefits at a level of resource intensity that allows policymakers to hold expenditure per beneficiary fixed.
For many scenarios, this represents a more realistic vision of the decision policymakers face, since cost-benefit ratios achieved with wildly different expenditure levels may require radically different target populations; a cost-equivalent comparison also relies less strongly on the additive separability of the policymaker's objective function.
Our approach yields an ex-post cost-equivalent comparison with relatively little power loss relative to a design that takes a single ex-ante guess about these ex-post costs, while permitting robustness checks that test sensitivity to the nature of the interpolation.

In applying this approach, we extend the small number of studies using cash transfers as one arm within multi-armed trials.  The largest extant literature is based on the comparison of cash aid to food aid \citep{leroy2010cash, schwab2013form, hoddinott2014impact, hidrobo2014cash, ahmed2016kinds},
uncovering a fairly consistent result that food aid leads to a larger change in total calories while cash aid leads to an improvement in the diversity of foods consumed, while benefits for targeted individuals can be limited and there can be adverse consequences for non-targeted individuals when market imperfections mean that cash transfers cause food price increases \citep{CunGioJayXXrestud,FilFriKanOni21restat}.\footnote{An interesting contribution to this food versus cash literature is \citet{hirvonen2020beneficiary}, who find that poor and geographically marginalized households may prefer food transfers because of a lack of consumer markets during times of shortfall.}$^,$\footnote{A related literature compares food transfers with \emph{vouchers} redeemable for the purchase of food. Recent work by \citet{BanHanOlkSatSum21nber} shows positive effects of shifting to vouchers, attributable to changes in the share of amount of assistance reaching targeted households and in the quality of food consumed.}  Efforts to compare more complex, multi-dimensional programs against cash include BRAC's Targeting the Ultra-Poor program \citep{chowdhury2016valuing}, microfranchising \citep{brudevold2017firm}, agricultural inputs \citep{brudevold2017firm}, and sustainable livelihoods \citep{sedlmayr2017cash}.  These studies have typically struggled with the question of how to anticipate costs and compliance well enough to realize an exact cost-equivalent comparison after the fact; we address this with our study design. Further, we extend the existing literature comparing in-kind interventions to cash by illuminating the extent of heterogeneity in the returns to cash versus kind, and by showing how the comparison of in-kind programming with a larger cash transfer on a benefit-cost basis would lead to different conclusion.

Beyond this, we add to the growing number of studies that show meaningful impacts of cash transfers on important life outcomes in the short term, in domains with plausible channels for long-term impact. These include evidence of impacts on child nutrition \citep{aguero2006impact, seidenfeld2014impact}, schooling \citep{skoufias2001conditional}, mental health \citep{baird2013income, samuels2016being}, teen pregnancy and HIV \citep{baird2011cash}, microenterprise outcomes \citep{de2012one}, consumer durables \citep{haushofer2016short}, and productive assets \citep{gertler2012investing}.  The evidence on the long-term impacts of cash transfers is more mixed, but some studies have found substantial impacts \citep{fernald200910, barham2014schooling, aizer2016long, hoynes2016long}.\footnote{For examples of studies that find dissipating long-term benefits, see \cite{ozler2016money},  \cite{araujo2017can}, and \cite{blattman2020long}.  For evidence from systematic reviews of cash transfers on schooling see \citep{molina2016long}, and on child health see \citep{pega2014unconditional, manley2013effective}.}  We contribute to the cash transfer literature by evaluating multiple transfer amounts and modalities in the same context, and illuminating the scope for beneficiary choice over cash-transfer modality to drive impacts.

The rest of the paper is organized as follows. The next section lays out the study design, including a detailed description of the interventions, sampling routine, costing principles, experimental structure, and primary and secondary outcomes. Section \ref{s:mainresults} presents the core empirical results of the benchmarking exercise, as well as the results of sub-experiments on cash transfer modalities. Section \ref{s:heterogeneity} estimates the extent of heterogeneity, both using pre-specified attributes as well as conducting a machine learning exercise to speak to optimal targeting rules. Section \ref{s:choice} discusses the choice experiment and evidence of the role of rigidity.  
Section \ref{s:conclusion} concludes with a discussion of how results from such cash-benchmarking studies can be used to bound the preferences over benefit/cost ratios across outcomes required to justify each program.

%% file: 2-Gikuriro-StudyDesign.tex
%!TEX root = ./Gikuriro.tex

In this section, we describe four dimensions of the study design:  the interventions, the study outcomes, the allocation of villages and individuals to treatment arms, and the approach to using experimental variation in cash transfer values to make cost-equivalent comparisons across modalities.

\subsection{Description of Interventions}

The Gikuriro program was developed by USAID, Catholic Relief Services (CRS), and the Netherlands Development Organization (SNV) to combat food insecurity among pregnant women and children, particularly during the critical first 1,000 days of life that play such a dominant role in later-life outcomes and cognition \citep{figlio2014effects}.  The resulting multi-faceted program brings together several components in order to attack this problem from multiple directions at once, and is a central pillar of the Government of Rwanda's approach to combatting malnutrition in rural Rwanda.\footnote{USAID's Global Health and Nutrition Strategy explicitly calls for multi-sectoral interventions that incorporate agriculture, WASH, education, and outreach to mothers in the first 1,000 days through the public health system.  The agency reports reaching 27 million children worldwide under the age of 5 in 2016 alone through such programs, which represent the prescribed USAID modality for Scaling up Nutrition (SUN) countries.}  Gikuriro combines an integrated nutrition program with a standard WASH curriculum (water, sanitation, and hygiene), and seeks to build the capacity of the health infrastructure providing services to mothers and newborns, particularly Community Health Workers (CHWs).  The program also seeks to improve livelihoods by providing additional assistance to eligible households, including (a) Village Nutrition Schools (VNS); (b) Farmer field learning schools (FFLS), which potentially includes distribution of small livestock, fortified seed, etc.; (c) Savings and Internal Lending Communities (SILCs); and (d) the Government of Rwanda's Community-Based Environmental Health Promotion Program (CBEHPP).  In addition, Gikuriro provided a program of Behavioral Change Communication (BCC), supporting participation in all components of the program including savings, agriculture, and nutrition, as well as hygiene.  This comprehensive approach seeks to build supply and demand for child health services simultaneously, and is fairly typical of the kinds of multi-sectoral child health programs implemented by USAID in many parts of the developing world.\footnote{Examples of similar integrated WASH/agriculture/child nutrition programs funded by USAID include SPRING in Bangladesh, RING in Ghana, Yaajende in Senegal, and ENGINE in Ethiopia.}

In terms of program expenditures, the most substantial component of Gikuriro is the Farmer Field Schools, which consume 41\% of the overall budget.  The main cost driver here was the distribution of seeds, as well as small livestock and poultry.\footnote{Narrative evaluation suggests that many of the chickens, goats, and rabbits distributed through the program fell ill and died soon after distribution, so the absence of supporting veterinary services were a problem.} Next most important were the village nutritional schools, with 19\% of the spend.  This component's core goal was to use demonstration plots to show how to use very small plots (or even gunny sacks for the land poor) to grow micro-gardens with a variety of nutrient rich greens that could be used to support child nutrition.  The other components of program expenditure were overall logistical program support (22\%), monitoring and evaluation (9\%), the SILCs (6\%), and the BCC and WASH trainings (4\%).  Our estimates suggest that a household participating in every component of Gikuriro would have actually received training and inputs worth \$70.13, of which \$5.06 is direct transfer of materials and inputs. 

To benchmark the impact of this program against cash we worked with GiveDirectly, a US-based 501(c)3 Non-Profit organization.  GiveDirectly specializes in sending mobile money transfers directly to the mobile phones of beneficiary households to provide large-scale household grants in developing countries including Kenya, Uganda, and Rwanda.  GiveDirectly's typical model has involved targeting households using mass-scale proxy targeting criteria such as roof quality.  GiveDirectly builds an in-country infrastructure that allows them to enroll and make transfers to households while simultaneously validating (via calls from a phone bank) that transfers have been received by the correct people and in a timely manner.  Their typical transfers are large and lump-sum, on the order of \$1,000, and the organization provides a programatically relevant counterfactual to standard development aid programs because it has a scalable business model that would in fact be capable of providing transfers to the tens of thousands of households that are served by the Gikuriro program.  Because of the nutritional focus of the Gikuriro intervention, GiveDirectly incorporated a `nudge' into the way the program was introduced \citep{benhassine2015turning}, distributing a flyer at the time of the intervention that emphasized the importance of child nutrition.  An English translation of this flyer is included in the Online Appendix.  Given observed impacts of cash transfers on other goods, e.g., productive assets and housing value, it is evident that households felt at liberty to spend the grants on items not directly related to child nutrition.

Rwanda may be a particularly interesting environment in which to pose the benchmarking question for several reasons.  First, child malnutrition rates overall are high---the prevalence of stunting among children
under age five in the 2014-15 Demographic and Health Survey was 37.9 percent, underweight 9.3 percent, and wasted 2.2 percent---though this represents an improvement in recent years \citep{dhs16rwanda}  Second, Rwanda is a country notable in Africa for its bureaucratic capacity and the public health infrastructure has been successful in delivering substantial improvements in child and maternal health outcomes \citep{NISR2015} through schemes such as Pay-for-Performance \citep{basinga2011effect}.  Hence, it may provide a relatively strong case in terms of interventions like Gikuriro that are led through the public health system and lean heavily on Community Health Workers (CHWs).  Third, the Government of Rwanda has been experimenting  extensively with cash transfer programs over the past few years, such as the inclusion of cash in the flagship \textit{Umurenge} poverty reduction program \citep{gahamanyi2015evaluating}, the \$50 million `Cash-to-poor' program supported by the World Bank, as well as a number of efforts to transition the support systems for the country's large population of refugees to cash transfers (such as a World Food Programme (WFP) program that is now supporting 15,000 refugees in Gihembe Camp using cash rather than traditional in-kind aid mechanisms \citep{taylor2016economic}).  Hence there should be the bureaucratic capacity to implement Gikuriro well, and there is both experience with and interest in cash transfers as a safety net modality in the country.\footnote{Given the framing provided by GiveDirectly and the unusually strong degree of social control exerted by local officials in the Rwandan context, it is certainly possible that our `unconditional' transfers have been more forcibly devoted to child consumption than they would have been in a different context.}

\subsection{Study outcomes and eligibility}

We pre-committed to five primary outcomes and seven secondary outcomes for the study.  The primary outcomes are consumption, dietary diversity, child growth (height-for-age and weight-for-age, as well as mid-upper arm circumference), and household non-land wealth.  The secondary outcomes are borrowing and saving, pregnancy and live birth rates, health knowledge, mortality, health-seeking behaviors, productive assets, and housing quality.  A more detailed description of all outcomes is provided in the Online Appendix.  Data were collected by Innovations for Poverty Action (IPA) teams through two survey waves; the baseline was conducted in August and September of 2016, and the endline was conducted in September and October of 2017.  The baseline included a comprehensive household survey module as well as anthropometric measurement of all children five and under in eligible households.\footnote{We weighed all children younger than 6 years once using a Seca 385 scale.  We measured length for children under two years with Seca 417 length boards and height for older children to the nearest 0.1 cm with Seca 213 stadiometers. These were measured twice and we took the average of the two measurements unless they differed by more than 0.7 cm in which case we took a third measurement and averaged the two closest measurements.} The endline consisted of these same measures plus blood-based anemia testing for the study children and for all women of childbearing age.\footnote{Using the guideline for anemia testing in population-based surveys, we used HemoCue Hb 301 system.}

Both implementers made contact with the study subjects and began enrollment immediately after baseline.  GD began implementation shortly after the baseline meaning that at endline individuals in that arm had experienced about 12 months of the household grants treatment (running up through the month before endline).  Gikuriro was slower than the cash program to begin implementation on the ground; in that arm households had typically experienced 8-9 months of household-level implementation at the time of the endline.\footnote{Since both programs had six months of notice that they would be implementing in the study sample in these two districts and began national-level implementation at the same time, this differential delay likely reflects a real difference in the relative ramp-up speeds of cash versus more complex programming.} The duration of the RCT component of the study was limited by the fact that local governments wanted to hit targets for the broader, national rollout of nutritional and WASH programming, and hence we were not able to maintain the control groups for more than one year.
We cannot therefore speak to the long-term impacts of the interventions.  Anticipating this issue, we took two approaches to measurement.  One of them was to try capture the stocks of intertemporal assets that would be the obvious conduits to future consumption benefits for the households.  The second was to emphasize outcomes such as dietary diversity and anemia that have the potential to respond quickly to changes in consumption patterns, while also retaining the more standard metrics of child malnutrition such as height for age (HAZ), weight for age (WAZ), and mid-upper arm circumference (MUAC).\footnote{Dietary diversity is an immediate indicator of improvements in consumption, and the clinical literature has shown that anemia tests respond within 3 months of improvements in diet \citep{habicht1990importance}.}  Further, a number of recent RCTs have shown that programs can have meaningful impacts on biometric outcomes over timeframes similar to that analyzed in this study, such as \cite{desai2015shine}, \cite{leroy2016tubaramure}, \cite{fink2017home}, and \cite{Null18lancetgh}.

A major practical issue with this kind of comparative study is how to define the the target group, given that each implementer would naturally do this differently.  We collectively resolved this by agreeing to a set of readily observable characteristics that both implementers deemed eligible for their funding and desirable to target, and then tasking the survey firm with listing all households in study villages and allocating them to the `eligible' or `ineligible' stratum.  Households were considered eligible if they contained a child known to the government growth monitoring system to be malnourished, or else were in the poorest two government poverty classifications (Ubudehe 1 or 2) and contained either children under the age of 5 or a pregnant/lactating woman.  All other households in study villages are classified as ineligible.  The survey firm passed the names of all eligible households to the implementing partners in an identical way, and we then randomly sampled up to 8 eligible and 4 ineligible households per village for the study.  The identity of the sampled households was not revealed to either implementer.  This structure allows us to use survey weights to estimate impacts that are representative either for eligible households (which we call the Intention to Treat) or for all households in study villages (the Total Causal Effect).  Impacts among ineligibles may arise either because the implementers treated some households outside of the IPA-defined eligible group, or because of spillovers from beneficiary to non-beneficiary households.  

Both implementers concurred quite closely with our definition of eligibility on the ground, and compliance was high:  we have 80 percent of the eligibles treated by Gikuriro  and 84 percent by GiveDirectly.\footnote{Because eligibility status was determined from records, in some cases the Ubudehe status of the households proved to be incorrect or unverifiable when they were approached by GD for treatment, and hence were not offered the program.  Gikuriro implemented a consultative process with community members that formed the basis of their targeting.}   The IPA-driven process of eligibility definition however encountered an average of 13.9 eligible households per village, much lower than the rate of 30 anticipated.  When Gikuriro began their actual program implementation, they continued their standard consultative process for beneficiary identification, which included the use of soft targeting information not available to the survey firm. Using this additional, richer information set to target, they identified and treated an average of 25.8 households in study villages in Kayonza district, and 26.97 in Nyabihu.  Since our first tranche of GiveDirectly treatments were only among IPA-defined eligible households, we found ourselves with a substantial discrepancy in the intensity of treatment across implementers in the ineligible group.  We responded to this asymmetry by drawing in an additional sample of the poorest ineligible households in GiveDirectly villages to receive cash grants so as to maintain parity in village-level treatment intensity.  Therefore, while the targeting of ineligible households differs across implementers, the treatment intensities across the two arms are identical by construction.

\subsection{Design of the Experiment}

Randomization occurred at the village level across 248 villages, using a blocked randomization where the blocks were formed by the combination of districts and village-level poverty scores within district, creating a total of 22 blocks with between 10 and 13 villages per block.  Fixed effects for these blocks are included in all analysis.  A computer was used by the researchers to conduct the randomization  based on a frame of villages agreed to by CRS and government officials.

Table \ref{t:research_design} presents a schematic of the design of the study.  74 villages were assigned to the Gikuriro intervention, 74 were assigned to the control group (no intervention), and 100 were assigned to GiveDirectly household grants.  The GiveDirectly villages were further split into four transfer amounts, randomized at the village level.  Three treatment amount arms, with 22 villages in each, received transfer amounts in a range around the anticipated cost of Gikuriro.  A final 34 villages were assigned to the `large' GiveDirectly transfer amount which was selected by GiveDirectly as the amount anticipated to maximize the cost effectiveness of cash.  The transfers actually received by households in the GD `main' arms were \$41.32, \$83.63, and \$116.91.  Then, the large GD arm actually transferred \$532 to households.   All transfer amounts were translated into Rwandan Francs at an exchange rate of 790 RwF per US dollar, and were rounded to the nearest hundred.\footnote{GiveDirectly believed that the most cost-effective use of these funds would be to attempt to equalize the amount transferred per household member, rather than to have households of very different sizes receiving the same transfer amount. To accomplish this, we scaled the transfer amounts within a village by household size, such that larger households received larger transfers, but leaving the mean transfer amount at the village level unaffected.  This formula first calculated the per-capita transfer for a village using household sizes and the desired average household transfer value. Second, it scaled household-level transfer amounts with household size, applying a minimum of 3 members and a maximum of 8 members, so as to achieve the intended mean transfer amount per household per village.}   Figure \ref{f:box_whisker} provides a box and whisker plot of the randomly assigned mean transfer amount per village relative to the actual amount received per household observed in the GD institutional data, and shows that the two correspond closely.

Within the GD arm we conducted an additional experiment that elicited choice over cash transfer modality in an incentivized way.  We wished to test the hypothesis that a regular monthly flow of transfers is likely to be a more effective way of delivering the kinds of nutrition and health outcomes central to Gikuriro \citep{haushofer2016short}, and to understand how much of this heterogeneity in benefits is understood by the recipients themselves.  Large Essential Heterogeneity in the sense of \cite{heckman2006understanding} would suggest that program choice could deliver superior overall impacts.  To do this, all respondents were given a menu illustrating the choice between a single lump-sum transfer delivered in any of the 12 months from August 2016 to July 2017 or a flow of monthly payments totaling the same amount.  They were informed that they would then be assigned to receive one or the other, and with probability of 1/8 would receive what they chose.  We then assigned 5/8ths of the cash sample to the flow arm (evenly divided over 12 monthly payments), and the remainder to receive an up-front lump sum cash transfer. While we do not use the choice arm as a treatment in itself, being able to observe the choice for all cash subjects allows to understand whether those who `got what they wanted' experience superior treatment effects.  The flow, lump sum, and choice arms are pooled together in Sections \ref{s:mainresults} and \ref{s:heterogeneity} when we analyze the overall effects of the cash transfer arm, and are only broken out separately in Section \ref{s:choice}.

\subsection{Cost Equivalence, Before and After the Fact}

 A high-quality estimate of the costs of each program is a critical building block of cost-adjusted comparisons.  The design-based approach to cost-equivalent comparisons across programs is hampered by the fact that we can only measure costs well when the programs have been fully implemented, but the head-to-head comparison would need to know these costs at the design phase.  Anticipating this issue, we costed both programs in detail  prior to and after the intervention period, following \cite{levin2001cost}. The ex-ante costing exercise  arrived at an ex-ante cost of \$119 per eligible household. We then randomized transfer amounts at the village level to bracket this anticipated cost.  Three smaller cash transfer costing \$77, \$119, and \$152 (with beneficiaries actually receiving \$41, \$84, and \$117, respectively) are used to form the cost-adjusted comparison.  The larger cash transfer cost \$567 and transferred \$517.  We pre-committed to a regression structure that linearly adjusts the cash cost to form the exact comparison to the final, ex-post costing figure.

The ex-post costing was conducted using the `ingredients method', valuing inputs at their opportunity costs \citep[see][for more discussion]{levin2001cost,dhaliwal2012research}.  The costing question is asked from the perspective of the donor (in this case, USAID), meaning that we consider the total money spent per beneficiary to achieve the benefits observed.   Overhead expenditures in the implementation chain are an inherent part of these costs, and so the lower transactions costs in getting mobile money to the beneficiary play an important role in their potential attractiveness.\footnote{Costs are inclusive of all direct costs, all indirect in-country management costs including transport, real estate, utilities, and the staffing required to manage the program, and all international overhead costs entailed in managing the Gikuriro program.  All administrative costs, including the appropriate share of the costs of maintaining international headquarters infrastructure, were included in the costing.   Beneficiary identification costs, incurred by the survey firm and identical across all arms of the study, are excluded from the cost-benefit calculation.  Monitoring and Evaluation costs, similarly, were excluded so as to be costing only the implementation component.} 

Since the Gikuriro program covers eight districts (e.g. much larger than the study population only) and many of the startup and administrative costs must be amortized over this whole beneficiary pool, we calculate cost per beneficiary in the full national program.  Because we did not want differences in scale to drive differential costs per beneficiary, we asked GiveDirectly to artificially scale up their operations and provide us with numbers reflecting the costs per beneficiary if they were running a national-scale program across eight districts, including 56,127 beneficiary households like Gikuriro.  We costed each GD arm separately, asking what the overhead rate would have been if GD had run a national program at the scale of Gikuriro giving only transfers of that amount.\footnote{Overhead costs as a percentage of the amount transferred decline sharply with transfer amount for GD because fixed costs represent a large share of their total overhead.}  This allows us to conduct the benefit/cost comparisons ‘at scale’, rather than having the artificial, multi-amount environment of the study contaminate the costing exercise across arms.

The ex-post costing exercise arrived at a cost per beneficiary household for Gikuriro of \$141.84.  Actual GD costs were \$66, \$111, and \$145, and \$567, meaning realized Gikuriro costs were within the range over which we randomized but 28 percent higher than the average of the three smaller arms.  Cost per beneficiary is not the natural counterpart to the Intention to Treat in cost/benefit terms, however, because the ITT ignores non-compliance and gives average outcomes in the entire eligible sample.  Hence, to calculate cost per eligible household (which is what was actually spent to generate the outcome observed), we need to adjust for non-compliance in costing.  We define `averted' costs, as those which are not spent on a household if they do not comply with treatment; and `non-averted' costs which will be expended whether or not the household complies.  For GiveDirectly all costs are averted, while for Gikuriro all variable costs except for the village-level behavior change component are averted.   
Gikuriro's non-averted costs are those conducted at the village level, namely (WASH and BCC), which drive 40 percent of the total cost of the program.  The averted costs are those arising from direct household treatment; for these we multiply the cost per beneficiary times the compliance rate to calculate the cost per eligible and cost per household overall.   Using this approach we can recover a cost-equivalent comparison even if the compliance rates are different across arms.

Table \ref{t:cost_compliance} provides the exact costing numbers arrived at by the ex-post exercise.  Gikuriro treatment rates are 80 percent among eligibles and 19 percent in the population as a whole.  Given an actual cost to USAID of \$141.84 per beneficiary and the role of averted costs, this gives a cost of \$124.49 per eligible household and \$28.02 per household in the village.  GD compliance rates among eligibles range from \%81-\%91 across arms, and among ineligibles are \%18-\%19.  Given this, GD costs per eligible household are \$54, \$96, \$121, and \$517, and costs per village household overall are \$12, \$21, \$27, and \$100.

We use these estimates to regression-adjust outcomes for the differential cost from Gikuriro, providing a way of testing the differential impact of the programs at identical cost to the donor (approach described in detail in the next section).  To compare cost-effectiveness across scale, we conduct F-tests for differences in the impact/cost ratio across all the arms of the study, each of which operated at a different cost.  We present comparative impacts across a wide range of outcomes, providing nuance as to the specific outcomes chosen by beneficiaries in the cash arm relative to those that the in-kind intervention drives.   We provide a number of ways to consider targeting heterogeneity, including standard interaction analysis, the use of machine learning matching algorithms to recover individual treatment effects, and analysis of the choice experiment conducted within the cash arm that speaks to the extent to which the beneficiaries themselves understood the heterogeneity in treatment effects.

%% file: 3-Gikuriro-Analysis.tex
%!TEX root = ./Gikuriro.tex

\subsection{Attrition and Balance of the Experiment}
Endline outcome measurement is subject to a number of distinct forms of attrition; we start our empirical analysis by considering each in turn.  The most straightforward of these is standard household-level attrition, meaning that a household sampled into the baseline survey attrited from the endline survey.  In Table \ref{t:attrition}, we see that overall rates of attrition at the household level were low, around 3.3 percent in the control.  We see the pattern typical in RCT studies where attrition is somewhat lower in the treatment groups (where both ongoing contact and a sense of reciprocity may keep individuals in the endline), but these differentials are small, from 0.89 percentage point in the GD `small' arm to 1.7 percentage point in the GD `large' arm; only the latter is significant, and only at the 10 percent level.  Looking at the other covariates of attrition in column 2 we see that attriters and non-attriting households are similar.  Hence we conclude that household-level attrition is unlikely to be a source of bias in the study.

When we turn to the analysis of individual-level outcomes in Columns (3)-(9) the picture is more complex because many of the primary and secondary outcomes are only measured for certain types of individuals (anthropometrics for children, birth outcomes only for those pregnant).  We analyze four types of individual missingness that may occur.  First, we compare the attrition of all household members from the roster in the household survey; both the rates and the differentials here are very similar to the household attrition problem suggesting that there has been little additional differential attrition of individuals.  Next we examine the anthropometric panel, whereby all children under 6 at baseline who were given anthropometrics at the baseline should have been followed up with at endline.  Here the absolute rates of attrition are a little more than double what they are for individuals overall, presumably because of the greater difficulty of finding and measuring children for this exercise.  More concerningly, the decline in attrition in the treatment groups now becomes strongly significant, particularly for Gikuriro villages (which perhaps is evidence of the superior monitoring of malnourished children taking place in those villages).  Given this significance, we follow our Pre-Analysis Plan in also presenting results for the anthropometric impacts that are corrected by inverse propensity weights to correct for the observable determinants of selection.  Third, we examine whether individuals who should have been anemia tested in the followup were; here we see no evidence of differential attrition across arms. Finally, we examine the likehood that a new household member appears (typically due to births subsequent to baseline), and find no significant differences.  Overall, then, differential selection across treatment arms is not a major problem for study outcomes other than anthropometrics.  We return to the issue of unequal attrition in anthropometrics in the following section.

Next we present the baseline comparison of primary and secondary outcomes as well as household-level control variables, using the attrited panel sample that will be the basis of the endline evaluation.  The regressions used here mimic as closely as possible the impact regressions, using fixed effects for randomization blocks, including a battery of baseline control variables, weighting to make the sample representative of all eligibles, and clustering standard errors at the village level to account for the design effect. All tables also present p-values adjusted for False Discovery Rates within outcome families (primary or secondary) using the technique of \cite{anderson2008multiple}; stars in the tables are based on clustered standard errors.  Looking first at balance on study outcomes, Tables \ref{t:balance_primary}, \ref{t:balance_secondary} show balance for primary and secondary outcomes, respectively.  Overall balance on secondary outcomes is excellent, but we do find evidence that we have superior outcomes on child anthropometrics in the GD Large arm.  HAZ and WAZ are about .2 standard deviations higher than the control in this arm, and the latter difference is significant at the 10\% level even after adjusting for multiple inference.  Given this imbalance, we present all core results on anthropometrics using an ANCOVA structure (controlling for baseline values), focusing on impacts on children born prior to baseline.  Table \ref{t:balance_RHS} examines balance over baseline covariates and finds all treatment arms to be comparable to the control across all covariates.

 Compliance with Gikuriro is complex because of the multi-dimensional nature of the program.  Table \ref{t:GK_Compliance_Determinants_eligibles} examines the determinants of participation with specific sub-components of Gikuriro, based on self-reports within the eligible population.   We have five forms of participation that can be ascribed directly to the program:  participation in three types of training (nutritional, cooking/hygiene, and agricultural extension), whether households have themselves harvested the nutrient-rich household vegetable plots that they were trained by the Farmer Field Schools to grow, and whether they received livestock directly from Gikuriro. Slightly more than half of eligible households report receiving nutrition or agricultural training, half receive hygiene training and harvest FFLS gardens, and a third receive livestock.  The two driving determinants that emerge from this table are that Gikuriro was successfully targeted at the poorer households even within the (relatively poor) eligible group, and that they met more success with households headed by younger individuals.  Conditional on this, however, they were not differentially successful at reaching households with children, or female-headed households.

\subsection{Basic Experimental Results on Eligibles}\label{s:itt}

Tables \ref{t:itt_primary} and \ref{t:itt_secondary} present the core intention-to-treat results of the study on primary and secondary outcomes for the eligible population.  The analysis is conducted as an ANCOVA, including the lagged dependent variable where available, fixed effects for the randomization blocks, and for each outcome including a set of baseline controls selected by post-double-LASSO \citep[as described in Online Appendix \ref{ss:BCH}]{BelCheHan14restud} for it by LASSO as described in the Online Appendix.  For parsimony, our main tables include a dummy for the Gikuriro treatment, another for the Large GD treatment, and then a single dummy that indicates the three smaller GD transfers.  Standard errors are clustered at the village level to reflect the design effect, and we also present ``sharpened'' $q$-values that adjust inference to control the False Discovery Rate within each table \citep{anderson2008multiple}. Observations are weighted to make the analysis representative of all eligible study households, incorporating both survey weights arising from the number of eligible households per village, and the intensive tracking weights.  Panel A  of each table analyzes household-level outcomes, and Panel B individual-level outcomes.

Taking the Gikuriro treatment first, we see no statistically significant impacts on primary outcomes at the household level.  Estimates are sufficiently precise to allow us to rule out impacts on consumption that would be sufficient to justify the program.  For example, the upper bound of a 95 percent confidence interval for Gikuriro's impact on household consumption is 0.086, ruling out consumption gains of more than 9 percent (equivalent to 0.064 standard deviations of consumption). Neither are there significant impacts on household wealth (point estimate 0.01; upper bound of 95 percent CI 0.36 equivalent to 0.086 standard deviations), or, in secondary outcomes, measures of physical asset ownership.    Dietary diversity, anthropometrics, and maternal anemia all move in the right direction but none of these changes is significant. 95 percent confidence intervals for HAZ and WAZ rule out impacts outside of the ranges of (-0.03, 0.13) and (-0.04, 0.12), respectively.\footnote{For comparison, these confidence intervals are sufficiently precise to rule out the estimated one-year HAZ impacts of 0.15 sd from a nutritional intervention from recent work in Bangladesh, but are essentially centered on the 0.05 sd one-year HAZ impacts of a combined nutrition and WASH intervention in the same study \citep{Lub08lancet}. A parallel trial in Kenya found its nutrition intervention to have a one-year impact of 0.11 sd, and its nutrition and WASH intervention to have a one-year impact of 0.12 standard deviations \citep{Null18lancetgh}.} Examining impacts on outcomes more directly on Gikuriro's causal path, household savings increase by a 109 percent, consistent with the creation of SILCs, while no impacts appear for health knowledge or sanitation practices.  Hence the program has been successful in moving an indicator closely related to one of its sub-components, but at least within the time-frame of the study these changes in savings have not translated into significant improvements in the core welfare outcomes.

We turn next to  to the impact of the three smaller (``Main'') GiveDirectly arms whose average cost per eligible is \$90.28, 72 percent of the cost of Gikuriro.  Transfers of this magnitude do not the primary household or child- and maternal-health outcomes, though we see quite a different set of secondary outcomes move, relative to Gikuriro's in-kind programming.  For instance, point estimates of household consumption impacts of 0.06, and corresponding 95 percent confidence interval upper bound of 0.24, rule out impacts equivalent to 0.18 standard deviations or greater; we similarly rule out HAZ and WAZ impacts in excess of 0.06 and 0.07 standard deviations of the reference population, respectively.   Instead of increasing savings, small GD transfers lead to a 77 percent pay-down of debt, and an increase in the value of productive and consumption assets, by 26 percent and 35 percent respectively.  
Thus far, then, the comparison of Gikuriro to cash breaks down into two distinct dimensions of improvement, each of which has a different and entirely plausible pathway to long-term improvements:  savings (Gikuriro), or debt reduction and asset investment (GiveDirectly).   \footnote{Table \ref{t:itt_primary_Each_GD} shows the experimental results with each of the three treatment cells within the `main' treatment broken out individually.  Because these cells are both small and feature treatment amounts that are similar, this more granular analysis does not turn up evidence of meaningful heterogeneity across the three smaller GD transfer amounts.}

When we examine the third column, however, a more transformative impact arising from the Large cash transfer is clearly apparent.  Not only do omnibus measures of consumption and wealth go up across the board, but metrics of consumption closely linked to child health improve.  The dietary diversity score increases by 0.55 food groups, or by 12 percent off a base of 4.77.  Figure \ref{f:foodgroups_endline} displays the fraction of each arm consuming each food group, and shows the treatment effects of the large arm to be most pronounced in fish, fruits, oils and fats, spices and condiments, and cereals.  Productive assets increase by 80 percent, consumer assets almost double in value, and home value increases substantially.  In the individual outcomes the benefits of this surge in consumption are evident as well; within the course of one year we see a 0.09 SD improvement in HAZ, a 0.07 SD improvement in WAZ, and a 0.13 SD improvement in MUAC, all significant at least at the 10 percent level, prior to adjustment for multiple inference.\footnote{These improvements should be viewed against the backdrop of a sharp deterioration in anthropometrics subsequent to birth that typically occurs in LDCs, leaving rural African children often two full SDs below the international norm by age 3 \citep{shrimpton2001worldwide, victora2010worldwide}.}  The ANCOVA specification should be particularly important in the analysis of the anthropometric indicators that showed signs of imbalance at baseline; indeed if we examine these outcomes in post-treatment levels we see substantially stronger apparent treatment effects.  Appendix Table \ref{t:itt_stunting_wasting} analyzes the binary anthropometric outcomes of stunting and wasting, and finds impacts very consonant with the continuous impacts on HAZ and WAZ.  The GD large transfer reduces stunting by 6 pp on a base of 50\% and wasting by 5 pp on a base of 16\%, both effects significant at the 10\% level.   Anemia falls slightly (not significant) and there is a substantial decrease in child mortality of almost 1 percentage point(or 70 percent off of the baseline value).  To contextualize these effects using unweighted numbers, the control group eligibles saw 13 cases of child mortality out of 2,596 children (0.5 percent) while the GD Large arm saw 2 cases out of 1,200 children (0.16 percent).   Hence the GD Large arm saw meaningful improvements in consumption and child health.  

In the final columns of each table we provide $p$-statistics on $F$-tests that the ratio of the benefits across arms differs from the ratio of their costs.  These statistics ask whether we can reject that the impacts scale in a manner similar to the costs.  For the comparison across the two GD arms we find two significant differences:  only in the case of debt reduction (where small transfers have a big effect and big transfers do not) and house quality (where small transfers have a negative and large transfers a positive effect) can we reject cost-proportional benefit scaling for cash transfers.  In comparing Gikuriro to the GD large arm we see more differences, with Gikuriro being more cost-effective in generating savings and GD Large superior in consumption, as well as productive and consumption assets. 

One of the most fundamental ideas in development economics is that poor households should have a single `shadow value' of cash which pulls down investment in all capital-hungry endeavors in a symmetric way.  The above findings are generally consistent with this view of the world, as an intervention that relaxes credit constraints leads to shifts in consumption patterns that are very broadly spread across domains.  This property means that small cash transfers are hard to detect because they move too many outcomes by too small an amount to be significant, while large cash transfers result in a broad-based increase in consumption in many dimensions.

Our pre-analysis plan states that for any outcomes where we find differential attrition, we would estimate a propensity to remain in the sample incorporating covariates, dummies for treatments, and their interactions on the right-hand side, and then re-weight the analysis by the product of the inverse of this probability and the standard sampling weight.  This procedure corrects the impacts for the observable determinants of attrition, and uses regression weighting to attempt to make the treatment and control samples comparable on important covariates even after attrition.  Because we primarily found significantly differential attrition for the anthropometric outcomes, in Table \ref{t:ITT_IPW} we present the results of this correction.  We interact with each treatment dummy the same right-hand side covariates we use the same controls in the anthropometric regressions:  gender, a linear, quadratic, and cubic for age in months, baseline household wealth, and a dummy for membership in Ubudehe poverty category 1.  The results in this table are virtually identical to Table \ref{t:itt_primary}, indicating that the types of children who attrited from the study are similar across arms and hence differential attrition is unlikely to be driving our impacts.

\subsection{Cost-Equivalent Benchmarking}

The core purpose of this experiment is comparative; namely  to exploit the randomized variation in transfer amounts to conduct an exact cost-equivalent benchmarking exercise.  Using the costing numbers emerging from the ex-post exercise, we use the observed costs, overhead rates, and compliance rates to calculate the donor cost per eligible household in each arm of the study.  We then estimate the impacts of Gikuriro on eligible households relative to an exact donor-cost-equivalent cash transfer in a specification that linearly interpolates cash transfer impacts with respect to the randomized transfer variation. 

To present our cost adjustment strategy more formally, let the subscript $i$ indicate the individual (or household, depending on unit of analysis), $c$ the cluster (village), and $b$ the randomization block.   
Outcomes are observed both at baseline $(Y_{icb1})$ and at endline $(Y_{icb2})$.    First, begin with the total GD donor cost per eligible within each transfer amount arm, denoted by $t_c$.  Subtract from this number the benchmarked Gikuriro cost per eligible household $C$ described above, and denote the difference $t_c-C=\tau_c$; this is the deviation (positive or negative) of each GD arm from the benchmarked Gikuriro cost.  Set $\tau_c$ to zero in the control and Gikuriro arms.  We can then run an ANCOVA impact regression as above, but controlling for a linear term in $\tau_c$, an indicator $T_c$ for receipt of either treatment, and an indicator $T_c^{GK}$ for receiving Gikuriro.  We estimate this specification, as illustrated in equation \eqref{eq:CostEquiv} below, on eligible households in our sample:

\begin{equation}\label{eq:CostEquiv}
Y_{icb2}=\alpha_b  + \delta^T T_c + \delta^{GK} T_c^{GK} + \beta X_{icb1} + \rho Y_{icb1} + \gamma_1 \tau_c  + \epsilon_{icb2}. 
\end{equation}

Because $\tau_c$ absorbs the deviation of the GD arm from the benchmarked Gikuriro cost, the coefficient on Gikuriro treatment, $\delta^{GK}$, will provide a direct benchmarking test:  this estimates the \emph{differential} impact of Gikuriro benchmarked against an exactly donor-cost-equivalent cash transfer.  In addition, subject to the assumption of linear transfer amount effects, the slope coefficient $\tau_c$ captures impacts arising from deviations in GD cost from Gikuriro cost, and the coefficient $\delta^T$ estimates the impact of GD at the cost of Gikuriro.  

A graphical representation of our strategy is provided in the left-hand panel of Figure \ref{f:CE_CEff_picture}, which plots the average outcome on the y-axis for all four GD treatment amounts (colored circles), for GK (gray diamond), and the control (gray circle).  The line represents the fitted average savings by GD transfer amount.  By predicting the outcome on this line at the exact cost of Gikuriro (hollow circle), the benchmarked differential is then the vertical difference between the Gikuriro impact and the projected cost-equivalent GD impact.

The results of the cost equivalent analysis  for primary and secondary outcomes are presented in Tables \ref{t:ce_primary} and \ref{t:ce_secondary}, respectively. 

The first column of these tables contains the heart of the comparative benchmarking exercise, comparing the effects of Gikuriro's in-kind programming to cash at exact donor-cost-equivalent levels. Gikuriro is significantly less effective than a cost-equivalent cash transfer at driving consumption levels, producing approximately 19 percent lower consumption.  At cost-equivalent levels, we find no statistically significant difference on other primary outcomes.  These null findings are powered to detect economically meaningful differences:  for instance, the 95 percent confidence interval for impacts on the household dietary diversity score rules out differences of a magnitude greater than approximately 0.25 in either direction---an increase of one quarter of a food type on average, a seemingly reasonable aspiration for an intervention seeking to change nutritional practices. Likewise, we find no statistically significant difference in child-growth outcomes between the interventions (95 percent confidence interval: -0.058, 0.098); by contrast, the addition of a nutrition component to a water, sanitation, and hand-washing intervention in Kenya produced a difference of 0.17 in weight-for-age z-scores \citep{Null18lancetgh}. 

We find statistically and economically significant differences in impacts at cost-equivalent levels across a range of secondary outcomes.  Gikuriro is less effective at driving the paydown of debt and the accumulation of assets, while the in-kind program is significantly more effective than cash at creating savings.  The differential effect of the programs on savings and borrowing is interesting, and suggests that while both interventions serve to improve the net stock of liquid wealth (savings net of borrowing), the focused push on SILC groups in Gikuriro drives this more strongly through the vehicle of new savings while households making their own choices are more strongly disposed to reduce debt instead. Which of these strategies makes more sense?  A simple comparison of interest rates is revealing.  Gikuriro SILCs were free to set their own interest rates, but typically paid about 5 percent per annum nominal.  Credit interest rates, by comparison, vary from an average of 22 percent in the MFI sector to upwards of 60 percent in informal credit markets.  Given that 32 percent of eligible households reported having both borrowing and savings at baseline (and 79 percent had either borrowing or saving) it seems that the desire to pay down debt might be warranted.

The second column of tables \ref{t:ce_primary} and \ref{t:ce_secondary} presents estimated coefficients on an indicator for assignment to any treatment (coefficient $\delta^T$ in equation \eqref{eq:CostEquiv}).  This intercept term estimates the impact of cash transfers at a cost equivalent to Gikuriro, although this precise amount was not included in the experiment.  Given that the mean transfer amount in the `small' arm is only slightly lower than the GK cost, this estimate looks generally similar to the second column of the ITT tables (the simple average experimental effect across the `small' transfer amounts).  At the exact cost of Gikuriro, we estimate that cash transfers would have led to a significant 73 percent decrease in the stock of debt, and a 30 percent and 40 percent increase in productive and consumption assets, respectively.

The third column of tables \ref{t:ce_primary} and \ref{t:ce_secondary} provides a direct estimate of the marginal effect of an additional 100 dollars in donor cost on primary and secondary outcomes, respectively.  As could be inferred from ITT estimates, this coefficient is strongly significant across a wide range of outcomes, particularly those most related to household consumption.  An extra 100 dollars leads to a 5 percent increase in consumption, a 9 percent increase in dietary diversity, a 19 percent increase in savings, an 13 percent and 14 percent increase in productive and consumption assets, respectively, and leads housing value to improve by 5 percent and the index of housing quality to increase by 0.1 SD.  In terms of anthropometrics, the change in value of transfer is positive but small in absolute magnitude for HAZ, WAZ, and MUAC, and does not survive correction for multiple inference.  An extra \$100 per beneficiary household---with eligible households containing an average of 2.7  children under the age of six---increases HAZ by 0.022 standard deviations.  Beyond this, none of the other individual outcomes respond to transfer amount in a manner that we can reject at 95 percent significance.  

While we pre-specified the simple linear functional form for interpolation of cash-transfer impacts to preserve statistical power, a natural question is the extent to which our benchmarking results are sensitive to the linear interpolation of cash-transfer impacts. 
To interrogate this, Tables \ref{t:ce_primary_robustness} and \ref{t:ce_secondary_robustness} present seven different ways of forming the cost-equivalent comparison.  Column 1 in these tables repeats the linear specification used elsewhere, column 2 uses a quadratic, and column 3 a cubic function in project cost, and the remaining columns use the linear specification but drop one of the cash transfer arms in each column.  The table reports only the differential parameter of Gikuriro over cost-equivalent cash reported in Column 1 of Tables \ref{t:ce_primary} and \ref{t:ce_secondary}, and the associated standard error. Overall, the results prove highly robust to specification; all of the outcomes significant in the main specification are significant in at least five out of six of the remaining specifications, except for the differential on the sanitation practices index.  These results also confirm the power gains arising from linear interpolation:  standard errors for the differential effect of Gikuriro are generally substantially smaller in the linear specification than in quadratic or cubic specifications.  To give a sense of magnitudes, to obtain a reduction in the variance of the cost-equivalent comparison for the consumption outcome equivalent to that arising from the linear specification, a researcher using a cubic specification would have to increase the sample size by 158 percent.  Although our transfer amounts (with three closely bunched together and one much larger) do not provide a great deal of power to test for non-linearity, this exercise suggests that our core results are robust to alternate ways of forming the cost equivalence comparison.\footnote{One new result that emerges from these tables is that in some specifications Gikuriro is superior to cost-equivalent cash at improving HAZ, significant at the 10\% level in three out of seven specifications.}

\subsection{Cost Equivalence versus Cost Effectiveness}

This study is designed to make a specific form of cost-equivalent comparison, namely the impact of a cash transfer intervention assessed at the exact cost of an in-kind intervention.  This comparison fixes the amount to be spent per beneficiary and asks which intervention is more effective.  A different but related question is that of cost effectiveness, where we compare programs that operate at different costs and ask which generates the greatest benefit per dollar spent---potentially making comparisons between programs that operate at radically different levels of resource intensity.  Fixed costs and indivisibility in program design mean that cost-benefit ratios do not represent alternatives that can be delivered for a given budget to a given population, but instead represent possible gains to an additive welfare function for policymakers who are indifferent to the size of the beneficiary population (for a given budget) or are willing to adjust their budget (for a given population).\footnote{Under the linear specification of equation \eqref{eq:CostEquiv}, the question of whether the benefits of cash scale proportionally to expenditure amounts to the hypothesis that $\delta^T=\tau_c C$. We fail to reject this null across all primary outcomes.}

The comparison between these two approaches is represented graphically in Figure \ref{f:CE_CEff_picture} for the primary study outcomes.  In the left-hand panel we plot comparative cost-equivalence; here the focus is on the value on the X-axis that is the ex-post cost of Gikuriro, and we are interested in seeing which is greater, the observed benefit of Gikuriro (represented by a black diamond) or the predicted benefit of cash at this cost (the hollow circle).  In the right-hand panel we connect the shaded circle that represents the outcome in the control group (at zero cost) with the outcome in each arm; because the plot represents outcomes in benefit/cost space, the intervention that features the steepest slope in this graph has the highest cost effectiveness.  Interestingly, this graphic illustrates that while in general there are not substantial differences between Gikuriro and cash at benchmarked cost, because the smallest cash arm is so inexpensive while producing outcomes that are generally as good as (or better than) more expensive treatments, for four out of the five outcomes represented the smallest cash transfer has the highest cost effectiveness.  The difference in cost effectiveness across arms is tested statistically for all primary and secondary outcomes in the final columns of Tables \ref{t:itt_primary} and \ref{t:itt_secondary}.

Table \ref{t:bcr_primary} provides a statistical analysis of the benefit-cost slope terms represented in the right-hand panel of Figure \ref{f:CE_CEff_picture}, pooling the three smaller GD arms together as done in the rest of the paper.  It presents the ITT coefficient from the prior tables divided by the cost of the arm, and so gives the household improvement generated per \$100 spent through that modality, as well as the corresponding standard error on this BCR.\footnote{Given that individual-level outcomes in this table represent average benefits for amount spent on the household, if we want the BCRs at the individual level we need to scale up the coefficients in this table by the average number of individuals per household for each outcome.}    The final three columns of this Table \ref{t:bcr_primary} provide statistical tests of the difference in  cost-effectiveness slopes and show how difficult it is to power a study to reject these; despite the relatively large sample size of this study, it is only for the consumption outcome that we are able to reject equality of CBRs across interventions, with the cash interventions outperforming Gikuriro.  Between the two cash arms we are unable to reject equal BCRs for any outcome, indicating benefits that scale linearly with cash transfer amounts.

\subsection{Total Causal Effects}
Because we randomly sampled from among ineligible as well as eligible households, we can also conduct an analysis representative of the population of study villages by pooling the strata together and using population sampling weights.  The average weight in the ineligible sample is 24.4 and in the eligible sample it is 2, meaning that while the unweighted eligible sample is larger, it is the ineligibles who will dominate the weighted sample.  Recall that the treatment effects on ineligible households may arise either from the treatment and targeting of the two interventions among ineligible households (with saturations set to be the same at just over 18 percent on average in both arms, but with targeting differing), or from spillovers between eligible and ineligible households.  With 11.4 percent of all households being defined as eligible, the treatment rate in the ineligible sample is 8.4 percent.  This means that the large majority of the additional sample included in the TCE analysis only receive impacts through spillover effects to untreated households.

These impacts are presented in Table \ref{t:tce_primary} and \ref{t:tce_secondary} for primary and secondary outcomes, respectively. Here, the overall picture is very different from the impact among eligibles.  For Gikuriro, we see improvements at the 99 percent significance level in the index of health knowledge, a core component of the program and one which was broadly targeted at the village population by the program (as reflected by our accounting of these costs as `non-averted').  Vaccinations, presumably provided by government health facilities but not a central focus of Gikuriro, also improve significantly.  So there is some real evidence of holistic benefits in health-related domains for the population of Gikuriro villages.  These changes, it is true, do not translate into observable improvements in health outcomes for children or adults within the timeframe of the study, but still suggest that Gikuriro implementers have been successful in driving community-level health knowledge.

With cash transfers, on the other hand, improvements appear to be more narrowly limited to the beneficiaries of the transfers.  The `small' transfers do not move any village-level outcome at the 5 percent significance level, even before adjustment for multiple inference.  The `large' transfers, so positive among beneficiaries, in general see negative signs across the consumption indicators, lead to a significant drop in savings at the village level, and are only positively associated with vaccination rates.   On net there is little evidence that the widespread benefits observed in the eligibles carry over to the broader population of the village when the transfers are targeted at a relatively small fraction of the households.

A comparison of the ITT and TCE, particularly for the GD Large arm, suggests that for a number of outcomes (such as dietary diversity and savings) we see a swing from positive effects on the eligibles to negative point estimates for the village as a whole.  This raises the question as to whether the GD treatment may be having negative spillover effects on ineligibles.  We are able to analyze spillovers experimentally within a subset of ineligible households.  Because the cash transfers were provided only to ineligible households who were in Ubudehe 1 and 2 and had 3 or more members (and thus might contain children), wealthier or childless households were never treated by GD.  Therefore, by comparing ineligibles either in Ubudehe 3 and 4 or with fewer than three members between the GD and control villages, we have a clean window into the spillovers effects of cash.\footnote{Individuals in all Ubudehe categories report receiving some benefits from Gikuriro, so we have no comparable ability to separate spillovers from treatment of ineligibles in that arm.}  The results of this analysis are presented in Tables \ref{t:spillovers_primary} and \ref{t:spillovers_secondary}.  The GD Small transfer has no detectable spillover effects on any outcome, and while the GD large does show negative spillovers on savings and positive spillovers on vaccination rates, no systematic picture emerges.  So, recognizing that this test is relatively low-powered, we conclude that in this case cash transfers do not appear to have resulted in meaningful spillovers on smaller, wealthier households in treatment villages.

\subsection{Benchmarked Total Causal Effects}

We can perform a similar cost-equivalent benchmarking exercise for the village-level TCE, adjusting now by cost per household in the overall village from Table \ref{t:cost_compliance} rather than the cost per eligible household.  This allows us to ask how the two programs differ in their impact on the village population when the same amount is spent by each program per household in the village.  This analysis is presented in tables and \ref{t:ce_tce_primary} and \ref{t:ce_tce_secondary}.  In general we lack the power to make many clear comparative statements across interventions in terms of TCEs.  Before multiple inference correction Gikuriro is significantly better at generating health knowledge in the population, but neither this result nor even any of the expenditure slope terms retain their significance once we look at the sharpened $q$-values.  Clearly, the implication from this result is not that these outcomes would be unmoved by large cash transfers to the whole population (as shown in \citep{egger2019general}), but rather that no arm in this study actually moved average village-level costs by a large enough sum to drive detectable population-level outcomes.

%% file: 4-Gikuriro-Heterogeneity.tex
%!TEX root = ./Gikuriro.tex

The average comparative impacts of cash transfers relative to Gikuriro's in-kind programming may mask substantial heterogeneity in relative impacts across groups.  This has (at least) two consequences.  First, it implies that decisions over intervention modality entail tradeoffs between population subgroups:  there can be both winners and losers from the switch from kind to cash. If policymakers place particular weight on certain subgroups, then these tradeoffs may change their preferences over modalities. Second, to the extent that this heterogeneity is predictable based on observed characteristics, policymakers may be able to improve on the average effects of both programs by finely targeting benefit types to individuals or groups on the basis of predicted relative impacts.  Here, we test for heterogeneity on pre-specified dimensions of heterogeneity anticipated to matter for relative efficacy, as well as by using machine learning methods to identify subgroups for whom cash is advantageous over in-kind programming, and vice-versa.   
We find that there is little heterogeneity on attributes selected ex ante, and that while machine-learning methods do deliver heterogeneity in impacts for each group, these bear little correlation across outcome dimensions, suggesting that targeting modalities on the basis of expected benefits is a promising approach for policymakers with narrow objective functions, but---at least in this instance---that even fine targeting offers only modest gains for broad welfare metrics.

Our pre-analysis plan highlighted two forms of heterogeneity that we anticipated would be important at the design phase; namely how baseline malnutrition and child age may moderate the impact of nutritional interventions.   Given that we have children who start the study outside of the first 1,000 days (those 2--5 years old at baseline), we might expect that the impact of the program on these more fully developed children would be smaller.  Similarly, we might expect that both of these interventions would be most effective for children who began the intervention most malnourished.

Our analysis generally reveals a lack of heterogeneity, in that impacts are not larger for children most malnourished at baseline (Table \ref{t:het_by_anthro}), or for children exposed to the treatments at younger ages (Table \ref{t:het_by_age}).  Figures \ref{f:HAZ_gd_large_fan} and \ref{f:anemia_gd_large_fan} provide fan regressions of impacts by child age and there is a suggestion that children exposed to large cash transfers in utero realize the largest benefits.\footnote{This pattern is similar to the medium-term results in \cite{ozler2016money}, who find unconditional transfers in Malawi to have the largest effect on children exposed in utero.} In general, however, our results are not suggestive of strong age- or malnutrition-driven heterogeneity of these interventions.\footnote{The longer-term literature has typically found impacts of large cash transfer programs on HAZ in the range of 0.2--0.45 standard deviations \citep{aguero2006impact,barham2014schooling}; comparison to the broader literature suggests that these impacts may grow over time.} 

While these pre-specified dimensions are not drivers of differential outcomes, it remains possible that there be substantial heterogeneity in the impacts of cash transfers relative to in-kind transfers. This heterogeneity may simply be poorly predicted by these pre-specified covariates.  To the extent that it \emph{is} predictable on the basis of observable attributes, however, then predictions of relative impacts that differ by subgroup might be used to target aid modalities to improve their impacts.  

Modern methods for using machine learning techniques to develop predictive models of heterogeneous causal effects  are well suited to this problem.  In this section, we deploy the \emph{causal forest} approach of \citet{WagAth18jasa} to estimate the relative effects of cash versus in-kind programming.  We first characterize the extent of heterogeneity across observable subgroups in each of the core outcome dimensions. Though there is substantial heterogeneity, we then demonstrate that it is limited in its correlation across outcome dimensions.  Turning to welfare, we show that for a simple metric of welfare gains, tradeoffs in impacts across outcome dimensions mean that even policymakers who are able to individually assign cash versus kind are limited in their ability to achieve broadly defined welfare gains by `personalized' choice of benefit type in this setting.

We estimate the extent of heterogeneity in treatment response in the sample by generalized random forest \citep{AthTibWag19annsat,AthWag19obs}.  To ensure that treatment effect estimates are centered on point estimates for the full sample as were reported in Section \ref{s:itt}, we begin by residualizing both the outcomes studied and an indicator for the GiveDirectly `Main' treatment on the basis of all covariates used in the ITT analysis of that section.  We then restrict attention only to the Gikuriro and GiveDirectly Main treatment arm; a regression of primary outcomes on the residualized GD-Main indicator alone in this sample replicates the point estimates of Table \ref{t:itt_primary}.  In this estimating sample, we allow the same candidate covariates used in the Post-Double Lasso to act as candidate effect moderators. 

Appendix Figure \ref{f:heterogeneity} displays estimates of treatment effect heterogeneity in groups defined by this approach.  Moderate point estimates often mask substantial heterogeneity across observed characteristics.  These estimates suggest that even where average effects are small, a policymaker who values a single dimension of these outcomes in particular can achieve substantial welfare gains by optimizing choice of modality.

However, welfare is surely not one-dimensional, and typically policymakers will value outcomes across multiple dimensions.  If this is the case, then the ability to optimize effectively will depend on the extent to which heterogeneity in the relative benefits of cash versus kind is correlated across dimensions. If individuals who benefit from cash in one outcome dimension also tend to benefit from cash in other dimensions, then it will be possible to assign treatments in ways that benefit them across the board. On the other hand, there may be no such free lunch:  if individuals' treatment responses differ across outcome dimensions, then there may be no assignment choice that uniformly raises these outcomes.  

In Table \ref{t:heterogeneity_crossoutcomes} we illustrate that the extent of these cross-outcome correlations in the sign of treatment responses are low. 
None of the correlations have an absolute value greater than .3, indicating that no single set of determinants predict benefits across a wide range of outcomes.  Perhaps unsurprisingly households that see larger wealth benefits see smaller consumption benefits (since transfers can either be saved or spent).  Interestingly there does seem to be a positive correspondence between the types of households effective in generating wealth and those that see child growth improve.\footnote{The determinants of impact heterogeneity across the three biometric measures of child growth concur very closely as shown in Online Appendix Figure \ref{f:heterogeneity_childhealth}; consequently we have collapsed the three child growth metrics down into a single variable for simplicity of presentation.} 
This is suggestive of a shared investment in long-term outcomes in response to cash transfers occurring through both the accumulation of physical and human capital in a subset of households.  
The limited concordance between treatment effect heterogeneity across outcome dimensions suggests that policymakers who value a multi-dimensional welfare metric will be constrained in their ability to target cash relative to in-kind transfers.  To illustrate this, consider a policymaker who values an additive social welfare metric combining consumption, wealth, dietary diversity, and child growth, with each measure standardized to have a variance of one.  

For policymakers who value only one outcome, optimizing targeting based on estimates like the ones above does deliver gains.  Table \ref{t:welfare_example} reports the impacts in standard deviation terms from optimal targeting.  These are modest in the case of targeting for consumption, but rise to 5.7, 8.6, and 10.3 percent of a standard deviations of the outcomes wealth, dietary diversity, and child growth, when targeting is based on these outcomes and they are the sole metric of benefit.  If optimal policies were perfectly aligned across these outcomes, the gains to personalized treatment assignment for a welfare metric that is the average of these outcomes would be equal to the average of the outcome-specific benefits.  But as Table \ref{t:welfare_example} shows, it is instead substantially smaller:  at 2.9 percent of a standard deviation of the composite welfare metric, this is approximately 44 percent of that theoretical maximum.  Tradeoffs across outcomes limit the extent to which policymakers can optimize cash versus kind regimes based on heterogeneous responses by households.

%% file: 5-GIkuriro-Choice.tex
A central argument against cash transfers, and particularly lump-sum unconditional transfers, is that they provide too large a temptation for misuse of resources.  From this behavioral perspective, the costs required to build additional rigidity into programs (by giving cash in smaller increments over time, or by giving benefits in-kind) can be justified if they deliver superior outcomes.  Even better, if those with behavioral problems are sophisticated about their own actions, then offering choice over program rigidity can permit the right individuals to `bind themselves to the mast' while allowing those without such problems to choose cheaper, lump-sum transfers \citep{ashraf2006tying}.  
We shed light on this behavioral angle by combining incentivized measures of present bias with an experiment on beneficiary choices over transfer timing embedded in the cash-transfer arm of the study.  

At baseline, we implemented an incentivized choice experiment over transfer modality within the cash arm.  All cash transfer subjects were shown a graphic representing the upcoming twelve months of the study, and for each month were asked to choose whether they would prefer to receive a lump sum cash transfer in that month, or flow transfers across all 12 months, with the payout amounts being equal.  Subjects were told that a small fraction of the sample (1 out of 12) would be randomly selected, a month would be randomly selected, and they would receive the actual choice that they had made in that month.  The remainder of the sample would then be randomly assigned to receive lump sum or flow transfers. The purpose of this exercise was not to be able to evaluate the choice arm per se (it is not adequately powered to be studied as an independent treatment), but rather to have a highly incentivized way of understanding beneficiary preferences across the two modalities.  This exercise can then be used either to test the impact of being assigned to lump sum versus flow transfers or as a tool for revealing preferences to understand whether allowing choice over transfer modality would improve outcomes.

 We combine the choice experiment, the direct randomization of transfer modality, and measures of behavioral heterogeneity to examine three core questions.  First, despite greater administrative complexity, are flow transfers more effective than lump-sum transfers at delivering child health?  This can be answered using the simple experimental comparison between those directly assigned to the flow and lump-sum arms.  Second, are the beneficiaries sufficiently sophisticated as to their own relative benefits to be able to choose the modality from which they would derive the most benefit?  This can be examined by seeing whether benefits are larger when beneficiaries are assigned to the arm that they selected in the choice exercise.  And third, are the comparative impacts of alternative modalities---cash versus kind, or lump-sum versus flow---heterogeneous according to behavioral measures?  Specifically, we examine whether more rigid interventions (such is in-kind over cash or flow transfers over lump sum) are disproportionately beneficial for individuals with self- or other-control problems, and whether such individuals who chose the flow arm are hurt when they are instead assigned to receive a lump sum transfer.

\subsection{Lump-sum versus flow transfers}

Beginning with the question of the impact of lump sum versus flow transfers, we can form a simple randomized comparison by dropping the households assigned to the choice arm and comparing those experimentally assigned to lump sum or flow to the control group.   Because of the large difference between the three smaller transfers and the large one, we analyze this experiment with four sets of dummies:  one for the main transfer cells, a second that measures the additional impact of lump-sum transfers in these cells, then a dummy for large transfers, and a dummy that measure the additional impact of large lump sum in the large arm.  We provide F-tests that give the p-values on the total effect of small lump sum and large lump sum.  

This analysis, presented in Tables \ref{t:lumpsum_flow_primary} and \ref{t:lumpsum_flow_secondary}, provides little support for the idea that flow transfers generate superior child health outcomes.  Looking first at the individual outcomes we see if anything lump-sum transfers generate slightly better child health outcomes (this is true for both WAZ and MUAC, while small flow transfers are modestly more effective at delivering HAZ).  Turning to the household analysis, we find evidence of interesting differences in the use of intertemporal assets; only when the smaller transfers are given as a flow do they lead to a reduction in debt loads.  Conversely, lump sum transfers are significantly better for generating savings.  Put together these suggest that households under the small transfer only avoid building debt when transfers come as a flow, but only save when transfers come as a lump sum.  For both small and large transfers there is a moderately greater increase in consumption assets when the transfers are lump-sum.   Taken as a whole, this analysis is indicative of muted overall differences between cash transfer modalities, with a slight edge for lump-sum transfers even on child health outcomes.  Given the greater ease and lower costs of making one-time transfers, the takeaway should be that we find no clear reason to incur the costs of flow transfers.

\subsection{Transfer modality choice}

Next we move to analysis of the choice experiment.  The choices made by households reveal an overall preference for lump-sum transfers.  In every one of the 12 months over which individuals made choices, roughly 65\% of the subjects chose lump sum.  Contrary to what we might expect if a large share of the sample were highly impatient or time inconsistent, we do not see any systematic decrease in the fraction choosing lump sum as we move from the immediate choice (month 1) to more distant months.  56\% of the subjects chose lump sum in every month and 74\% ever chose lump sum in any month.  

The differential impacts of `getting what you wanted' can be thought of as a test of \emph{essential heterogeneity}  \citep{heckman2006understanding}, in that if impacts were heterogeneous across modalities and this was well understood by beneficiaries, then we would expect to see systematically superior outcomes when the assignment was consonant with the choice.  Improved outcomes when subjects get what they choose would provide a powerful argument in favor of beneficiary-driven flexibility in modalities.  

This analysis must account both for the (endogenous) choice and the (randomized) treatment status.  The regression includes a dummy for choosing lump sum, a dummy for being assigned to the lump sum treatment, and then a dummy for `got what I wanted'.\footnote{Given this structure, the lump sum dummy gives the impact of being assigned to LS and not wanting it, the dummy for LS choice is the impact of choosing LS and not getting it, and the last dummy gives the average difference between getting and not getting one's choice.}  The outcome of this analysis, presented in tables \ref{t:choice_primary} and \ref{t:choice_secondary}, is very clear (and is the same as seen if we instead partition the analysis to look at those who originally chose LS and flow separately).  The results indicate that the receipt of the desired treatment modality has no detectable benefit in terms of the main study outcomes, and indeed appears to result in significantly worse HAZ.  This result is potentially quite important, suggesting that no meaningful essential heterogeneity exists across transfer modalities in this sample. The two potential candidate explanations for this result would be either that no heterogeneity exists, or that it exists but is not understood by the participants.  Given the lack of strong differences between lump sum and flow transfers when randomly assigned from the prior section, the former explanation may be the most reasonable.

\subsection{Does rigidity help those with self- or other-control problems?}\label{ss:ControlProbs}

We added features to the design of the experiment that let us speak to behavioral dimensions of the cash benchmarking question.  One of the most powerful motivations for in-kind programs is the idea that beneficiaries lack the control to manage large sums of cash effectively.  Whether because of behavioral self-control problems \citep{thaler2004save, ashraf2006tying} or intra-household other-control problems \citep{anderson2002economics, ashraf2009spousal}, any such failure will lead to the inability to translate one-shot lump sum transfers into the type of steady, sustained increases in consumption that are likely to deliver improvements in child nutrition and health.  In this sense, changing the cash transfer modality away from lump sum transfers and towards a steady flow of smaller monthly payments can be seen as analogous to the types of intentional rigidity that underlie in-kind programs. This logic would suggest that program rigidity would prove helpful those with control issues.  If this use of flow transfers as a way of `binding oneself to the mast' is understood by sophisticated beneficiaries, then we should see the choice to receive flow transfers, and the benefit of receiving flow over lump sum, to be particularly pronounced among individuals who are both time inconsistent and sophisticated about this fact.

To measure the relevant behavioral heterogeneity, we conducted an incentivized Convex Time Budget (CTB) exercise \citep{AndSpr12aer_estimating, GinGolSilYan18ej} that provides an individual-level measure of discounting ($\delta$) and time inconsistency ($\beta$) for each individual. This exercise asks participants to allocate a sum of money between being received immediately and received 30 days in the future, varying both the interest rate and the starting point for the 30 days (today or 90 days from now).  To measure self-control, we define a dummy variable for `impatient' and a dummy variable for `time inconsistent' based on a straightforward use of the data from the Convex Time Budget exercise.\footnote{Specifically, we count an individual as impatient if they elect to receive money today rather than in 30 days despite being paid double to wait, and as inconsistent if they allocate more money to `soon' when they are faced with the same choices in 90 versus 120 days.   We estimated structural utility models for the CTB using three different methods:  Logit, Tobit, and Non-Linear Least Squares. The estimates from these different models correlate poorly with each other and with survey-based measures of impatience and time inconsistency, so we have elected to use the CTB questions in the simplest way possible. Using these measures 25\% of the sample is impatient, and 67\% is inconsistent.  Results are similar if we instead use a survey-based measure of self-control problems.}  To measure other-control, we use the responses to a set of baseline questions that measure the extent to which an individual reports financial threats from others around them.\footnote{An individual was counted as lacking other control if they:  (a) did not want to keep money around the house because of a lack of trust of others, because of thieves, or relatives who might steal from them, or (b) reported that they had a spouse with whom they had conflict about money, or who was `wasteful' or `irresponsible'.  By this metric 27\% of our sample lacks other control.}

We have three ways to address this question.  First, do individuals with self- or other-control problems benefit from the rigidity of receiving a program in kind rather than in cash?  This can be asked by comparing the interaction effects of these attributes with indicators for the Gikuriro and the smaller GD arms (omitting the large GD arm for comparability).  Second, within the cash transfer arm do they benefit from the comparative rigidity of receiving transfers as flows rather than as a single lump sum (similar analysis, now comparing lump sum to flow within the GD arm).  Finally, we can focus on the most concentrated case for rigidity as a potential commitment device:  are individuals with control problems most damaged by being having rigidity removed when they had indicated they need it?  We can pose this question by looking among those who chose flow transfers for heterogeneity in whether they were then assigned to lump sum or flow.

In the end, despite these numerous ways of posing questions about the benefits of rigidity, we find little evidence of behavioral heterogeneity.  In Appendix Tables \ref{t:Het_behavior_GK_vs_GD} and \ref{t:Het_behavior_LS_Flow} we present the result of first two analyses described above.  Focusing on the F-statistics at the bottom that test for differential impacts across the two interventions on those with self- or other-control problems, in both the comparison of Gikuriro to cash as well as the comparison of Lump Sum versus Flow we never find the behavioral attributes to have significantly differential effects for the more rigid program relative to the more flexible one.  In \ref{t:Het_behavior_choose_Flow} we conduct the third analysis described above; even in the group we might most expect to be signalling a desire for rigidity (chose flow) we find no harm from being moved to a less rigid modality (lump sum).  Further, the simple determinants of choice do not support the idea that this decision is driven by behavioral parameters; if we construct a measure of `sophistication' as being those individuals who are time inconsistent per the CTB exercise and also report in the survey having concerns about their own control over money, thee individuals are no more likely to choose flow transfers than individuals who are time inconsistent and naive.\footnote{We consider an individual to be sophisticated if they are time inconsistent by the above measure and respond yes to any of the three questions at baseline:  doesn’t keep cash on hand because doesn’t trust self with money, regrets a purchase made in the past month, or reports has problems with self wasting money on things the house doesn’t need.  Just over 10\% of time inconsistent individuals are sophisticated by this measure; about 66\% of both groups choose Flow.}  Those who report other control problems also choose lump sum at virtually identical rates to those who do not.   The takeaway is that at least by these metrics of individual heterogeneity, we see no evidence to support the `rigidity as commitment' idea.

%% file: 6-Gikuriro-Conclusion.tex
This study uses a large-scale randomized experiment to pose a number of questions in comparative cost-effectiveness.  Most centrally, we establish an approach to ask whether it is better to run complex multi-dimensional programs or simply to provide cash grants of equal ex-post cost, when these costs are not known with perfect certainty before the trial. We combine randomization of cash-transfer amounts with linear interpolation of their effects to make exact ex-post cost-equivalent comparisons, and demonstrate that, while point estimates do not generally depend on linearity, this approach delivers power gains equivalent to more than a doubling of sample size relative to alternative methods of interpolation.

Our application of this approach to the critical battle against child malnutrition highlights an even more basic lesson: evidently, regardless of modality, it simply costs more than \$140 per household to deliver clinically relevant impacts on child malnutrition outcomes within a year.\footnote{Gikuriro's costs were substantially lower than a related program in neighboring Burundi.}  
Even at this relatively low cost, however, the programs do trigger meaningfully different responses in the household use of intertemporal assets, with cost-equivalent cash transfers leading to higher levels of consumption, the pay-down of debt, and growth in asset investment, while Gikuriro households save more than those receiving cost-equivalent cash transfers.  

Our cost-equivalent estimates allow us to make the trade-off across outcomes very explicit.  
For example, a policy preference for Gikuriro is equivalent to asserting that the greater savings induced by Gikuriro is worth foregoing the higher consumption flows and productive asset stocks obtained under a cost-equivalent cash transfer.  In dollar terms, the estimates of Tables \ref{t:ce_primary} and \ref{t:ce_secondary} mean that moving from the predicted mean consumption of a cost-equivalent cash transfer to Gikuriro implies giving up \$28.32 in monthly consumption and adding \$8.35 in debt stocks, in exchange for an increase of \$18.70 in savings stocks. 
This kind of precision can help policymakers be much more exact as to the types of trade-offs required to justify one type of intervention over another.  

Given the modest impacts of both programs at costs equivalent to Gikuriro, comparisons across expenditure levels highlight a key tradeoff.  When we compare the cost-equivalent programs to a cash transfer of almost five times the amount, we see that larger sums of money can not only powerfully improve  overall consumption and dietary diversity, but also lead to modest improvements in child growth.  Transfer amounts in this study are not well-powered to test the linearity assumption (three small transfers of similar size, one much larger), however we generally find outcomes scaling with transfer amount in a simple way. Randomized variation in transfer values lets us form a number of interesting counterfactuals; for example the smallest transfer at which the benefit of cash would exceed Gikuriro; for savings this number is \$694, and for HAZ it is \$277.\footnote{Evidence from the sister study in Rwanda \citep{mcintosh2020hugukadukore}, which featured more intermediate treatment amounts, did in fact find evidence of a non-linear effect, with outcomes involving investment and time use peaking at transfers a little over \$400.  Nonetheless, in this study for no primary outcome are we able to reject a constant benefit/cost ratio across different cash amounts.}  Policymakers seeking to move child growth outcomes face difficult tradeoffs between the depth and breadth of their interventions; the comparative ease with which the resource intensity of cash transfers can be adapted makes this counterfactual modality particularly capable of revealing these tradeoffs.

Finally, we are able to test how the program impacts differ according to the sub-populations targeted.  Overall this analysis provides evidence of surprisingly homogeneous returns from these interventions, whether looking across pre-identified study subgroups,  using machine learning to identify optimal targeting rules that span multiple outcomes, or by introducing self-targeting  through beneficiary choice.   
The general lack of heterogeneity or scale effects, the lack of evidence that flow transfers are superior, absence of essential heterogeneity, and the uncorrelated individual-level benefits across different dimensions of impact all suggest that it is reasonable for policymakers to use simple lump-sum cash transfers and to target them using preference weights based on individual attributes or inequality.

There are some important limitations to what can be learned from the study.  First, the cross-village experiment lets us measure the impact of those parts of the Gikuriro intervention that are implemented at the village or household level only.  Village-level assignment means that some of our tests involving transfer amounts are not highly powered statistically.  Most importantly, the time frame of the study (13 months, but only 8--9 months after treatment with Gikuriro) means that we are measuring endline outcomes more quickly than would be ideal, particularly for anthropometrics that may respond slowly to improvements in nutrition.  Gikuriro made substantial investments in local capacity around health and sanitation and this infrastructure may drive future benefits in a way not captured in our study.  However, given that the targeting criteria for these programs will cause children to age out of eligibility within a few years, it remains important to show that such programs can generate benefits quickly for a given set of vulnerable children, as do the larger cash transfers we study.

Given the nuance of our findings, it is hard to square them with any simple idea of cash transfers as a kind of uni-dimensional `index fund'.  While business investment may have a single, cardinal objective---financial profit---development policy is undertaken with many goals in mind, and a perfect reconciling of these competing benefits would require a clear statement of trade-offs in this multi-dimensional space. Perhaps a clearer way of expressing the counterfactual provided by unconditional cash is that it gives us a statement of the priorities that the beneficiaries themselves hold when credit constraints are relaxed, and thereby motivates us to be clearer about the logic underlying paternalistic development programs.  While beneficiary decisions may not be `optimal' in terms of long-term social welfare (for example, due to high discount rates, to self- or other-control problem, or to resource and information constraints), the impact of unconditional cash is nonetheless a powerful statement of the outcomes that the beneficiaries themselves want changed.  For us to argue that a program is justified in using resources to drive outcomes different from the ones the beneficiaries would choose, we should have a clear reason why they fail to arrive at the welfare-maximizing outcome themselves.  This is a view of benchmarking that quantifies tradeoffs rather than picking a winner.

%% file: 7-Gikuriro-Tables.tex
\begin{table}[!hbp]
\caption{Research design}
\label{t:research_design}
\begin{center}
\begin{footnotesize}
\input{tables/research_design}
\floatfoot{
	Notes: Table gives the number of observations across the arms of the study.  The first row is the number of villages, the unit of assignment, and the remaining rows are numb er of households.}
\end{footnotesize}
\end{center}
\end{table}

\clearpage

\begin{table}[!hbtp]
\caption{Intervention costs and compliance rates}  
\label{t:cost_compliance}
\begin{footnotesize}
\input{tables/cost_compliance.tex}
\floatfoot{
	Notes: Table gives costs for each of the study arms.  The first row is the cost per beneficiary that emerged from the ex-post costing exercise.  The second row gives the share of program costs averted among non-compliers, and the third row the compliance rate.  Using these the fourth row provides the average amount spent per study eligible household.  The fifth and sixth rows provide cost numbers for the average household in the village population.}
\end{footnotesize} 
\end{table}

\clearpage

\begin{table}
\caption{ITT estimates:  Primary outcomes}
\label{t:itt_primary}
\begin{footnotesize}
\begin{center}
\input{tables/itt_primary}

\end{center}

\vskip-4ex 

\floatfoot{
	Notes: Table presents Intention to Treat impacts on primary outcomes, with the three study arms presented in rows and the three smaller GiveDirectly transfers pooled into the `Main' treatment.  Regressions include but do not report the lagged dependent variable, fixed effects for randomization blocks, and a set of LASSO-selected baseline covariates, and are weighted be representative of the eligible population. Standard errors (in parentheses) are clustered at the household level to reflect the design effect.	Asterices denote significance at the 10, 5, and 1 percent levels, and are based on clustered standard errors, in parentheses.  \citet{anderson2008multiple} sharpened $q$-values presented in brackets. Variables marked with a $\dag$ are in inverse hyperbolic sines.  Reported $p$-values in final two columns derived from $F$-tests of hypotheses that cost-benefit ratios are equal between GD Main and Large transfer amounts (GD=GDL), and between Gikuriro and GD Large (GK=GDL).}
\end{footnotesize}
\end{table}

\begin{table}
\caption{ITT estimates:  Secondary outcomes}
\label{t:itt_secondary}
\begin{footnotesize}
\begin{center}
\resizebox{0.925\linewidth}{!}{
\input{tables/itt_secondary}

}
\end{center}

\vskip-4ex 

\floatfoot{
	Notes: Table presents Intention to Treat impacts on secondary outcomes, with the three study arms presented in rows and the three smaller GiveDirectly transfers pooled into the `Main' treatment.  Regressions include but do not report the lagged dependent variable, fixed effects for randomization blocks, and a set of LASSO-selected baseline covariates, and are weighted be representative of the eligible population. Standard errors (in parentheses) are clustered at the household level to reflect the design effect.	Asterices denote significance at the 10, 5, and 1 percent levels, and are based on clustered standard errors, in parentheses.  \citet{anderson2008multiple} sharpened $q$-values presented in brackets. Variables marked with a $\dag$ are in inverse hyperbolic sines.  Reported $p$-values in final two columns derived from $F$-tests of hypotheses that cost-benefit ratios are equal between GD Main and Large transfer amounts (GD=GDL), and between Gikuriro and GD Large (GK=GDL).}
\end{footnotesize}
\end{table}

\begin{table}
\caption{Cost-equivalent comparisons:  Primary outcomes}
\label{t:ce_primary}
\begin{footnotesize}
\begin{center}
\input{tables/ce_primary.tex}

\end{center}

\vskip-4ex 

\floatfoot{
  Notes: First column is a dummy for Gikuriro treatment, giving the differential effect of Gikuriro over cash at equivalent cost.  Second column is a dummy for either treatment, giving the impact of cash at the cost of Gikuriro. Third column is the cost slope, measured as the dollar-value deviation (in hundreds of dollars) of the treatment received from the cost of Gikuriro.   Asterices denote significance at the 10, 5, and 1 percent levels, and are based on clustered standard errors, in parentheses.  \citet{anderson2008multiple} sharpened $q$-values presented in brackets. Variables marked with a $\dag$ are in inverse hyperbolic sines.
	
}
\end{footnotesize}
\end{table}

\begin{table}
\caption{Cost-equivalent comparisons: Secondary outcomes}
\label{t:ce_secondary}

\begin{footnotesize}
\begin{center}
\resizebox{0.875\linewidth}{!}{
    \input{tables/ce_secondary.tex}

}
\end{center}

\vskip-4ex 

\floatfoot{
  Notes: First column is a dummy for Gikuriro treatment, giving the differential effect of Gikuriro over cash at equivalent cost.  Second column is a dummy for either treatment, giving the impact of cash at the cost of Gikuriro. Third column is the cost slope, measured as the dollar-value deviation (in hundreds of dollars) of the treatment received from the cost of Gikuriro.   Asterices denote significance at the 10, 5, and 1 percent levels, and are based on clustered standard errors, in parentheses.  \citet{anderson2008multiple} sharpened $q$-values presented in brackets. Variables marked with a $\dag$ are in inverse hyperbolic sines.}
\end{footnotesize}
\end{table}

\begin{table}
\caption{Benefit-cost ratios for primary outcomes}
\label{t:bcr_primary}
\begin{footnotesize}
\input{tables/BCR_primary}

\floatfoot{
Notes: Benefit-cost ratios derived form ITT estimates and estimated costs per eligible household.  $p$-values report tests of equal BCR between (a) Gikuriro and GD-Main; (b) GD-Main and GD-Large; and (c) Gikuriro and GD-large.  Household-level BCRs; the average eligible household contains 5.2 members, 1.5 anthro-eligible children, 1.7 children eligible for anemia testing, and 1.2 adult women eligible for anemia testing.  Per-person BCRs should be scaled up by these numbers.
}
\end{footnotesize}
\end{table}

\clearpage

\begin{table}[hbtp]
\caption{Village-level Total Causal Effects: Primary outcomes}
\label{t:tce_primary}
\begin{footnotesize}
\begin{center}
\input{tables/tce_primary.tex}

\end{center}

\vskip-4ex 

\floatfoot{
	Notes: Analysis pools eligible and ineligible households and is weighted to be representative of the population in study villages.  Asterices denote significance at the 10, 5, and 1 percent levels, and are based on clustered standard errors, in parentheses.  \citet{anderson2008multiple} sharpened $q$-values presented in brackets. Variables marked with a $\dag$ are in inverse hyperbolic sines.  Final two columns present $p$-values on tests of the equality of benefit-cost ratios between arms.
	
}
\end{footnotesize}
\end{table}

\cleardoublepage

\begin{table}
\caption{Correlation between Household-level Impact Domains}
\label{t:heterogeneity_crossoutcomes}
\begin{footnotesize}
\input{tables/heterogeneity_rank_corrs}
\end{footnotesize}

\floatfoot{
    Notes: Table shows correlation coefficients between estimated subgroup mean treatment effects across outcome domains.  Treatment effects estimated by causal forest.  Child growth is a standardized index of HAZ, WAZ, and MUAC.
}

\end{table}

\clearpage

\begin{table}
\caption{Welfare Gains from Targeting Modality}
\label{t:welfare_example}

\begin{tabular}{l c }
\toprule
Target & Gains (standard deviation of outcome) \\
\midrule
Consumption & 0.016   \\
Wealth & 0.057 \\
Dietary diversity &   0.086 \\
Child growth &  0.103  \\ 
Composite welfare metric & 0.029 \\
\bottomrule
\end{tabular}

\floatfoot{
Notes: Table reports gains relative to control from offering households the treatment (Gikuriro or GD Main) that is expected to deliver the bigger impact, based on causal forest estimates.  All outcomes are standardized to have mean of zero and variance of one in the control group at endline.  The first four rows represent welfare gains if treatment assignment is optimized for that outcome only; the final row represents average gains across the four dimension if households are assigned the treatment that is expected to deliver the largest average impact across these dimensions.
}

\end{table}

%% file: tables/research_design.tex
\begin{tabular}{l *{7}{C{0.075\textwidth}} }
\toprule 
	& & & \multicolumn{4}{c}{GiveDirectly} &  \\
\cmidrule(lr){4-7} 
	&  & &  Lower transfer & Middle transfer & Upper transfer & Large transfer  &  \\
	& Control & Gikuriro & (\$66) & (\$111) &  (\$145) & (\$566) & Total \\ 
\midrule %
\multicolumn{8}{l}{\emph{Panel A. Village-level randomization}} \\[2ex]

Villages & 74 & 74 & 22 & 22 & 22 & 34 & 248 \\
Ineligible households & 298 & 297 & 88 & 87 & 88 & 137 & 995 \\ 
Eligible households & 521 & 541 & 165 & 154 & 167 & 246 & 1,794 \\[2ex]
\multicolumn{8}{l}{\emph{Panel B.  Household-level randomization of cash payment modality among eligibles}} \\[2ex]
Flow transfers & & & 83 & 87 & 104 & 147 & 421 \\
Lump-sum transfers & & & 51 & 50 & 41 & 68 & 210 \\ 
Choice & & & 31 & 17 & 22 & 31 & 101 \\ 
\bottomrule 
\end{tabular}

%% file: tables/cost_compliance.tex
\begin{tabular}{p{0.33\textwidth}ccccc}
\toprule
 & & \multicolumn{4}{c}{GiveDirectly} \\ 
\cmidrule(lr){3-6}
    &   \multicolumn{1}{c}{Gikuriro} & \multicolumn{1}{c}{Lower} & \multicolumn{1}{c}{Middle} & \multicolumn{1}{c}{Upper} & \multicolumn{1}{c}{Large} \\
\midrule
Cost to USAID per beneficiary & \$141.84 & \$66.02 & \$111.09 & \$145.43 & \$566.55 \\[1ex] 
Share averted if untreated & 0.60 & 1.00 & 1.00 & 1.00 & 1.00 \\[1ex] 
Compliance rate among eligibles & 0.80 & 0.81 & 0.86 & 0.83 & 0.91 \\[1ex] 
Cost to USAID per eligible household & \$124.49 & \$53.58 & \$95.86 & \$121.24 & \$517.44 \\[1ex] 
Compliance rate in population & 0.19 & 0.18 & 0.19 & 0.18 & 0.18 \\[1ex] 
Cost to USAID per village household & \$28.02 & \$12.20 & \$20.83 & \$26.69 & \$99.56 \\[1ex] 
\bottomrule
\end{tabular}

%% file: tables/itt_primary.tex
\begin{tabular}{l *{3}{S} ScSSS}
\toprule
 & & \multicolumn{2}{c}{GiveDirectly} &  \multicolumn{1}{c}{\raisebox{-1ex}[0pt]{Control}} & & &\multicolumn{2}{c}{$ p$-values: B/C ratios} \\ 
\cmidrule(lr){3-4} \cmidrule(lr){8-9}
\multicolumn{1}{c}{\text{ }} & \multicolumn{1}{c}{\text{Gikuriro}} & \multicolumn{1}{c}{\text{Main}} & \multicolumn{1}{c}{\text{Large}} & \multicolumn{1}{c}{\text{Mean (SD)}} & \multicolumn{1}{c}{\text{Obs.}} & \multicolumn{1}{c}{\text{$ R^2$}} & \multicolumn{1}{c}{\text{GD=GDL}} & \multicolumn{1}{c}{\text{GK=GDL}}\\
\midrule
\multicolumn{9}{l}{\emph{Panel A.  Household outcomes}}  \\ 
\addlinespace[1ex] \multirow[t]{ 3 }{0.2\textwidth}{Consumption$^\dag$ }  &     -0.11 &      0.06 &      0.30\ensuremath{^{***}} &     10.69 &      1750 &      0.14 &      0.93 &      0.04 \\ 
 & (0.10)  & (0.09)  & (0.11)  & (1.34)  \\  & [    0.41]  & [    0.57]  & [    0.02]  \\ \addlinespace[1ex] 
\multirow[t]{ 3 }{0.2\textwidth}{Household dietary diversity score }  &      0.19 &      0.17 &      0.55\ensuremath{^{***}} &      4.77 &      1751 &      0.18 &      0.60 &      0.59 \\ 
 & (0.12)  & (0.15)  & (0.13)  & (1.84)  \\  & [    0.38]  & [    0.41]  & [    0.00]  \\ \addlinespace[1ex] 
\multirow[t]{ 3 }{0.2\textwidth}{Household non-land wealth$^\dag$ }  &      0.01 &      0.00 &      0.40 &     13.04 &      1751 &      0.22 &      0.74 &      0.60 \\ 
 & (0.18)  & (0.21)  & (0.28)  & (4.24)  \\  & [    0.60]  & [    0.60]  & [    0.38]  \\ \addlinespace[1ex] 
\multicolumn{9}{l}{\emph{Panel B.  Individual outcomes}}  \\ 
\addlinespace[1ex] \multirow[t]{ 3 }{0.2\textwidth}{Height-for-age }  &      0.05 &     -0.02 &      0.09\ensuremath{^{**}} &     -1.97 &      2125 &      0.71 &      0.33 &      0.47 \\ 
 & (0.04)  & (0.04)  & (0.05)  & (1.10)  \\  & [    1.00]  & [    1.00]  & [    0.78]  \\ \addlinespace[1ex] 
\multirow[t]{ 3 }{0.2\textwidth}{Weight-for-age }  &      0.04 &      0.01 &      0.07\ensuremath{^{*}} &     -1.04 &      2104 &      0.68 &      0.97 &      0.54 \\ 
 & (0.04)  & (0.03)  & (0.04)  & (0.98)  \\  & [    1.00]  & [    1.00]  & [    0.78]  \\ \addlinespace[1ex] 
\multirow[t]{ 3 }{0.2\textwidth}{Mid-upper arm circumference }  &      0.02 &     -0.01 &      0.13\ensuremath{^{*}} &     -0.59 &      1629 &      0.51 &      0.63 &      0.85 \\ 
 & (0.06)  & (0.07)  & (0.08)  & (0.95)  \\  & [    1.00]  & [    1.00]  & [    0.78]  \\ \addlinespace[1ex] 
\multirow[t]{ 3 }{0.2\textwidth}{Child anemia }  &      0.00 &      0.02 &     -0.01 &      0.18 &      2372 &      0.07 &      0.30 &      0.77 \\ 
 & (0.02)  & (0.02)  & (0.04)  & (0.39)  \\  & [    1.00]  & [    1.00]  & [    1.00]  \\ \addlinespace[1ex] 
\multirow[t]{ 3 }{0.2\textwidth}{Maternal anemia }  &     -0.02 &     -0.00 &     -0.03 &      0.12 &      1581 &      0.11 &      0.91 &      0.48 \\ 
 & (0.03)  & (0.03)  & (0.03)  & (0.33)  \\  & [    1.00]  & [    1.00]  & [    1.00]  \\ \addlinespace[1ex] 
\bottomrule
\end{tabular}

%% file: tables/itt_secondary.tex
\begin{tabular}{l *{3}{S} ScSSS}
\toprule
 & & \multicolumn{2}{c}{GiveDirectly} &  \multicolumn{1}{c}{\raisebox{-1ex}[0pt]{Control}} & & &\multicolumn{2}{c}{$ p$-values: B/C ratios} \\ 
\cmidrule(lr){3-4} \cmidrule(lr){8-9}
\multicolumn{1}{c}{\text{ }} & \multicolumn{1}{c}{\text{Gikuriro}} & \multicolumn{1}{c}{\text{Main}} & \multicolumn{1}{c}{\text{Large}} & \multicolumn{1}{c}{\text{Mean (SD)}} & \multicolumn{1}{c}{\text{Obs.}} & \multicolumn{1}{c}{\text{$ R^2$}} & \multicolumn{1}{c}{\text{GD=GDL}} & \multicolumn{1}{c}{\text{GK=GDL}}\\
\midrule
\multicolumn{9}{l}{\emph{Panel A.  Household outcomes}}  \\ 
\addlinespace[1ex] \multirow[t]{ 3 }{0.2\textwidth}{Stock of borrowing$^\dag$ }  &     0.067 &    -0.765\ensuremath{^{**}} &    -0.341 &      7.39 &      1751 &      0.12 &      0.01 &      0.63 \\ 
 & (    0.350)  & (    0.316)  & (    0.397)  & (4.82)  \\  & [    0.71]  & [    0.05]  & [    0.64]  \\ \addlinespace[1ex] 
\multirow[t]{ 3 }{0.2\textwidth}{Stock of saving$^\dag$ }  &     1.115\ensuremath{^{***}} &    -0.128 &     0.656\ensuremath{^{**}} &      5.88 &      1751 &      0.16 &      0.46 &      0.00 \\ 
 & (    0.324)  & (    0.345)  & (    0.329)  & (4.87)  \\  & [    0.01]  & [    0.71]  & [    0.14]  \\ \addlinespace[1ex] 
\multirow[t]{ 3 }{0.2\textwidth}{Health knowledge index }  &    -0.076 &     0.153 &     0.075 &      2.89 &      1751 &      0.04 &      0.64 &      0.78 \\ 
 & (    0.368)  & (    0.321)  & (    0.468)  & (4.01)  \\  & [    0.71]  & [    0.71]  & [    0.71]  \\ \addlinespace[1ex] 
\multirow[t]{ 3 }{0.2\textwidth}{Sanitation practices index }  &    -0.280 &     0.146 &     0.087 &     -0.68 &      1751 &      0.07 &      0.54 &      0.13 \\ 
 & (    0.219)  & (    0.227)  & (    0.284)  & (2.71)  \\  & [    0.37]  & [    0.71]  & [    0.71]  \\ \addlinespace[1ex] 
\multirow[t]{ 3 }{0.2\textwidth}{Productive assets$^\dag$ }  &     0.020 &     0.257\ensuremath{^{**}} &     0.800\ensuremath{^{***}} &     11.22 &      1751 &      0.29 &      0.21 &      0.06 \\ 
 & (    0.100)  & (    0.100)  & (    0.116)  & (1.81)  \\  & [    0.71]  & [    0.05]  & [    0.00]  \\ \addlinespace[1ex] 
\multirow[t]{ 3 }{0.2\textwidth}{Consumption assets$^\dag$ }  &    -0.367 &     0.354\ensuremath{^{*}} &     0.932\ensuremath{^{***}} &      8.70 &      1751 &      0.35 &      0.32 &      0.01 \\ 
 & (    0.240)  & (    0.206)  & (    0.243)  & (4.08)  \\  & [    0.28]  & [    0.22]  & [    0.00]  \\ \addlinespace[1ex] 
\multirow[t]{ 3 }{0.2\textwidth}{House value$^\dag$ }  &    -0.023 &    -0.029 &     0.196\ensuremath{^{***}} &     13.81 &      1654 &      0.34 &      0.27 &      0.16 \\ 
 & (    0.056)  & (    0.061)  & (    0.061)  & (0.87)  \\  & [    0.71]  & [    0.71]  & [    0.01]  \\ \addlinespace[1ex] 
\multirow[t]{ 3 }{0.2\textwidth}{Housing quality index }  &    -0.195 &    -0.217\ensuremath{^{*}} &     0.211 &     -0.17 &      1751 &      0.10 &      0.05 &      0.08 \\ 
 & (    0.146)  & (    0.132)  & (    0.174)  & (1.46)  \\  & [    0.36]  & [    0.24]  & [    0.37]  \\ \addlinespace[1ex] 
\multicolumn{9}{l}{\emph{Panel B.  Individual outcomes}}  \\ 
\addlinespace[1ex] \multirow[t]{ 3 }{0.2\textwidth}{Child mortality }  &    -0.006 &    -0.004 &    -0.009\ensuremath{^{**}} &      0.01 &      2687 &      0.01 &      0.73 &      0.39 \\ 
 & (    0.005)  & (    0.006)  & (    0.004)  & (0.11)  \\  & [    1.00]  & [    1.00]  & [    1.00]  \\ \addlinespace[1ex] 
\multirow[t]{ 3 }{0.2\textwidth}{Pregnancy }  &    -0.031 &    -0.035 &    -0.007 &      0.20 &      2552 &      0.08 &      0.25 &      0.18 \\ 
 & (    0.025)  & (    0.031)  & (    0.027)  & (0.40)  \\  & [    1.00]  & [    1.00]  & [    1.00]  \\ \addlinespace[1ex] 
\multirow[t]{ 3 }{0.2\textwidth}{Live birth }  &     0.103 &     0.091 &    -0.068 &      0.68 &       411 &      0.13 &      0.12 &      0.09 \\ 
 & (    0.079)  & (    0.072)  & (    0.081)  & (0.47)  \\  & [    1.00]  & [    1.00]  & [    1.00]  \\ \addlinespace[1ex] 
\multirow[t]{ 3 }{0.2\textwidth}{Birth in facility }  &    -0.046 &     0.069 &    -0.062 &      0.84 &       293 &      0.16 &      0.11 &      0.59 \\ 
 & (    0.059)  & (    0.052)  & (    0.099)  & (0.37)  \\  & [    1.00]  & [    1.00]  & [    1.00]  \\ \addlinespace[1ex] 
\multirow[t]{ 3 }{0.2\textwidth}{Any vaccinations in past year }  &     0.010 &    -0.010 &    -0.005 &      0.72 &      1291 &      0.26 &      0.75 &      0.72 \\ 
 & (    0.034)  & (    0.032)  & (    0.039)  & (0.45)  \\  & [    1.00]  & [    1.00]  & [    1.00]  \\ \addlinespace[1ex] 
\multirow[t]{ 3 }{0.2\textwidth}{Completed vaccination schedule }  &     0.011 &    -0.013 &     0.006 &      0.58 &      1291 &      0.17 &      0.67 &      0.79 \\ 
 & (    0.039)  & (    0.035)  & (    0.038)  & (0.49)  \\  & [    1.00]  & [    1.00]  & [    1.00]  \\ \addlinespace[1ex] 
\multirow[t]{ 3 }{0.2\textwidth}{Disease burden }  &    -0.020 &    -0.031 &    -0.018 &      0.54 &      2680 &      0.05 &      0.33 &      0.61 \\ 
 & (    0.032)  & (    0.030)  & (    0.033)  & (0.50)  \\  & [    1.00]  & [    1.00]  & [    1.00]  \\ \addlinespace[1ex] 
\multirow[t]{ 3 }{0.2\textwidth}{Diarrheal prevalence }  &    -0.003 &    -0.000 &    -0.007 &      0.09 &      2680 &      0.04 &      0.95 &      0.89 \\ 
 & (    0.015)  & (    0.016)  & (    0.015)  & (0.29)  \\  & [    1.00]  & [    1.00]  & [    1.00]  \\ \addlinespace[1ex] 
\bottomrule
\end{tabular}

%% file: tables/ce_primary.tex
\begin{tabular}{l *{3}{S} ScS}
\toprule
\multicolumn{1}{c}{\text{ }} & \multicolumn{1}{c}{\text{\makecell[b]{Gikuriro:\\ Differential}}} & \multicolumn{1}{c}{\text{\makecell[b]{Cost-equivalent\\GD impact}}} & \multicolumn{1}{c}{\text{\makecell[b]{Transfer\\Cost}}} & \multicolumn{1}{c}{\text{\makecell[b]{Control\\Mean}}} & \multicolumn{1}{c}{\text{Observations}} & \multicolumn{1}{c}{\text{$ R^2$}}\\
\midrule
\multicolumn{7}{l}{\emph{A.  Household outcomes}}  \\ 
\addlinespace[1ex] \multirow[t]{ 3 }{0.2\textwidth}{Consumption$^\dag$ }  &     -0.19\ensuremath{^{**}} &      0.08 &      0.05\ensuremath{^{**}} &     10.69 &      1750 &      0.14 \\ 
 & (0.08)  & (0.09)  & (0.02)  \\  & [    0.06]  & [    0.49]  & [    0.06]  \\ \addlinespace[1ex] 
\multirow[t]{ 3 }{0.2\textwidth}{Household dietary diversity score }  &     -0.01 &      0.20 &      0.09\ensuremath{^{**}} &      4.77 &      1751 &      0.18 \\ 
 & (0.13)  & (0.14)  & (0.03)  \\  & [    0.71]  & [    0.28]  & [    0.06]  \\ \addlinespace[1ex] 
\multirow[t]{ 3 }{0.2\textwidth}{Household non-land wealth$^\dag$ }  &     -0.03 &      0.04 &      0.09 &     13.04 &      1751 &      0.22 \\ 
 & (0.20)  & (0.21)  & (0.07)  \\  & [    0.71]  & [    0.71]  & [    0.30]  \\ \addlinespace[1ex] 
\multicolumn{7}{l}{\emph{B.  Individual outcomes}}  \\ 
\addlinespace[1ex] \multirow[t]{ 3 }{0.2\textwidth}{Height-for-age }  &      0.06 &     -0.01 &      0.02\ensuremath{^{**}} &     -1.97 &      2125 &      0.71 \\ 
 & (0.04)  & (0.04)  & (0.01)  \\  & [    0.96]  & [    1.00]  & [    0.40]  \\ \addlinespace[1ex] 
\multirow[t]{ 3 }{0.2\textwidth}{Weight-for-age }  &      0.02 &      0.02 &      0.01 &     -1.04 &      2104 &      0.68 \\ 
 & (0.04)  & (0.03)  & (0.01)  \\  & [    1.00]  & [    1.00]  & [    0.96]  \\ \addlinespace[1ex] 
\multirow[t]{ 3 }{0.2\textwidth}{Mid-upper arm circumference }  &      0.02 &      0.00 &      0.03 &     -0.59 &      1629 &      0.51 \\ 
 & (0.06)  & (0.06)  & (0.02)  \\  & [    1.00]  & [    1.00]  & [    0.96]  \\ \addlinespace[1ex] 
\multirow[t]{ 3 }{0.2\textwidth}{Child anemia }  &     -0.02 &      0.02 &     -0.01 &      0.22 &      2372 &      0.07 \\ 
 & (0.03)  & (0.02)  & (0.01)  \\  & [    1.00]  & [    1.00]  & [    1.00]  \\ \addlinespace[1ex] 
\multirow[t]{ 3 }{0.2\textwidth}{Maternal anemia }  &     -0.02 &     -0.00 &     -0.00 &      0.12 &      1581 &      0.10 \\ 
 & (0.02)  & (0.03)  & (0.01)  \\  & [    1.00]  & [    1.00]  & [    1.00]  \\ \addlinespace[1ex] 
\bottomrule
\end{tabular}

%% file: tables/ce_secondary.tex
\begin{tabular}{l *{3}{S} ScS}
\toprule
\multicolumn{1}{c}{\text{ }} & \multicolumn{1}{c}{\text{\makecell[b]{Gikuriro:\\ Differential}}} & \multicolumn{1}{c}{\text{\makecell[b]{Cost-equivalent\\GD impact}}} & \multicolumn{1}{c}{\text{\makecell[b]{Transfer\\Cost}}} & \multicolumn{1}{c}{\text{\makecell[b]{Control\\Mean}}} & \multicolumn{1}{c}{\text{Observations}} & \multicolumn{1}{c}{\text{$ R^2$}}\\
\midrule
\multicolumn{7}{l}{\emph{A.  Household outcomes}}  \\ 
\addlinespace[1ex] \multirow[t]{ 3 }{0.2\textwidth}{Stock of borrowing$^\dag$ }  &     0.793\ensuremath{^{***}} &    -0.726\ensuremath{^{**}} &     0.094 &      7.39 &      1751 &      0.12 \\ 
 & (    0.277)  & (    0.309)  & (    0.077)  \\  & [    0.01]  & [    0.04]  & [    0.19]  \\ \addlinespace[1ex] 
\multirow[t]{ 3 }{0.2\textwidth}{Stock of saving$^\dag$ }  &     1.184\ensuremath{^{***}} &    -0.070 &     0.188\ensuremath{^{**}} &      5.88 &      1751 &      0.16 \\ 
 & (    0.333)  & (    0.327)  & (    0.080)  \\  & [    0.01]  & [    0.54]  & [    0.04]  \\ \addlinespace[1ex] 
\multirow[t]{ 3 }{0.2\textwidth}{Health knowledge index }  &    -0.227 &     0.151 &    -0.023 &      2.89 &      1751 &      0.04 \\ 
 & (    0.329)  & (    0.316)  & (    0.103)  \\  & [    0.44]  & [    0.54]  & [    0.54]  \\ \addlinespace[1ex] 
\multirow[t]{ 3 }{0.2\textwidth}{Sanitation practices index }  &    -0.421\ensuremath{^{**}} &     0.141 &    -0.014 &     -0.68 &      1751 &      0.07 \\ 
 & (    0.213)  & (    0.220)  & (    0.069)  \\  & [    0.06]  & [    0.44]  & [    0.54]  \\ \addlinespace[1ex] 
\multirow[t]{ 3 }{0.2\textwidth}{Productive assets$^\dag$ }  &    -0.283\ensuremath{^{***}} &     0.302\ensuremath{^{***}} &     0.125\ensuremath{^{***}} &     11.22 &      1751 &      0.29 \\ 
 & (    0.097)  & (    0.096)  & (    0.026)  \\  & [    0.01]  & [    0.01]  & [    0.00]  \\ \addlinespace[1ex] 
\multirow[t]{ 3 }{0.2\textwidth}{Consumption assets$^\dag$ }  &    -0.764\ensuremath{^{***}} &     0.397\ensuremath{^{**}} &     0.139\ensuremath{^{**}} &      8.70 &      1751 &      0.35 \\ 
 & (    0.256)  & (    0.199)  & (    0.058)  \\  & [    0.01]  & [    0.06]  & [    0.04]  \\ \addlinespace[1ex] 
\multirow[t]{ 3 }{0.2\textwidth}{House value$^\dag$ }  &    -0.017 &    -0.007 &     0.049\ensuremath{^{***}} &     13.81 &      1654 &      0.34 \\ 
 & (    0.051)  & (    0.059)  & (    0.014)  \\  & [    0.54]  & [    0.55]  & [    0.01]  \\ \addlinespace[1ex] 
\multirow[t]{ 3 }{0.2\textwidth}{Housing quality index }  &    -0.008 &    -0.187 &     0.104\ensuremath{^{**}} &     -0.17 &      1751 &      0.10 \\ 
 & (    0.151)  & (    0.126)  & (    0.048)  \\  & [    0.56]  & [    0.12]  & [    0.05]  \\ \addlinespace[1ex] 
\multicolumn{7}{l}{\emph{B.  Individual outcomes}}  \\ 
\addlinespace[1ex] \multirow[t]{ 3 }{0.2\textwidth}{Child mortality }  &    -0.002 &    -0.004 &    -0.001 &      0.01 &      2687 &      0.01 \\ 
 & (    0.005)  & (    0.006)  & (    0.001)  \\  & [    1.00]  & [    1.00]  & [    1.00]  \\ \addlinespace[1ex] 
\multirow[t]{ 3 }{0.2\textwidth}{Pregnancy }  &     0.002 &    -0.033 &     0.007 &      0.20 &      2552 &      0.08 \\ 
 & (    0.027)  & (    0.029)  & (    0.007)  \\  & [    1.00]  & [    1.00]  & [    1.00]  \\ \addlinespace[1ex] 
\multirow[t]{ 3 }{0.2\textwidth}{Live birth }  &     0.024 &     0.079 &    -0.037\ensuremath{^{**}} &      0.68 &       411 &      0.13 \\ 
 & (    0.067)  & (    0.069)  & (    0.017)  \\  & [    1.00]  & [    1.00]  & [    1.00]  \\ \addlinespace[1ex] 
\multirow[t]{ 3 }{0.2\textwidth}{Birth in facility }  &    -0.104\ensuremath{^{**}} &     0.058 &    -0.029 &      0.84 &       293 &      0.16 \\ 
 & (    0.052)  & (    0.051)  & (    0.023)  \\  & [    1.00]  & [    1.00]  & [    1.00]  \\ \addlinespace[1ex] 
\multirow[t]{ 3 }{0.2\textwidth}{Any vaccinations in past year }  &     0.018 &    -0.008 &    -0.000 &      0.72 &      1291 &      0.26 \\ 
 & (    0.032)  & (    0.031)  & (    0.009)  \\  & [    1.00]  & [    1.00]  & [    1.00]  \\ \addlinespace[1ex] 
\multirow[t]{ 3 }{0.2\textwidth}{Completed vaccination schedule }  &     0.021 &    -0.010 &     0.003 &      0.58 &      1291 &      0.17 \\ 
 & (    0.036)  & (    0.033)  & (    0.009)  \\  & [    1.00]  & [    1.00]  & [    1.00]  \\ \addlinespace[1ex] 
\multirow[t]{ 3 }{0.2\textwidth}{Disease burden }  &     0.012 &    -0.031 &     0.004 &      0.54 &      2680 &      0.05 \\ 
 & (    0.033)  & (    0.029)  & (    0.008)  \\  & [    1.00]  & [    1.00]  & [    1.00]  \\ \addlinespace[1ex] 
\multirow[t]{ 3 }{0.2\textwidth}{Diarrheal prevalence }  &    -0.003 &    -0.000 &    -0.002 &      0.09 &      2680 &      0.04 \\ 
 & (    0.014)  & (    0.015)  & (    0.004)  \\  & [    1.00]  & [    1.00]  & [    1.00]  \\ \addlinespace[1ex] 
\bottomrule
\end{tabular}

%% file: tables/BCR_primary.tex
\begin{tabular}{l *{3}{S[table-format=03.3]} SSS}
\toprule
 & & \multicolumn{2}{c}{GiveDirectly} & \multicolumn{3}{c}{$ p$-values} \\ 
\cmidrule(lr){3-4} \cmidrule(lr){5-7}
\multicolumn{1}{c}{\text{ }} & \multicolumn{1}{c}{\text{Gikuriro}} & \multicolumn{1}{c}{\text{Main}} & \multicolumn{1}{c}{\text{Large}} & \multicolumn{1}{c}{\text{(a)}} & \multicolumn{1}{c}{\text{(b)}} & \multicolumn{1}{c}{\text{(c)}}\\
\midrule
\multicolumn{7}{l}{\emph{Panel A.  Household outcomes}}  \\ 
\addlinespace[1ex] \multirow[t]{ 2 }{0.25\textwidth}{Consumption$^\dag$ }  &    -0.090 &     0.066 &     0.058 &      0.05 &      0.93 &      0.04 \\ 
 & (    0.077)  & (    0.100)  & (    0.021)  \\ \addlinespace[1ex] 
\multirow[t]{ 2 }{0.25\textwidth}{Household dietary diversity score }  &     0.155 &     0.186 &     0.106 &      0.83 &      0.60 &      0.59 \\ 
 & (    0.098)  & (    0.162)  & (    0.024)  \\ \addlinespace[1ex] 
\multirow[t]{ 2 }{0.25\textwidth}{Household non-land wealth$^\dag$ }  &     0.007 &     0.005 &     0.077 &      0.99 &      0.74 &      0.60 \\ 
 & (    0.148)  & (    0.235)  & (    0.054)  \\ \addlinespace[1ex] 
\multicolumn{7}{l}{\emph{Panel B.  Individual outcomes}}  \\ 
\addlinespace[1ex] \multirow[t]{ 2 }{0.25\textwidth}{Height-for-age }  &     0.042 &    -0.021 &     0.018 &      0.11 &      0.33 &      0.47 \\ 
 & (    0.036)  & (    0.044)  & (    0.009)  \\ \addlinespace[1ex] 
\multirow[t]{ 2 }{0.25\textwidth}{Weight-for-age }  &     0.031 &     0.012 &     0.013 &      0.61 &      0.97 &      0.54 \\ 
 & (    0.032)  & (    0.037)  & (    0.007)  \\ \addlinespace[1ex] 
\multirow[t]{ 2 }{0.25\textwidth}{Mid-upper arm circumference }  &     0.018 &    -0.007 &     0.026 &      0.71 &      0.63 &      0.85 \\ 
 & (    0.045)  & (    0.072)  & (    0.015)  \\ \addlinespace[1ex] 
\multirow[t]{ 2 }{0.25\textwidth}{Child anemia }  &     0.003 &     0.026 &    -0.002 &      0.38 &      0.30 &      0.77 \\ 
 & (    0.018)  & (    0.028)  & (    0.007)  \\ \addlinespace[1ex] 
\multirow[t]{ 2 }{0.25\textwidth}{Maternal anemia }  &    -0.019 &    -0.002 &    -0.005 &      0.46 &      0.91 &      0.48 \\ 
 & (    0.022)  & (    0.032)  & (    0.006)  \\ \addlinespace[1ex] 
\bottomrule
\end{tabular}

%% file: tables/tce_primary.tex
\begin{tabular}{l *{3}{S} ScSSS}
\toprule
 & & \multicolumn{2}{c}{GiveDirectly} &  \multicolumn{1}{c}{\raisebox{-1ex}[0pt]{Control}} & & &\multicolumn{2}{c}{$ p$-values: B/C ratios} \\ 
\cmidrule(lr){3-4} \cmidrule(lr){8-9}
\multicolumn{1}{c}{\text{ }} & \multicolumn{1}{c}{\text{Gikuriro}} & \multicolumn{1}{c}{\text{Main}} & \multicolumn{1}{c}{\text{Large}} & \multicolumn{1}{c}{\text{Mean}} & \multicolumn{1}{c}{\text{Obs.}} & \multicolumn{1}{c}{\text{$ R^2$}} & \multicolumn{1}{c}{\text{GD=GDL}} & \multicolumn{1}{c}{\text{GK=GDL}}\\
\midrule
\multicolumn{9}{l}{\emph{Panel A.  Household outcomes}}  \\ 
\addlinespace[1ex] \multirow[t]{ 3 }{0.2\textwidth}{Consumption$^\dag$ }  &     -0.14 &     -0.14 &     -0.06 &     10.55 &      2717 &      0.15 &      0.18 &      0.20 \\ 
 & (0.10)  & (0.10)  & (0.17)  \\  & [    0.60]  & [    0.60]  & [    0.81]  \\ \addlinespace[1ex] 
\multirow[t]{ 3 }{0.2\textwidth}{Household dietary diversity score }  &      0.12 &      0.00 &     -0.28 &      4.29 &      2718 &      0.22 &      0.64 &      0.11 \\ 
 & (0.13)  & (0.13)  & (0.19)  \\  & [    0.60]  & [    0.85]  & [    0.60]  \\ \addlinespace[1ex] 
\multirow[t]{ 3 }{0.2\textwidth}{Household non-land wealth$^\dag$ }  &     -0.19 &     -0.32 &     -0.46 &     14.02 &      2718 &      0.29 &      0.40 &      0.79 \\ 
 & (0.23)  & (0.28)  & (0.30)  \\  & [    0.60]  & [    0.60]  & [    0.60]  \\ \addlinespace[1ex] 
\multicolumn{9}{l}{\emph{Panel B.  Individual outcomes}}  \\ 
\addlinespace[1ex] \multirow[t]{ 3 }{0.2\textwidth}{Height-for-Age }  &     -0.01 &     -0.01 &      0.05 &     -1.75 &      2618 &      0.73 &      0.66 &      0.62 \\ 
 & (0.05)  & (0.05)  & (0.06)  \\  & [    1.00]  & [    1.00]  & [    1.00]  \\ \addlinespace[1ex] 
\multirow[t]{ 3 }{0.2\textwidth}{Weight-for-Age }  &     -0.06 &     -0.01 &      0.02 &     -0.87 &      2594 &      0.73 &      0.62 &      0.17 \\ 
 & (0.05)  & (0.04)  & (0.05)  \\  & [    1.00]  & [    1.00]  & [    1.00]  \\ \addlinespace[1ex] 
\multirow[t]{ 3 }{0.2\textwidth}{Mid-Upper Arm Circ }  &     -0.07 &     -0.02 &      0.06 &     -0.61 &      1981 &      0.57 &      0.57 &      0.17 \\ 
 & (0.07)  & (0.07)  & (0.09)  \\  & [    1.00]  & [    1.00]  & [    1.00]  \\ \addlinespace[1ex] 
\bottomrule
\end{tabular}

%% file: tables/heterogeneity_rank_corrs.tex
\begin{tabular}{l c c c c} 
\toprule 
 & Consumption & Wealth & Dietary diversity & Child growth \\ 
\midrule 
Consumption &  &  &  &  \\ 
Wealth & -0.29 &  &  &  \\ 
Dietary diversity & 0.07 & 0.07 &  &  \\ 
Child growth & 0.04 & 0.28 & 0.15 &  \\ 
\bottomrule 
\end{tabular}

%% file: 8-Gikuriro-Figures.tex
\begin{figure}[htbp]
\caption{Cost Equivalence versus Cost Effectiveness}\label{f:CE_CEff_picture}
\begin{center}
\includegraphics[width=.875\textwidth]{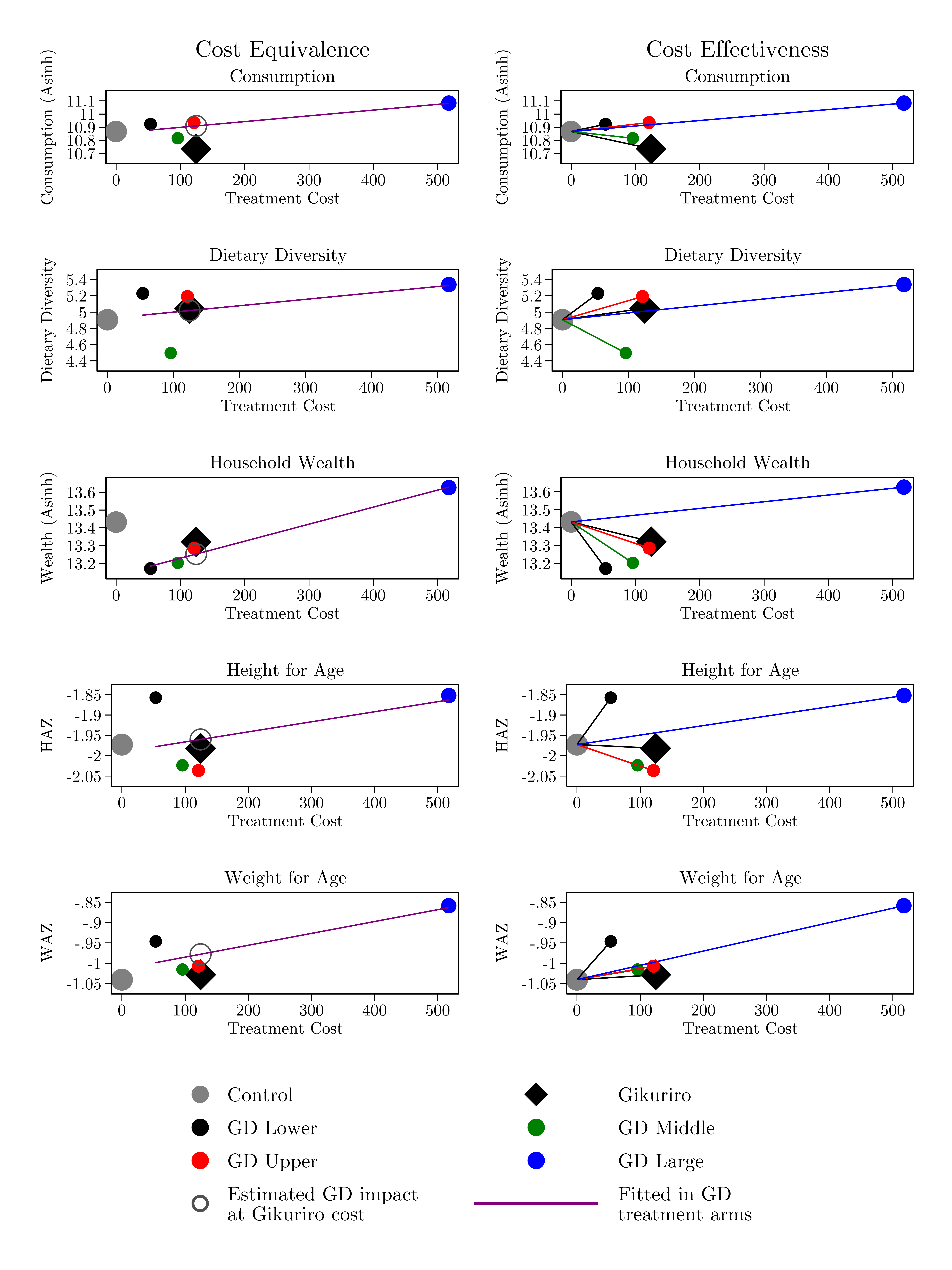}	
\end{center}

\vskip-5ex

\floatfoot{
	Notes:  Figures in left column visualize a cost-equivalent comparison (with no covariate adjustment). Dots represent mean outcomes in each treatment arm. Purple line represents a (population-weighted) regression of outcomes on treatment cost in the cash-transfer arms only.  Hollow circle represents the point on that regression line for which expenditure per beneficiary is equivalent to the ex-post cost of Gikuriro; cost-equivalent comparison compares this to the diamond, which is the mean outcome in the Gikuriro arm.  By contrast, figures in the right column illustrate a cost-benefit approach to comparing treatments (with no covariate adjustment).  Here, the slope of the ray extending from the control group to the  relevant treatment-specific mean represents the benefit-cost ratio; policymakers following this approach would favor the arm with the steepest slope.
}
\end{figure}

\begin{figure}[htbp]
\caption{Impacts on Dietary Diversity}\label{f:foodgroups_endline}
\begin{center}
\includegraphics[scale=.6]{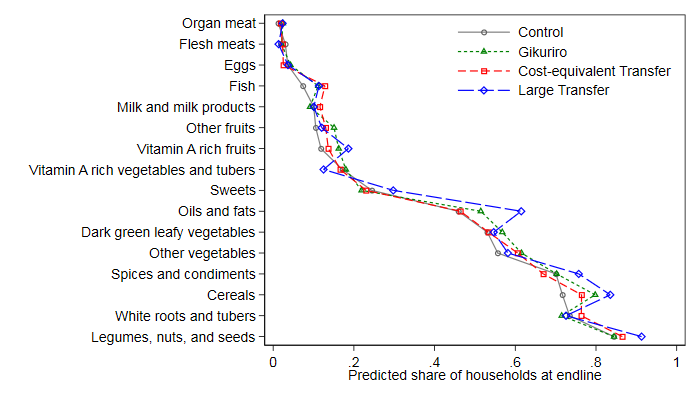}	
\end{center}

\floatfoot{
  Notes:  Figure presents estimates of shares of households consuming goods that comprise the dietary diversity score in the past 7 days, by treatment arm.  Control group represents unadjusted mean consumption rate of each food type.  Other treatment arms represent the value in the control group, with the estimated impact of that treatment added, using the regression specification used to estimate intent-to-treat program impacts in Section \ref{s:itt}.  Food types sorted by the share of control-group households who consume each item.  
}

\end{figure}

%% file: 9-Gikuriro-Appendix-Tables.tex
\begin{landscape}

\begin{table}\caption{Attrition}
\label{t:attrition}

\begin{footnotesize}
\begin{center}
\input{tables/attrition.tex}
\end{center}

\vskip-4ex 

\floatfoot{
	Notes:  First two columns analyze household-level attrition from the baseline sample of households.  (3) and (4) examine individual-level attrition from the roster of household members.  (5) and (6) examine attrition from the baseline sample of child anthropometrics.  (7) and (8) analyze attrition among individuals who should have received anemia testing, and (9) analyzes the probability of new household members appearing in rosters.
	
}
\end{footnotesize}
\end{table}

\end{landscape}

\begin{table}\caption{Balance on Primary Outcomes}
\label{t:balance_primary}

\begin{footnotesize}
\begin{center}
\input{tables/balance_primary.tex}

\end{center}

\vskip-4ex 

\floatfoot{
	Notes:  Columns present coefficients and standard errors from a regression of each baseline outcome on treatment indicators, with fixed effects for blocks.   Asterices denote significance at the 10, 5, and 1 percent levels, and are based on clustered standard errors, in parentheses.  \citet{anderson2008multiple} sharpened $q$-values presented in brackets. Variables marked with a $\dag$ are in inverse hyperbolic sines. 
	
}
\end{footnotesize}
\end{table}

\clearpage 

\begin{table}\caption{Regressions with Granular GD Treatment Cells}
\label{t:itt_primary_Each_GD}

	\begin{footnotesize}
	\begin{center}
	\input{tables/itt_primary_Each_GD.tex}

	\end{center}
\vskip-4ex 

\floatfoot{
	Notes:  Analysis includes dummies for each of the four GD transfer amounts separately.  Asterices denote significance at the 10, 5, and 1 percent levels, and are based on clustered standard errors, in parentheses.  \citet{anderson2008multiple} sharpened $q$-values presented in brackets. Variables marked with a $\dag$ are in inverse hyperbolic sines.
}
\end{footnotesize}
\end{table}

\clearpage

\begin{table}
\caption{ITT Impacts on Stunting and Wasting}
\label{t:itt_stunting_wasting}
	\begin{footnotesize}
\begin{center}
\resizebox{0.97\textwidth}{!}{	
	\input{tables/itt_stunting_wasting}

}
\end{center}
\vskip-4ex 

\floatfoot{
	Notes:  Table reports the Intention to Treat Impacts of the study arms on the binary outcomes of stunting and wasting (HAZ and WAZ respectively <-2).  Asterices denote significance at the 10, 5, and 1 percent levels, and are based on clustered standard errors, in parentheses.  Variables marked with a $\dag$ are in inverse hyperbolic sines. 
}
\end{footnotesize}
\end{table}

\clearpage

\begin{table}
\caption{Robustness of Linearity in Primary Cost Equivalence Adjustment}
\label{t:ce_primary_robustness}
	\begin{footnotesize}
\begin{center}
\resizebox{0.97\textwidth}{!}{	
	\input{tables/ce_primary_robustness}

}
\end{center}
\vskip-4ex 

\floatfoot{
	Notes:  Table reports the coefficient on the differential effect of Gikuriro over cost-equivalent cash using seven different specifications.  Column 1 is the linear adjustment reported elsewhere.  Column 2 includes a quadratic, and column 3 a quadratic and cubic term in the cost deviations from Gikuriro.  Columns 4-7 leave out one of the cash treatment arms and repeat the linear cost adjustment.  Asterices denote significance at the 10, 5, and 1 percent levels, and are based on clustered standard errors, in parentheses.  Variables marked with a $\dag$ are in inverse hyperbolic sines. 
}
\end{footnotesize}
\end{table}

\begin{table}
\caption{Robustness of Linearity in Secondary Cost Equivalence Adjustment}
\label{t:ce_secondary_robustness}
	\begin{footnotesize}
\begin{center}
\resizebox{0.97\textwidth}{!}{	
	\input{tables/ce_secondary_robustness}

}
\end{center}
\vskip-4ex 

\floatfoot{
	Notes:  Table reports the coefficient on the differential effect of Gikuriro over cost-equivalent cash using seven different specifications.  Column 1 is the linear adjustment reported elsewhere.  Column 2 includes a quadratic, and column 3 a quadratic and cubic term in the cost deviations from Gikuriro.  Columns 4-7 leave out one of the cash treatment arms and repeat the linear cost adjustment.  Asterices denote significance at the 10, 5, and 1 percent levels, and are based on clustered standard errors, in parentheses.  Variables marked with a $\dag$ are in inverse hyperbolic sines. 
}
\end{footnotesize}
\end{table}

\clearpage

\begin{table}[hbtp]
\caption{Village-level total causal effects: Secondary outcomes}
\label{t:tce_secondary}
\begin{footnotesize}
\begin{center}
\resizebox{0.9\textwidth}{!}{
\input{tables/tce_secondary.tex}

}
\end{center}

\vskip-4ex 

\floatfoot{
	Notes:  Analysis pools eligible and ineligible households and is weighted to be representative of the population in study villages.  Asterices denote significance at the 10, 5, and 1 percent levels, and are based on clustered standard errors, in parentheses.  \citet{anderson2008multiple} sharpened $q$-values presented in brackets. Variables marked with a $\dag$ are in inverse hyperbolic sines.  Final two columns present $p$-values on tests of the equality of benefit-cost ratios between arms.
}
\end{footnotesize}
\end{table}

\begin{table}\caption{Estimated spillover effects of the GD intervention on primary outcomes}
\label{t:spillovers_primary}

	\begin{footnotesize}
	\begin{center}
	\input{tables/spillovers_primary.tex}

	\end{center}
\vskip-4ex 

\floatfoot{
	Notes:  Analysis of spillover effects of cash within the strata (non-poor or no kids) not eligible to be treated by GD Weighted regression, using product of sampling weight and probability that an observation is \emph{not} treated, if in control village.  Households treated by GD in villages assigned to that arm receive a weight of zero.   $p$-values for test of equality of ratio between effect sizes and costs per eligible individual in the village.
}
\end{footnotesize}
\end{table}

\begin{table}\caption{Estimated spillover effects of the GD intervention on secondary outcomes}
\label{t:spillovers_secondary}

\begin{footnotesize}
\begin{center}
\resizebox{0.75\linewidth}{!}{
\input{tables/spillovers_secondary.tex}

}
\end{center}
	
\vskip-4ex 

\floatfoot{
	Notes:  Weighted regression, using product of sampling weight and probability that an observation is \emph{not} treated, if in control village.  Households treated by GD in villages assigned to that arm receive a weight of zero.   $p$-values for test of equality of ratio between effect sizes and costs per eligible individual in the village.
}
\end{footnotesize}
\end{table}

\begin{table}
\caption{Comparison of Lump Sum and Flow Transfers:  Primary}
\label{t:lumpsum_flow_primary}
	\begin{footnotesize}
	\begin{center}
	\resizebox{\linewidth}{!}{
	\input{tables/lumpsum_flow_primary.tex}

	}
	\end{center}
\vskip-4ex 

\floatfoot{
	Notes:  Analysis uses only the Control and GD arms; the first and third columns are dummies for GD transfer amount cells and the second and fourth are dummies measuring the additional effect of lump sum transfers within each cell.   Asterices denote significance at the 10, 5, and 1 percent levels, and are based on clustered standard errors, in parentheses.  \citet{anderson2008multiple} sharpened $q$-values presented in brackets. Variables marked with a $\dag$ are in inverse hyperbolic sines. 
}
\end{footnotesize}
\end{table}

\begin{table}
\caption{Comparison of Lump Sum and Flow Transfers:  Secondary}
\label{t:lumpsum_flow_secondary}

	\begin{footnotesize}
	\begin{center}
	\resizebox{0.9\textwidth}{!}{
	\input{tables/lumpsum_flow_secondary.tex}

	}
	\end{center}
\vskip-4ex 

\floatfoot{
	Notes:  Analysis uses only the Control and GD arms; the first and third columns are dummies for GD transfer amount cells and the second and fourth are dummies measuring the additional effect of lump sum transfers within each cell.   Asterices denote significance at the 10, 5, and 1 percent levels, and are based on clustered standard errors, in parentheses.  \citet{anderson2008multiple} sharpened $q$-values presented in brackets. Variables marked with a $\dag$ are in inverse hyperbolic sines. 
}
\end{footnotesize}
\end{table}

\begin{table}
\caption{Effect of Transfer Modality Choice:  Primary}
\label{t:choice_primary}
	\begin{footnotesize}
	\begin{center}
	\input{tables/choice_primary.tex}

	\end{center}
\vskip-4ex 

\floatfoot{
	Notes:  Analysis uses only the GD arm.  First column is a dummy for getting the chosen transfer modality, the second column is an (endogenous) indicator for choosing lump sum, and the third column is a dummy for actually receiving the lump sum treatment.  Asterices denote significance at the 10, 5, and 1 percent levels, and are based on clustered standard errors, in parentheses.  \citet{anderson2008multiple} sharpened $q$-values presented in brackets. Variables marked with a $\dag$ are in inverse hyperbolic sines. 
}
\end{footnotesize}
\end{table}

\begin{table}
\caption{Effect of Transfer Modality Choice:  Secondary}
\label{t:choice_secondary}

\begin{footnotesize}
\begin{center}
\resizebox{0.85\textwidth}{!}{	
	\input{tables/choice_secondary.tex}

}
\end{center}

\vskip-4ex 

\floatfoot{
	Notes:  Analysis uses only the GD arm.  First column is a dummy for getting the chosen transfer modality, the second column is an (endogenous) indicator for choosing lump sum, and the third column is a dummy for actually receiving the lump sum treatment.  Asterices denote significance at the 10, 5, and 1 percent levels, and are based on clustered standard errors, in parentheses.  \citet{anderson2008multiple} sharpened $q$-values presented in brackets. Variables marked with a $\dag$ are in inverse hyperbolic sines. 
}
\end{footnotesize}
\end{table}

\cleardoublepage

\begin{figure}[!hp]
\caption{Heterogeneity in cash impacts relative to in-kind programming}
\label{f:heterogeneity}

\begin{center}
\subfloat{
   \includegraphics[width=0.35\textwidth]{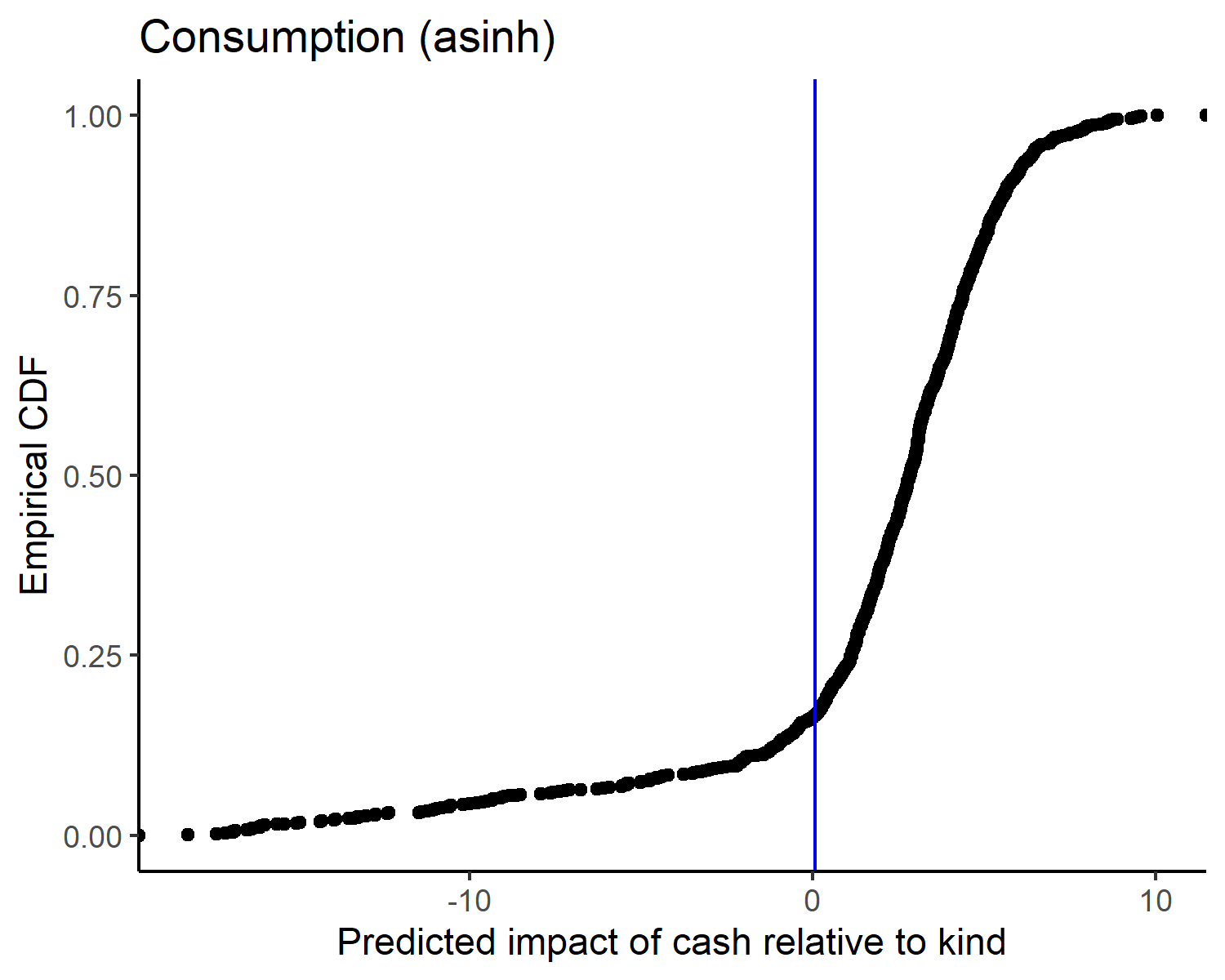}
}
\subfloat{
   \includegraphics[width=0.35\textwidth]{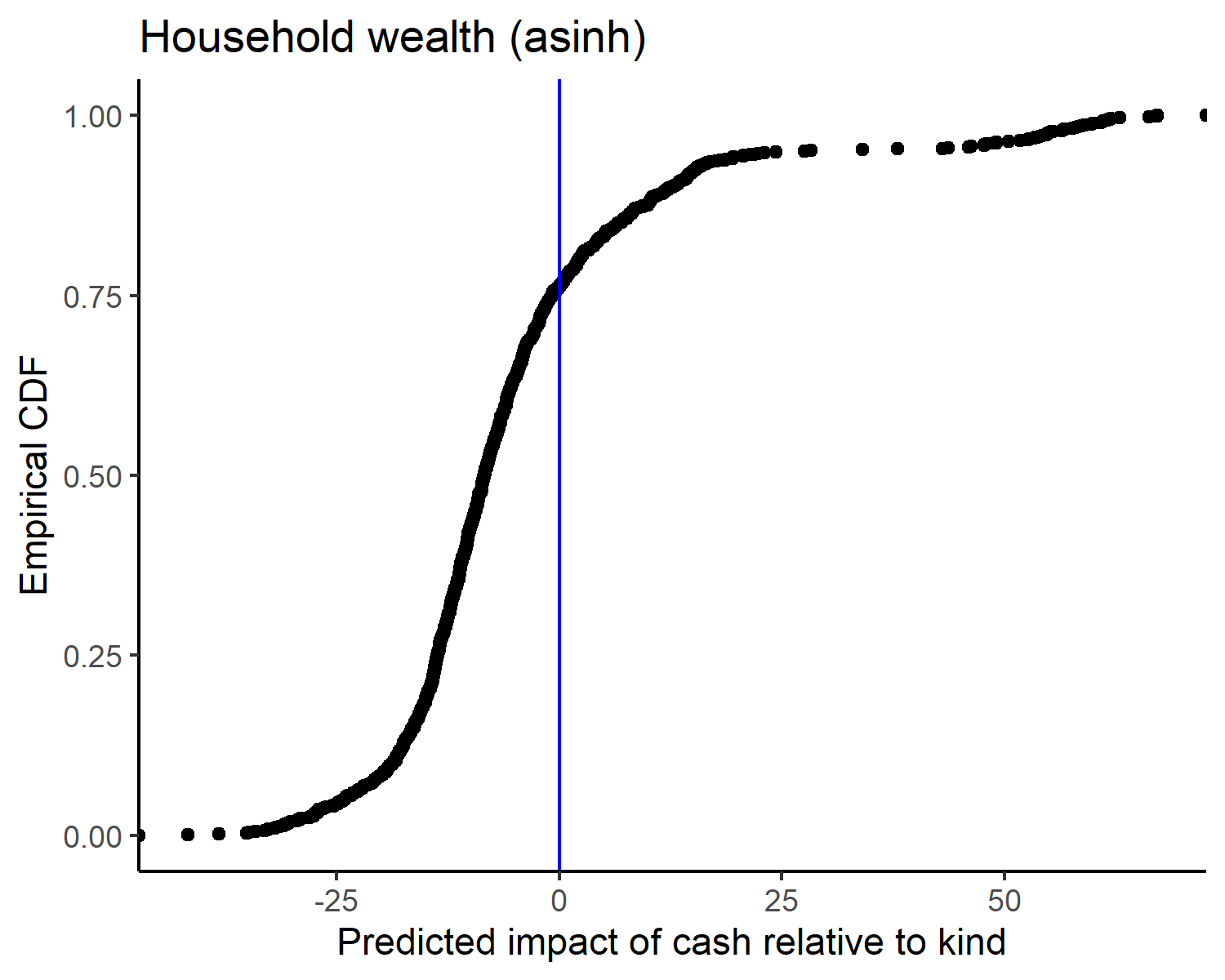} 
} \\ 
\subfloat{
	\includegraphics[width=0.35\textwidth]{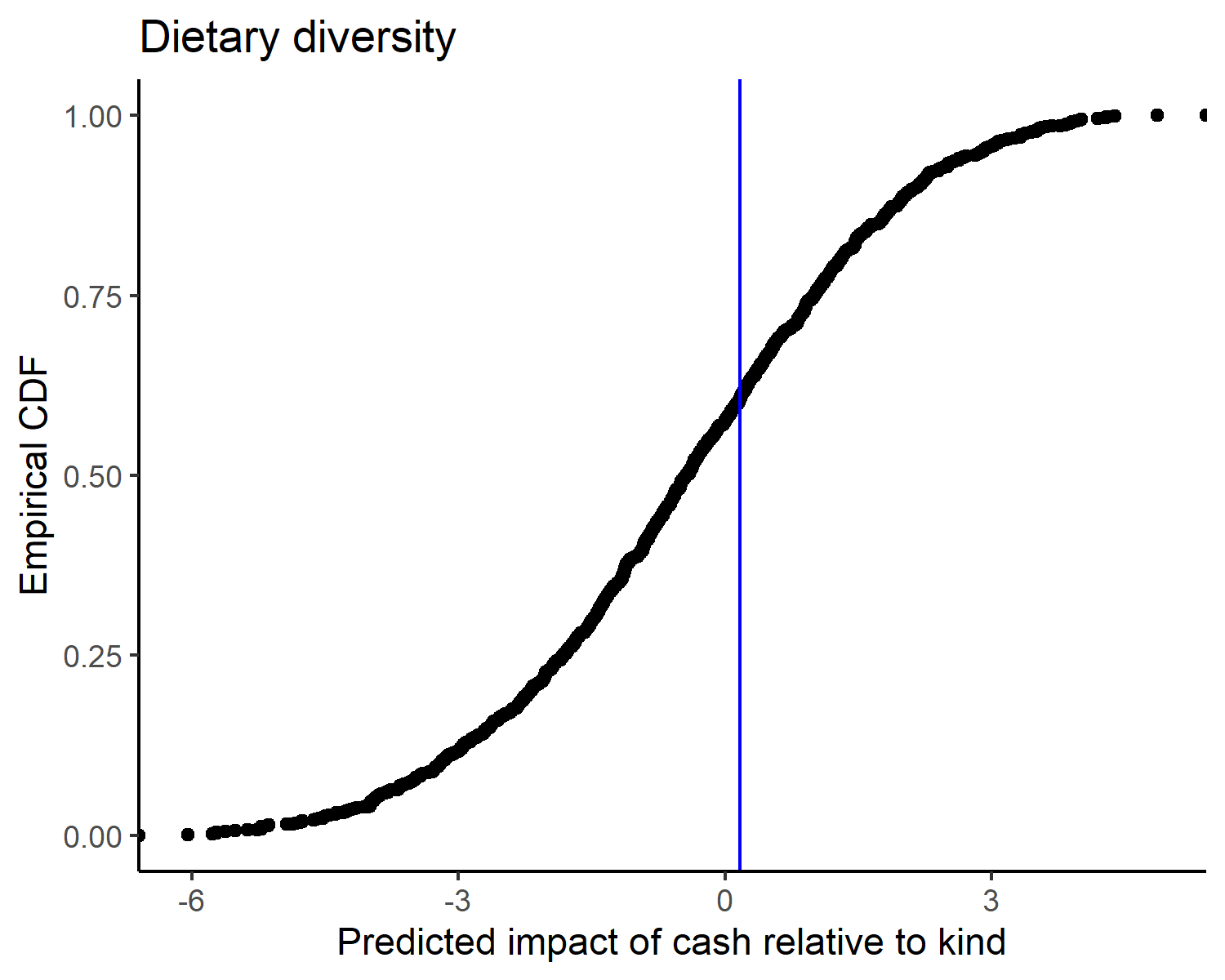}
}
\subfloat{
	\includegraphics[width=0.35\textwidth]{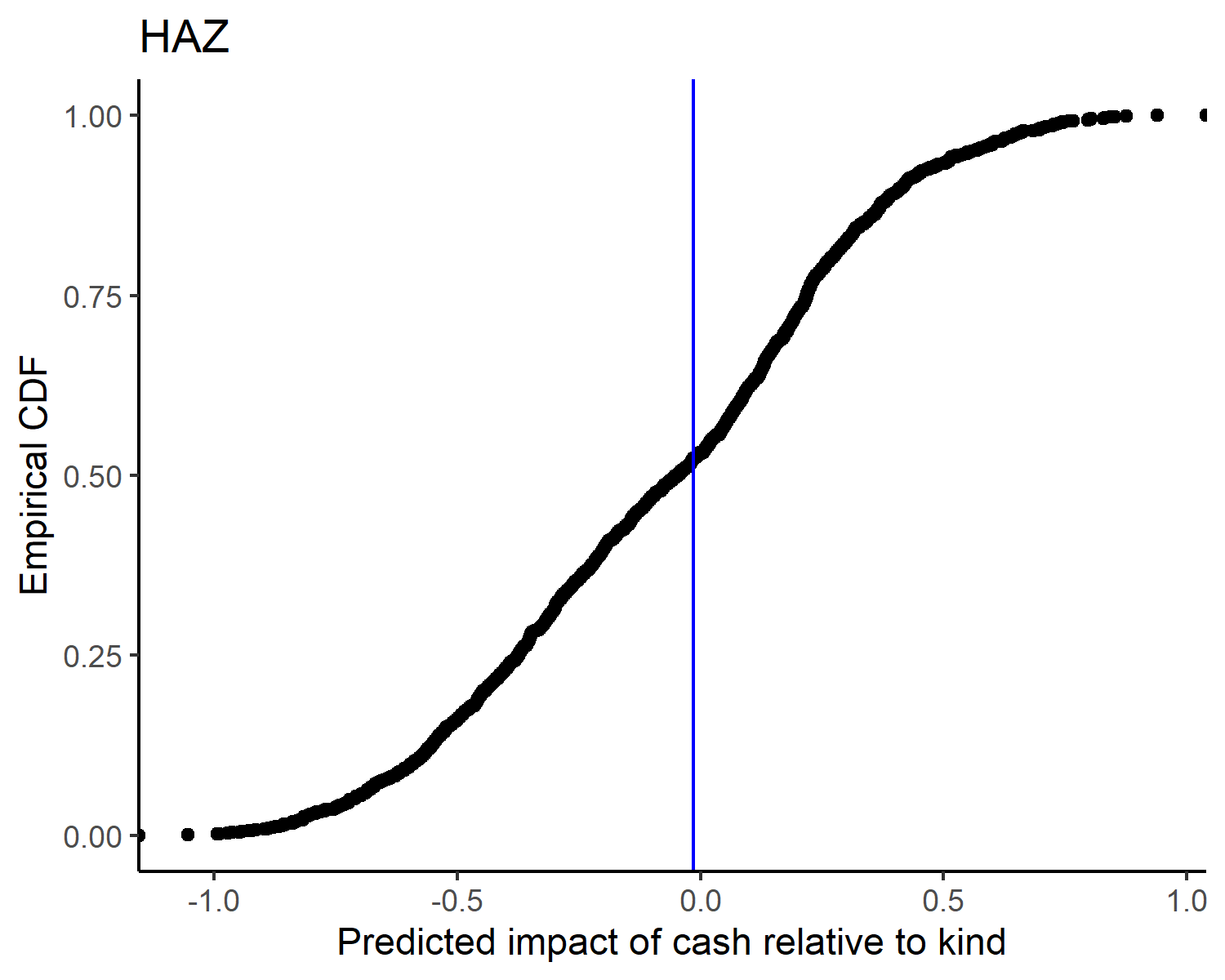}
} \\
\subfloat{
	\includegraphics[width=0.35\textwidth]{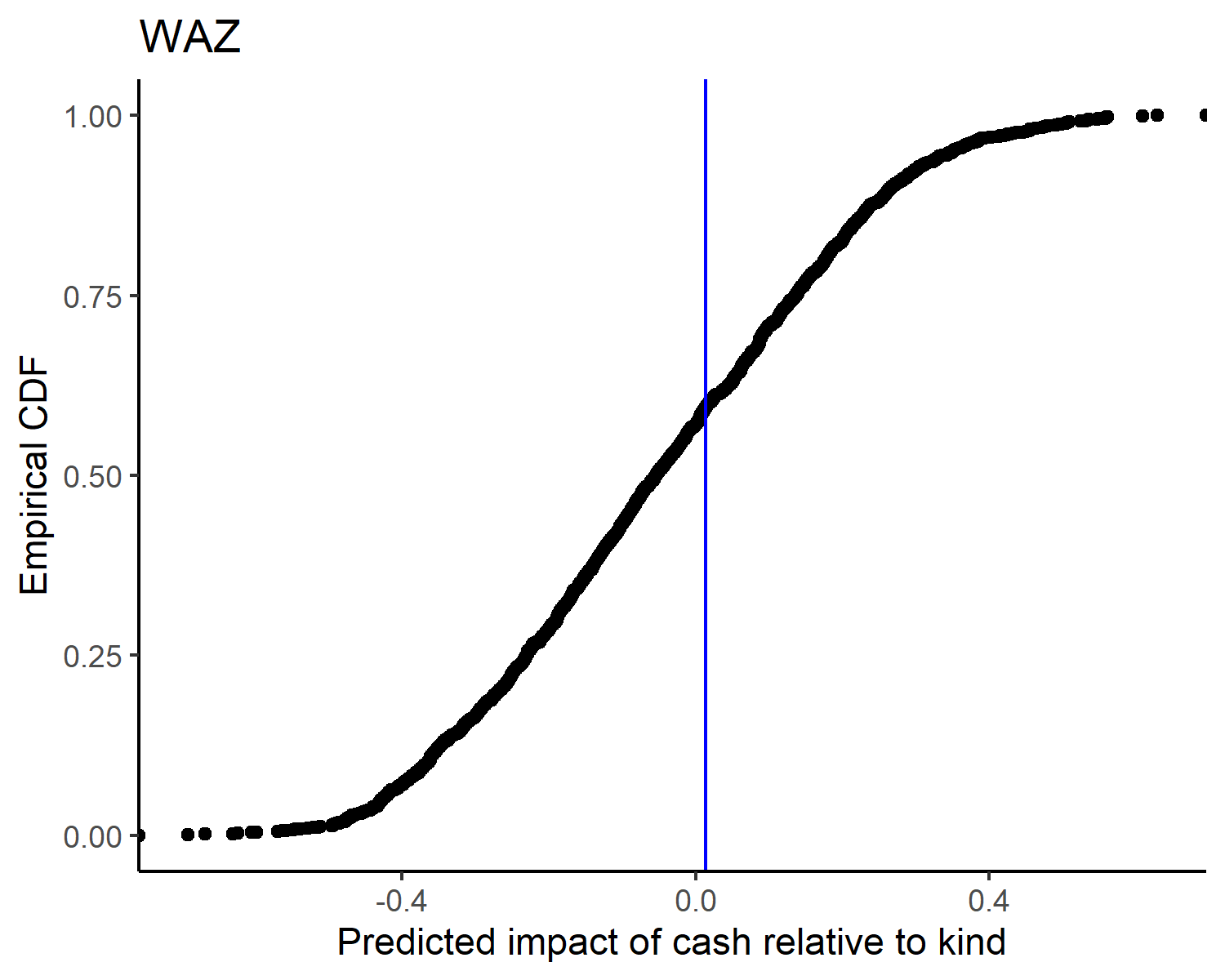}
} 
\subfloat{
	\includegraphics[width=0.35\textwidth]{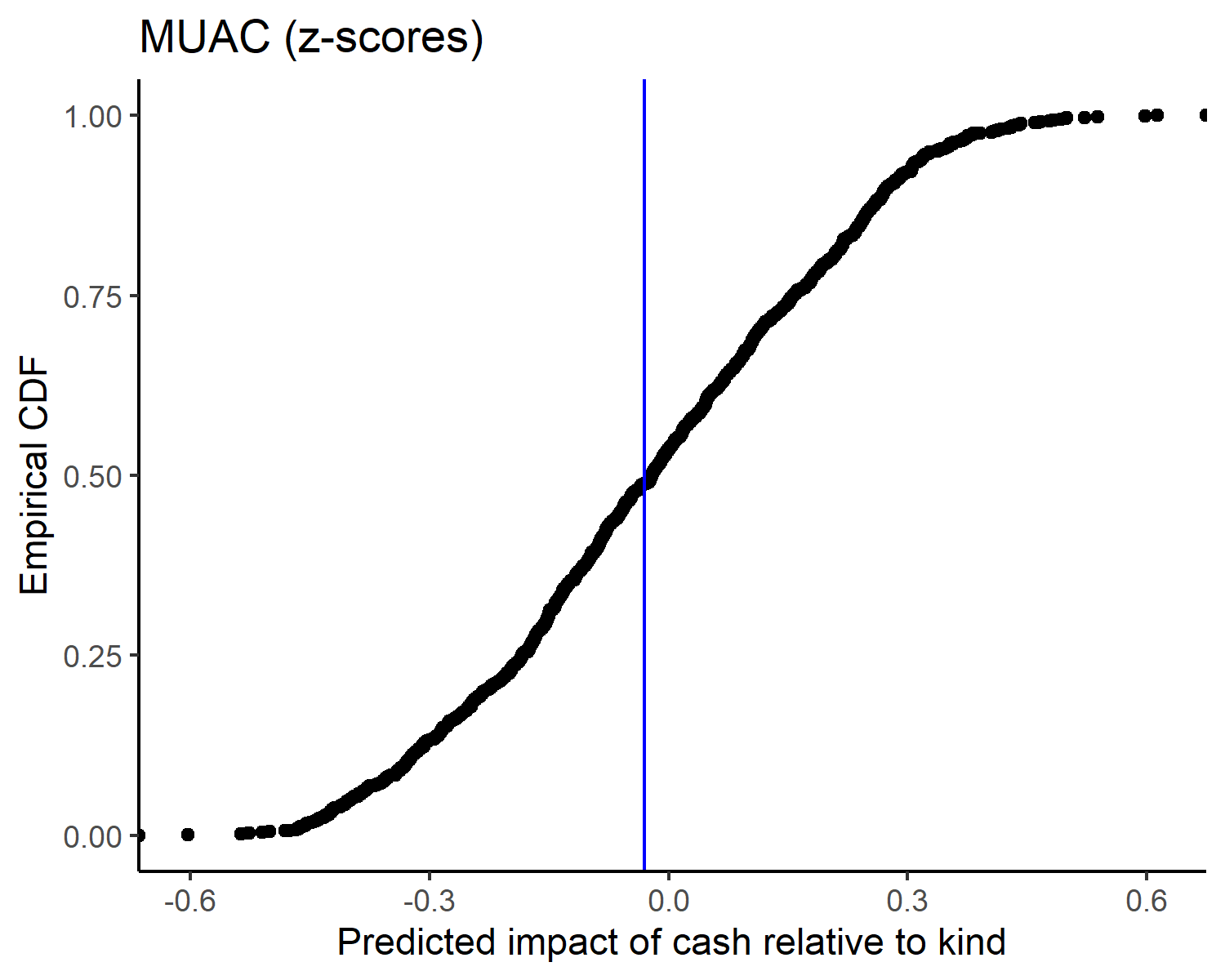}
}
\\ 
\subfloat{
	\includegraphics[width=0.35\textwidth]{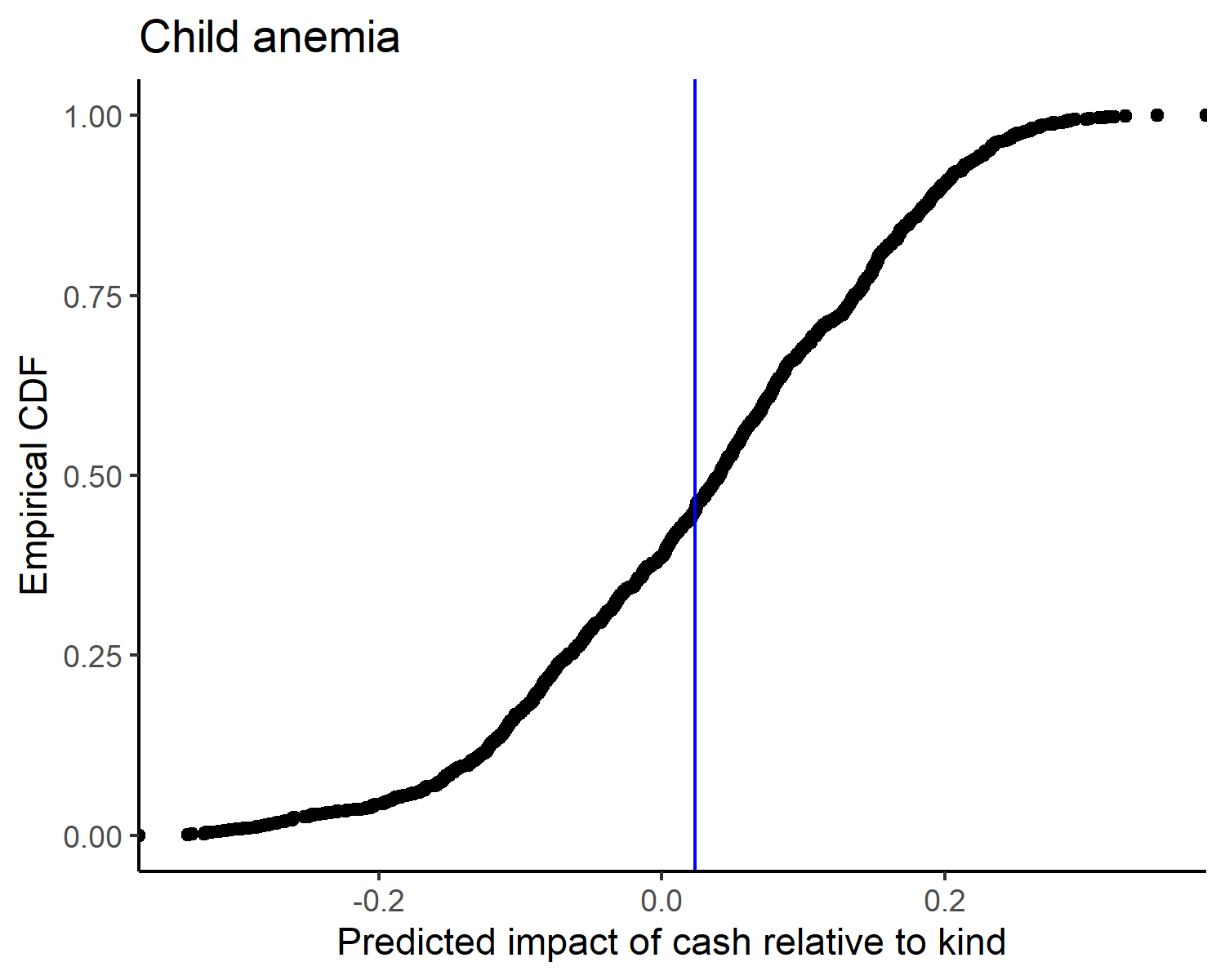}
}
\end{center}

\floatfoot{
    Notes:  Panels display CDFs of subgroup average impacts of receiving GD Main transfer as opposed to Gikuriro, estimated by causal forest.  Estimates presented for all primary household and child-level outcomes.
}

\end{figure}

\cleardoublepage

%% file: tables/attrition.tex
{
\def\sym#1{\ifmmode^{#1}\else\(^{#1}\)\fi}
\begin{tabular}{l*{9}{c}}
\hline\hline
                    &\multicolumn{1}{c}{(1)}&\multicolumn{1}{c}{(2)}&\multicolumn{1}{c}{(3)}&\multicolumn{1}{c}{(4)}&\multicolumn{1}{c}{(5)}&\multicolumn{1}{c}{(6)}&\multicolumn{1}{c}{(7)}&\multicolumn{1}{c}{(8)}&\multicolumn{1}{c}{(9)}\\
                    &\multicolumn{1}{c}{Household}&\multicolumn{1}{c}{Household}&\multicolumn{1}{c}{Roster}&\multicolumn{1}{c}{Roster}&\multicolumn{1}{c}{Anthro}&\multicolumn{1}{c}{Anthro}&\multicolumn{1}{c}{Anemia}&\multicolumn{1}{c}{Anemia}&\multicolumn{1}{c}{New Member}\\
                    &        b/se   &        b/se   &        b/se   &        b/se   &        b/se   &        b/se   &        b/se   &        b/se   &        b/se   \\
\hline
Gikuriro            &      -0.013   &      -0.011   &      -0.010   &     -0.0086   &      -0.037***&      -0.038***&     -0.0044   &     -0.0046   &    -0.00050   \\
                    &    (0.0092)   &    (0.0085)   &    (0.0083)   &    (0.0079)   &     (0.013)   &     (0.012)   &     (0.021)   &     (0.016)   &    (0.0081)   \\
GD Main             &     -0.0089   &     -0.0097   &     -0.0077   &     -0.0076   &      -0.024*  &      -0.025** &     -0.0069   &     -0.0013   &     -0.0027   \\
                    &     (0.010)   &     (0.010)   &    (0.0093)   &    (0.0097)   &     (0.013)   &     (0.013)   &     (0.019)   &     (0.016)   &    (0.0088)   \\
GD Large            &      -0.017*  &      -0.015   &      -0.015*  &      -0.013   &      -0.022   &      -0.020   &       0.027   &      0.0056   &     -0.0051   \\
                    &    (0.0097)   &    (0.0093)   &    (0.0086)   &    (0.0088)   &     (0.016)   &     (0.015)   &     (0.026)   &     (0.023)   &    (0.0096)   \\
Agricultural        &               &      -0.015   &               &      -0.020   &               &      -0.037*  &               &       0.023   &               \\
                    &               &     (0.012)   &               &     (0.013)   &               &     (0.020)   &               &     (0.023)   &               \\
Wage Worker         &               &      0.0076   &               &       0.016   &               &       0.019   &               &      0.0038   &               \\
                    &               &    (0.0093)   &               &    (0.0099)   &               &     (0.015)   &               &     (0.016)   &               \\
Microenterprise     &               &     -0.0083   &               &     -0.0085   &               &    -0.00060   &               &      -0.039*  &               \\
                    &               &    (0.0075)   &               &    (0.0063)   &               &     (0.017)   &               &     (0.022)   &               \\
Savings Group       &               &     0.00040   &               &     -0.0051   &               &      -0.015   &               &     -0.0020   &               \\
                    &               &    (0.0085)   &               &    (0.0071)   &               &     (0.012)   &               &     (0.017)   &               \\
Village Eligibility Ratio&               &       0.021   &               &       0.015   &               &       0.032   &               &     -0.0030   &               \\
                    &               &     (0.053)   &               &     (0.048)   &               &     (0.050)   &               &     (0.060)   &               \\
Age of Head         &               &    -0.00034   &               &    -0.00025   &               &     0.00013   &               &     -0.0017***&               \\
                    &               &   (0.00022)   &               &   (0.00019)   &               &   (0.00034)   &               &   (0.00052)   &               \\
Schooling of Head   &               &        0.11   &               &        0.21   &               &       0.053   &               &        0.13***&               \\
                    &               &      (0.14)   &               &      (0.21)   &               &      (0.11)   &               &     (0.025)   &               \\
Dependency Ratio    &               &       0.035   &               &       0.027   &               &       0.058   &               &       0.029   &               \\
                    &               &     (0.030)   &               &     (0.032)   &               &     (0.056)   &               &     (0.063)   &               \\
Household Size      &               &     -0.0036   &               &     -0.0031   &               &     -0.0060   &               &      -0.040***&               \\
                    &               &    (0.0023)   &               &    (0.0023)   &               &    (0.0040)   &               &    (0.0058)   &               \\
Poorest Category    &               &      0.0097   &               &     -0.0055   &               &      0.0065   &               &      -0.017   &               \\
                    &               &     (0.011)   &               &    (0.0098)   &               &     (0.018)   &               &     (0.021)   &               \\
Next Poorest Category&               &      0.0061   &               &     -0.0048   &               &      0.0059   &               &      0.0064   &               \\
                    &               &    (0.0085)   &               &    (0.0094)   &               &     (0.014)   &               &     (0.017)   &               \\
female              &               &               &               &      0.0011   &               &      0.0012   &               &      -0.045***&               \\
                    &               &               &               &    (0.0026)   &               &    (0.0088)   &               &     (0.014)   &               \\
\hline
Control group mean  &       0.033   &       0.033   &       0.030   &       0.030   &       0.071   &       0.071   &        0.80   &        0.80   &       0.057   \\
Observations        &        1793   &        1793   &        9189   &        9189   &        2265   &        2265   &        5192   &        5192   &        9509   \\
$R^2$               &      0.0019   &       0.023   &      0.0014   &       0.028   &      0.0049   &       0.025   &     0.00068   &       0.068   &    0.000054   \\
\hline\hline
\end{tabular}
}

%% file: tables/balance_primary.tex
\begin{tabular}{l *{3}{S} ScS}
\toprule
\multicolumn{1}{c}{\text{ }} & \multicolumn{1}{c}{\text{\makecell[b]{Gikuriro\\Village}}} & \multicolumn{1}{c}{\text{\makecell[b]{GD Main\\Village}}} & \multicolumn{1}{c}{\text{\makecell[b]{GD Large\\Village}}} & \multicolumn{1}{c}{\text{\makecell[b]{Control\\Mean}}} & \multicolumn{1}{c}{\text{Observations}} & \multicolumn{1}{c}{\text{$ R^2$}}\\
\midrule
\multicolumn{7}{l}{\emph{A.  Household outcomes}}  \\ 
\addlinespace[1ex] \multirow[t]{ 3 }{0.2\textwidth}{Consumption$^\dag$ }  &     0.053 &     0.047 &    -0.103 &     10.39 &      1751 &      0.05 \\ 
 & (    0.116)  & (    0.122)  & (    0.130)  \\  & [    1.00]  & [    1.00]  & [    1.00]  \\ \addlinespace[1ex] 
\multirow[t]{ 3 }{0.2\textwidth}{Household dietary diversity score }  &    -0.064 &    -0.071 &    -0.058 &      4.16 &      1751 &      0.10 \\ 
 & (    0.137)  & (    0.139)  & (    0.172)  \\  & [    1.00]  & [    1.00]  & [    1.00]  \\ \addlinespace[1ex] 
\multirow[t]{ 3 }{0.2\textwidth}{Household non-land wealth$^\dag$ }  &     0.027 &    -0.042 &    -0.288 &     12.94 &      1751 &      0.06 \\ 
 & (    0.227)  & (    0.214)  & (    0.254)  \\  & [    1.00]  & [    1.00]  & [    1.00]  \\ \addlinespace[1ex] 
\multicolumn{7}{l}{\emph{B.  Individual outcomes}}  \\ 
\addlinespace[1ex] \multirow[t]{ 3 }{0.2\textwidth}{Height-for-Age }  &     0.046 &     0.080 &     0.215\ensuremath{^{**}} &     -1.93 &      2187 &      0.02 \\ 
 & (    0.086)  & (    0.097)  & (    0.098)  \\  & [    1.00]  & [    1.00]  & [    0.13]  \\ \addlinespace[1ex] 
\multirow[t]{ 3 }{0.2\textwidth}{Weight-for-Age }  &     0.023 &     0.041 &     0.187\ensuremath{^{***}} &     -1.06 &      2180 &      0.02 \\ 
 & (    0.069)  & (    0.070)  & (    0.069)  \\  & [    1.00]  & [    1.00]  & [    0.07]  \\ \addlinespace[1ex] 
\multirow[t]{ 3 }{0.2\textwidth}{Mid-Upper Arm Circ }  &     0.015 &     0.025 &     0.070 &     -0.72 &      1987 &      0.04 \\ 
 & (    0.068)  & (    0.068)  & (    0.081)  \\  & [    1.00]  & [    1.00]  & [    1.00]  \\ \addlinespace[1ex] 
\bottomrule
\end{tabular}

%% file: tables/itt_primary_Each_GD.tex
\begin{tabular}{l *{3}{S} ScSSS}
\toprule
 & & \multicolumn{4}{c}{GiveDirectly} &  \multicolumn{1}{c}{\raisebox{-1ex}[0pt]{Control}} \\ 
\cmidrule(lr){3-6}
\multicolumn{1}{c}{\text{ }} & \multicolumn{1}{c}{\text{Gikuriro}} & \multicolumn{1}{c}{\text{Small}} & \multicolumn{1}{c}{\text{Mid}} & \multicolumn{1}{c}{\text{Upper}} & \multicolumn{1}{c}{\text{Large}} & \multicolumn{1}{c}{\text{Mean}} & \multicolumn{1}{c}{\text{Obs.}} & \multicolumn{1}{c}{\text{$ R^2$}}\\
\midrule
\multicolumn{9}{l}{\emph{Panel A.  Household outcomes}}  \\ 
\addlinespace[1ex] \multirow[t]{ 3 }{0.2\textwidth}{Consumption$^\dag$ }  &     -0.11 &      0.10 &     -0.02 &      0.10 &      0.30\ensuremath{^{***}} &     10.69 &      1750 &      0.14 \\ 
 & (0.10)  & (0.11)  & (0.12)  & (0.12)  & (0.11)  \\  & [    0.48]  & [    0.66]  & [    0.96]  & [    0.66]  & [    0.04]  \\ \addlinespace[1ex] 
\multirow[t]{ 3 }{0.2\textwidth}{Household dietary diversity score }  &      0.20 &      0.40\ensuremath{^{**}} &     -0.25 &      0.36\ensuremath{^{**}} &      0.56\ensuremath{^{***}} &      4.77 &      1751 &      0.19 \\ 
 & (0.12)  & (0.19)  & (0.22)  & (0.15)  & (0.13)  \\  & [    0.29]  & [    0.12]  & [    0.48]  & [    0.10]  & [    0.00]  \\ \addlinespace[1ex] 
\multirow[t]{ 3 }{0.2\textwidth}{Household non-land wealth$^\dag$ }  &      0.01 &      0.06 &      0.04 &     -0.09 &      0.40 &     13.04 &      1751 &      0.22 \\ 
 & (0.19)  & (0.40)  & (0.26)  & (0.26)  & (0.28)  \\  & [    0.96]  & [    0.96]  & [    0.96]  & [    0.96]  & [    0.41]  \\ \addlinespace[1ex] 
\multicolumn{9}{l}{\emph{Panel B.  Individual outcomes}}  \\ 
\addlinespace[1ex] \multirow[t]{ 3 }{0.2\textwidth}{Height-for-age }  &      0.05 &      0.04 &     -0.04 &     -0.05 &      0.09\ensuremath{^{**}} &     -1.97 &      2125 &      0.71 \\ 
 & (0.04)  & (0.05)  & (0.06)  & (0.05)  & (0.05)  \\  & [    1.00]  & [    1.00]  & [    1.00]  & [    1.00]  & [    1.00]  \\ \addlinespace[1ex] 
\multirow[t]{ 3 }{0.2\textwidth}{Weight-for-age }  &      0.04 &      0.01 &     -0.01 &      0.03 &      0.07\ensuremath{^{*}} &     -1.04 &      2104 &      0.68 \\ 
 & (0.04)  & (0.05)  & (0.05)  & (0.04)  & (0.04)  \\  & [    1.00]  & [    1.00]  & [    1.00]  & [    1.00]  & [    1.00]  \\ \addlinespace[1ex] 
\multirow[t]{ 3 }{0.2\textwidth}{Mid-upper arm circumference }  &      0.02 &     -0.04 &      0.07 &     -0.06 &      0.13\ensuremath{^{*}} &     -0.59 &      1629 &      0.51 \\ 
 & (0.06)  & (0.08)  & (0.12)  & (0.07)  & (0.08)  \\  & [    1.00]  & [    1.00]  & [    1.00]  & [    1.00]  & [    1.00]  \\ \addlinespace[1ex] 
\multirow[t]{ 3 }{0.2\textwidth}{Child anemia }  &      0.00 &      0.05 &      0.01 &      0.02 &     -0.01 &      0.18 &      2372 &      0.07 \\ 
 & (0.02)  & (0.06)  & (0.03)  & (0.02)  & (0.04)  \\  & [    1.00]  & [    1.00]  & [    1.00]  & [    1.00]  & [    1.00]  \\ \addlinespace[1ex] 
\multirow[t]{ 3 }{0.2\textwidth}{Maternal anemia }  &     -0.02 &     -0.03 &     -0.00 &      0.03 &     -0.03 &      0.12 &      1581 &      0.11 \\ 
 & (0.03)  & (0.03)  & (0.03)  & (0.05)  & (0.03)  \\  & [    1.00]  & [    1.00]  & [    1.00]  & [    1.00]  & [    1.00]  \\ \addlinespace[1ex] 
\bottomrule
\end{tabular}

%% file: tables/itt_stunting_wasting.tex
\begin{tabular}{l *{3}{S} ScSSS}
\toprule
 & & \multicolumn{2}{c}{GiveDirectly} &  \multicolumn{1}{c}{\raisebox{-1ex}[0pt]{Control}} & & &\multicolumn{2}{c}{$ p$-values: B/C ratios} \\ 
\multicolumn{1}{c}{\text{ }} & \multicolumn{1}{c}{\text{Gikuriro}} & \multicolumn{1}{c}{\text{Main}} & \multicolumn{1}{c}{\text{Large}} & \multicolumn{1}{c}{\text{Mean}} & \multicolumn{1}{c}{\text{Obs.}} & \multicolumn{1}{c}{\text{$ R^2$}} & \multicolumn{1}{c}{\text{GD=GDL}} & \multicolumn{1}{c}{\text{GK=GDL}}\\
\midrule
\multirow[t]{ 3 }{0.2\textwidth}{Stunted }  &      0.01 &      0.02 &     -0.06\ensuremath{^{*}} &      0.50 &      2360 &      0.04 &      0.37 &      0.41 \\ 
 & (0.04)  & (0.04)  & (0.04)  \\  & [    1.00]  & [    1.00]  & [    0.35]  \\ \addlinespace[1ex] 
\multirow[t]{ 3 }{0.2\textwidth}{Wasted }  &      0.00 &     -0.01 &     -0.05\ensuremath{^{*}} &      0.16 &      2347 &      0.03 &      0.96 &      0.50 \\ 
 & (0.02)  & (0.02)  & (0.03)  \\  & [    1.00]  & [    1.00]  & [    0.35]  \\ \addlinespace[1ex] 
\bottomrule
\end{tabular}

%% file: tables/ce_primary_robustness.tex
\begin{tabular}{l *{7}{S} ScS}
\toprule
\multicolumn{1}{c}{\text{ }} & \multicolumn{1}{c}{\text{\makecell[b]{Base\\Linear}}} & \multicolumn{1}{c}{\text{\makecell[b]{Quad-\\ratic}}} & \multicolumn{1}{c}{\text{\makecell[b]{Cubic\\}}} & \multicolumn{1}{c}{\text{\makecell[b]{Drop\\lower}}} & \multicolumn{1}{c}{\text{\makecell[b]{Drop\\mid}}} & \multicolumn{1}{c}{\text{\makecell[b]{Drop\\upper}}} & \multicolumn{1}{c}{\text{\makecell[b]{Drop\\huge}}}\\
\midrule
\multicolumn{8}{l}{\emph{A.  Household outcomes}}  \\ 
\addlinespace[1ex] \multirow[t]{ 2 }{0.2\textwidth}{Consumption$^\dag$ }  &    -0.193\ensuremath{^{**}} &    -0.164 &    -0.240\ensuremath{^{*}} &    -0.158\ensuremath{^{*}} &    -0.226\ensuremath{^{***}} &    -0.175\ensuremath{^{**}} &    -0.185\ensuremath{^{*}} \\ 
 & (    0.079)  & (    0.102)  & (    0.127)  & (    0.086)  & (    0.086)  & (    0.084)  & (    0.103)  \\ \addlinespace[1ex] 
\multirow[t]{ 2 }{0.2\textwidth}{Household dietary diversity score }  &    -0.010 &     0.107 &    -0.278 &     0.139 &    -0.208\ensuremath{^{*}} &     0.047 &     0.109 \\ 
 & (    0.132)  & (    0.179)  & (    0.173)  & (    0.155)  & (    0.125)  & (    0.155)  & (    0.183)  \\ \addlinespace[1ex] 
\multirow[t]{ 2 }{0.2\textwidth}{Household non-land wealth$^\dag$ }  &    -0.032 &     0.066 &     0.117 &     0.010 &    -0.028 &    -0.095 &     0.104 \\ 
 & (    0.199)  & (    0.246)  & (    0.294)  & (    0.192)  & (    0.245)  & (    0.228)  & (    0.255)  \\ \addlinespace[1ex] 
\multicolumn{8}{l}{\emph{B.  Individual outcomes}}  \\ 
\addlinespace[1ex] \multirow[t]{ 2 }{0.2\textwidth}{Height-for-age }  &     0.060 &     0.098\ensuremath{^{*}} &     0.062 &     0.081\ensuremath{^{*}} &     0.030 &     0.047 &     0.095\ensuremath{^{*}} \\ 
 & (    0.040)  & (    0.053)  & (    0.062)  & (    0.047)  & (    0.043)  & (    0.046)  & (    0.053)  \\ \addlinespace[1ex] 
\multirow[t]{ 2 }{0.2\textwidth}{Weight-for-age }  &     0.024 &     0.009 &    -0.032 &     0.023 &     0.005 &     0.038 &     0.002 \\ 
 & (    0.039)  & (    0.048)  & (    0.055)  & (    0.043)  & (    0.043)  & (    0.043)  & (    0.049)  \\ \addlinespace[1ex] 
\multirow[t]{ 2 }{0.2\textwidth}{Mid-upper arm circumference }  &     0.017 &     0.037 &     0.108 &     0.018 &     0.055 &    -0.004 &     0.035 \\ 
 & (    0.062)  & (    0.084)  & (    0.087)  & (    0.079)  & (    0.061)  & (    0.076)  & (    0.079)  \\ \addlinespace[1ex] 
\multirow[t]{ 2 }{0.2\textwidth}{Anemia }  &    -0.018 &    -0.003 &    -0.016 &    -0.005 &    -0.027 &    -0.019 &    -0.002 \\ 
 & (    0.026)  & (    0.025)  & (    0.029)  & (    0.021)  & (    0.032)  & (    0.031)  & (    0.026)  \\ \addlinespace[1ex] 
\multirow[t]{ 2 }{0.2\textwidth}{Anemia }  &    -0.018 &    -0.052 &    -0.059 &    -0.035 &    -0.017 &    -0.003 &    -0.052 \\ 
 & (    0.021)  & (    0.036)  & (    0.049)  & (    0.026)  & (    0.024)  & (    0.017)  & (    0.036)  \\ \addlinespace[1ex] 
\bottomrule
\end{tabular}

%% file: tables/ce_secondary_robustness.tex
\begin{tabular}{l *{7}{S} ScS}
\toprule
\multicolumn{1}{c}{\text{ }} & \multicolumn{1}{c}{\text{\makecell[b]{Base\\Linear}}} & \multicolumn{1}{c}{\text{\makecell[b]{Quad-\\ratic}}} & \multicolumn{1}{c}{\text{\makecell[b]{Cubic\\}}} & \multicolumn{1}{c}{\text{\makecell[b]{Drop\\lower}}} & \multicolumn{1}{c}{\text{\makecell[b]{Drop\\mid}}} & \multicolumn{1}{c}{\text{\makecell[b]{Drop\\upper}}} & \multicolumn{1}{c}{\text{\makecell[b]{Drop\\huge}}}\\
\midrule
\multicolumn{8}{l}{\emph{A.  Household outcomes}}  \\ 
\addlinespace[1ex] \multirow[t]{ 2 }{0.2\textwidth}{Stock of borrowing$^\dag$ }  &     0.793\ensuremath{^{***}} &     0.917\ensuremath{^{**}} &     0.923\ensuremath{^{**}} &     0.847\ensuremath{^{***}} &     0.758\ensuremath{^{***}} &     0.767\ensuremath{^{***}} &     0.964\ensuremath{^{***}} \\ 
 & (    0.277)  & (    0.360)  & (    0.411)  & (    0.322)  & (    0.278)  & (    0.293)  & (    0.371)  \\ \addlinespace[1ex] 
\multirow[t]{ 2 }{0.2\textwidth}{Stock of saving$^\dag$ }  &     1.184\ensuremath{^{***}} &     1.071\ensuremath{^{**}} &     0.597 &     1.253\ensuremath{^{***}} &     0.958\ensuremath{^{***}} &     1.326\ensuremath{^{***}} &     1.037\ensuremath{^{**}} \\ 
 & (    0.333)  & (    0.488)  & (    0.595)  & (    0.392)  & (    0.350)  & (    0.337)  & (    0.500)  \\ \addlinespace[1ex] 
\multirow[t]{ 2 }{0.2\textwidth}{Health knowledge index }  &    -0.227 &    -0.112 &    -0.148 &    -0.140 &    -0.243 &    -0.285 &    -0.162 \\ 
 & (    0.329)  & (    0.380)  & (    0.442)  & (    0.343)  & (    0.357)  & (    0.366)  & (    0.379)  \\ \addlinespace[1ex] 
\multirow[t]{ 2 }{0.2\textwidth}{Sanitation practices index }  &    -0.421\ensuremath{^{**}} &    -0.397 &    -0.742\ensuremath{^{**}} &    -0.365 &    -0.569\ensuremath{^{**}} &    -0.352 &    -0.397 \\ 
 & (    0.213)  & (    0.264)  & (    0.304)  & (    0.224)  & (    0.247)  & (    0.240)  & (    0.261)  \\ \addlinespace[1ex] 
\multirow[t]{ 2 }{0.2\textwidth}{Productive assets$^\dag$ }  &    -0.283\ensuremath{^{***}} &    -0.241\ensuremath{^{*}} &    -0.283\ensuremath{^{*}} &    -0.246\ensuremath{^{**}} &    -0.309\ensuremath{^{***}} &    -0.287\ensuremath{^{***}} &    -0.218\ensuremath{^{*}} \\ 
 & (    0.097)  & (    0.129)  & (    0.157)  & (    0.112)  & (    0.103)  & (    0.104)  & (    0.132)  \\ \addlinespace[1ex] 
\multirow[t]{ 2 }{0.2\textwidth}{Consumption assets$^\dag$ }  &    -0.764\ensuremath{^{***}} &    -0.868\ensuremath{^{***}} &    -1.117\ensuremath{^{***}} &    -0.776\ensuremath{^{***}} &    -0.872\ensuremath{^{***}} &    -0.669\ensuremath{^{**}} &    -0.854\ensuremath{^{***}} \\ 
 & (    0.256)  & (    0.304)  & (    0.337)  & (    0.275)  & (    0.276)  & (    0.279)  & (    0.304)  \\ \addlinespace[1ex] 
\multirow[t]{ 2 }{0.2\textwidth}{House value$^\dag$ }  &    -0.017 &     0.082 &     0.128 &     0.029 &    -0.008 &    -0.069 &     0.086 \\ 
 & (    0.051)  & (    0.073)  & (    0.092)  & (    0.059)  & (    0.057)  & (    0.054)  & (    0.074)  \\ \addlinespace[1ex] 
\multirow[t]{ 2 }{0.2\textwidth}{Housing quality index }  &    -0.008 &    -0.112 &    -0.204 &    -0.043 &    -0.028 &     0.056 &    -0.189 \\ 
 & (    0.151)  & (    0.168)  & (    0.162)  & (    0.160)  & (    0.161)  & (    0.175)  & (    0.163)  \\ \addlinespace[1ex] 
\multicolumn{8}{l}{\emph{B.  Individual outcomes}}  \\ 
\addlinespace[1ex] \multirow[t]{ 2 }{0.2\textwidth}{Child mortality }  &    -0.002 &     0.002 &     0.010\ensuremath{^{**}} &    -0.001 &     0.001 &    -0.006 &     0.003 \\ 
 & (    0.005)  & (    0.006)  & (    0.004)  & (    0.006)  & (    0.004)  & (    0.006)  & (    0.006)  \\ \addlinespace[1ex] 
\multirow[t]{ 2 }{0.2\textwidth}{Pregnancy }  &     0.002 &    -0.017 &    -0.052 &     0.000 &    -0.013 &     0.017 &    -0.023 \\ 
 & (    0.027)  & (    0.041)  & (    0.047)  & (    0.035)  & (    0.027)  & (    0.029)  & (    0.041)  \\ \addlinespace[1ex] 
\multirow[t]{ 2 }{0.2\textwidth}{Live birth }  &     0.024 &     0.011 &    -0.098 &     0.048 &    -0.021 &     0.080 &    -0.005 \\ 
 & (    0.067)  & (    0.089)  & (    0.110)  & (    0.078)  & (    0.069)  & (    0.075)  & (    0.088)  \\ \addlinespace[1ex] 
\multirow[t]{ 2 }{0.2\textwidth}{Birth in facility }  &    -0.104\ensuremath{^{**}} &    -0.141\ensuremath{^{**}} &    -0.177\ensuremath{^{***}} &    -0.122\ensuremath{^{**}} &    -0.134\ensuremath{^{**}} &    -0.060 &    -0.156\ensuremath{^{***}} \\ 
 & (    0.052)  & (    0.061)  & (    0.068)  & (    0.058)  & (    0.055)  & (    0.063)  & (    0.060)  \\ \addlinespace[1ex] 
\multirow[t]{ 2 }{0.2\textwidth}{Any vaccinations in past year }  &     0.018 &     0.060 &     0.089 &     0.039 &     0.025 &    -0.003 &     0.060 \\ 
 & (    0.032)  & (    0.042)  & (    0.054)  & (    0.034)  & (    0.037)  & (    0.035)  & (    0.041)  \\ \addlinespace[1ex] 
\multirow[t]{ 2 }{0.2\textwidth}{Completed vaccination schedule }  &     0.021 &     0.055 &     0.081 &     0.037 &     0.027 &     0.003 &     0.056 \\ 
 & (    0.036)  & (    0.045)  & (    0.054)  & (    0.038)  & (    0.039)  & (    0.040)  & (    0.045)  \\ \addlinespace[1ex] 
\multirow[t]{ 2 }{0.2\textwidth}{Disease burden }  &     0.012 &    -0.024 &    -0.034 &    -0.006 &     0.010 &     0.027 &     0.120\ensuremath{^{***}} \\ 
 & (    0.033)  & (    0.040)  & (    0.045)  & (    0.035)  & (    0.036)  & (    0.036)  & (    0.039)  \\ \addlinespace[1ex] 
\multirow[t]{ 2 }{0.2\textwidth}{Diarrheal prevalence }  &    -0.003 &     0.010 &     0.001 &     0.007 &    -0.010 &    -0.006 &     0.122\ensuremath{^{***}} \\ 
 & (    0.014)  & (    0.023)  & (    0.032)  & (    0.017)  & (    0.017)  & (    0.014)  & (    0.038)  \\ \addlinespace[1ex] 
\bottomrule
\end{tabular}

%% file: tables/tce_secondary.tex
\begin{tabular}{l *{3}{S} ScSSS}
\toprule
 & & \multicolumn{2}{c}{GiveDirectly} &  \multicolumn{1}{c}{\raisebox{-1ex}[0pt]{Control}} & & &\multicolumn{2}{c}{$ p$-values: B/C ratios} \\ 
\cmidrule(lr){3-4} \cmidrule(lr){8-9}
\multicolumn{1}{c}{\text{ }} & \multicolumn{1}{c}{\text{Gikuriro}} & \multicolumn{1}{c}{\text{Main}} & \multicolumn{1}{c}{\text{Large}} & \multicolumn{1}{c}{\text{Mean}} & \multicolumn{1}{c}{\text{Obs.}} & \multicolumn{1}{c}{\text{$ R^2$}} & \multicolumn{1}{c}{\text{GD=GDL}} & \multicolumn{1}{c}{\text{GK=GDL}}\\
\midrule
\multicolumn{9}{l}{\emph{Panel A.  Household outcomes}}  \\ 
\addlinespace[1ex] \multirow[t]{ 3 }{0.2\textwidth}{Stock of borrowing$^\dag$ }  &     0.141 &    -0.392 &    -0.262 &      5.35 &      2715 &      0.11 &      0.35 &      0.56 \\ 
 & (    0.408)  & (    0.394)  & (    0.428)  \\  & [    1.00]  & [    1.00]  & [    1.00]  \\ \addlinespace[1ex] 
\multirow[t]{ 3 }{0.2\textwidth}{Stock of saving$^\dag$ }  &    -0.222 &    -0.427 &    -0.755\ensuremath{^{**}} &      6.03 &      2718 &      0.15 &      0.46 &      0.98 \\ 
 & (    0.370)  & (    0.395)  & (    0.380)  \\  & [    1.00]  & [    1.00]  & [    1.00]  \\ \addlinespace[1ex] 
\multirow[t]{ 3 }{0.2\textwidth}{Health knowledge index }  &     1.448\ensuremath{^{***}} &     0.692\ensuremath{^{*}} &    -0.388 &     -0.12 &      2718 &      0.05 &      0.05 &      0.00 \\ 
 & (    0.392)  & (    0.405)  & (    0.545)  \\  & [    0.01]  & [    1.00]  & [    1.00]  \\ \addlinespace[1ex] 
\multirow[t]{ 3 }{0.2\textwidth}{Sanitation practices index }  &     0.165 &    -0.269 &     0.304 &      0.29 &      2718 &      0.07 &      0.14 &      0.68 \\ 
 & (    0.202)  & (    0.232)  & (    0.345)  \\  & [    1.00]  & [    1.00]  & [    1.00]  \\ \addlinespace[1ex] 
\multirow[t]{ 3 }{0.2\textwidth}{Productive assets$^\dag$ }  &    -0.081 &    -0.190 &     0.222 &     12.14 &      2718 &      0.30 &      0.08 &      0.26 \\ 
 & (    0.141)  & (    0.145)  & (    0.173)  \\  & [    1.00]  & [    1.00]  & [    1.00]  \\ \addlinespace[1ex] 
\multirow[t]{ 3 }{0.2\textwidth}{Consumption assets$^\dag$ }  &     0.102 &    -0.028 &     0.354 &      9.85 &      2718 &      0.32 &      0.66 &      0.99 \\ 
 & (    0.223)  & (    0.238)  & (    0.361)  \\  & [    1.00]  & [    1.00]  & [    1.00]  \\ \addlinespace[1ex] 
\multirow[t]{ 3 }{0.2\textwidth}{House value$^\dag$ }  &    -0.003 &     0.009 &     0.062 &     13.96 &      2531 &      0.39 &      0.96 &      0.76 \\ 
 & (    0.072)  & (    0.065)  & (    0.074)  \\  & [    1.00]  & [    1.00]  & [    1.00]  \\ \addlinespace[1ex] 
\multirow[t]{ 3 }{0.2\textwidth}{Housing quality index }  &    -0.172 &    -0.007 &    -0.019 &      0.20 &      2718 &      0.16 &      0.98 &      0.28 \\ 
 & (    0.174)  & (    0.155)  & (    0.192)  \\  & [    1.00]  & [    1.00]  & [    1.00]  \\ \addlinespace[1ex] 
\multicolumn{9}{l}{\emph{Panel B.  Individual outcomes}}  \\ 
\addlinespace[1ex] \multirow[t]{ 3 }{0.2\textwidth}{Child Mortality }  &     0.004 &    -0.001 &    -0.003 &      0.01 &      3373 &      0.02 &      0.69 &      0.38 \\ 
 & (    0.006)  & (    0.002)  & (    0.002)  \\  & [    1.00]  & [    1.00]  & [    1.00]  \\ \addlinespace[1ex] 
\multirow[t]{ 3 }{0.2\textwidth}{Pregnancy }  &     0.022 &     0.003 &    -0.003 &      0.12 &      4137 &      0.11 &      0.83 &      0.25 \\ 
 & (    0.022)  & (    0.019)  & (    0.022)  \\  & [    1.00]  & [    1.00]  & [    1.00]  \\ \addlinespace[1ex] 
\multirow[t]{ 3 }{0.2\textwidth}{Live Birth }  &     0.006 &    -0.047 &    -0.041 &      0.70 &       594 &      0.13 &      0.66 &      0.81 \\ 
 & (    0.078)  & (    0.092)  & (    0.093)  \\  & [    1.00]  & [    1.00]  & [    1.00]  \\ \addlinespace[1ex] 
\multirow[t]{ 3 }{0.2\textwidth}{Birth in Facility }  &    -0.029 &     0.053 &    -0.007 &      0.90 &       416 &      0.16 &      0.30 &      0.66 \\ 
 & (    0.069)  & (    0.061)  & (    0.088)  \\  & [    1.00]  & [    1.00]  & [    1.00]  \\ \addlinespace[1ex] 
\multirow[t]{ 3 }{0.2\textwidth}{Any Vaccinations in past year }  &     0.115\ensuremath{^{*}} &     0.076 &     0.159\ensuremath{^{***}} &      0.73 &      1479 &      0.31 &      0.37 &      0.16 \\ 
 & (    0.060)  & (    0.056)  & (    0.062)  \\  & [    0.41]  & [    1.00]  & [    0.08]  \\ \addlinespace[1ex] 
\multirow[t]{ 3 }{0.2\textwidth}{Completed Vaccinations }  &     0.199\ensuremath{^{***}} &     0.080 &     0.210\ensuremath{^{***}} &      0.48 &      1479 &      0.17 &      0.51 &      0.01 \\ 
 & (    0.064)  & (    0.064)  & (    0.064)  \\  & [    0.03]  & [    1.00]  & [    0.03]  \\ \addlinespace[1ex] 
\multirow[t]{ 3 }{0.2\textwidth}{Disease Burden }  &     0.018 &    -0.027 &    -0.002 &      0.50 &      3366 &      0.06 &      0.53 &      0.69 \\ 
 & (    0.052)  & (    0.047)  & (    0.055)  \\  & [    1.00]  & [    1.00]  & [    1.00]  \\ \addlinespace[1ex] 
\multirow[t]{ 3 }{0.2\textwidth}{Diarrheal Prevalence }  &     0.007 &     0.034 &     0.030 &      0.07 &      3366 &      0.05 &      0.19 &      0.93 \\ 
 & (    0.018)  & (    0.022)  & (    0.025)  \\  & [    1.00]  & [    1.00]  & [    1.00]  \\ \addlinespace[1ex] 
\bottomrule
\end{tabular}

%% file: tables/spillovers_primary.tex
\begin{tabular}{l *{2}{S} ScS}
\toprule
\multicolumn{1}{c}{\text{ }} & \multicolumn{1}{c}{\text{\makecell[b]{GD Main\\Village}}} & \multicolumn{1}{c}{\text{\makecell[b]{GD Large\\Village}}} & \multicolumn{1}{c}{\text{\makecell[b]{Control\\Mean}}} & \multicolumn{1}{c}{\text{Observations}} & \multicolumn{1}{c}{\text{$ R^2$}}\\
\midrule
\multicolumn{6}{l}{\emph{A.  Household outcomes}}  \\ 
\addlinespace[1ex] \multirow[t]{ 3 }{0.2\textwidth}{Consumption$^\dag$ }  &    -0.112 &    -0.158 &     10.95 &       477 &      0.19 \\ 
 & (    0.140)  & (    0.204)  \\  & [    0.91]  & [    0.91]  \\ \addlinespace[1ex] 
\multirow[t]{ 3 }{0.2\textwidth}{Household dietary diversity score }  &    -0.004 &    -0.454 &      4.77 &       477 &      0.24 \\ 
 & (    0.186)  & (    0.279)  \\  & [    0.97]  & [    0.91]  \\ \addlinespace[1ex] 
\multirow[t]{ 3 }{0.2\textwidth}{Household non-land wealth$^\dag$ }  &    -0.463 &    -0.635 &     14.09 &       477 &      0.23 \\ 
 & (    0.391)  & (    0.479)  \\  & [    0.91]  & [    0.91]  \\ \addlinespace[1ex] 
\multicolumn{6}{l}{\emph{B.  Individual outcomes}}  \\ 
\addlinespace[1ex] \multirow[t]{ 3 }{0.2\textwidth}{Height-for-Age }  &     0.040 &     0.073 &     -1.58 &       267 &      0.77 \\ 
 & (    0.083)  & (    0.082)  \\  & [    1.00]  & [    1.00]  \\ \addlinespace[1ex] 
\multirow[t]{ 3 }{0.2\textwidth}{Weight-for-Age }  &    -0.005 &    -0.031 &     -0.69 &       267 &      0.75 \\ 
 & (    0.065)  & (    0.082)  \\  & [    1.00]  & [    1.00]  \\ \addlinespace[1ex] 
\multirow[t]{ 3 }{0.2\textwidth}{Mid-Upper Arm Circ }  &    -0.004 &     0.086 &     -0.22 &       188 &      0.62 \\ 
 & (    0.115)  & (    0.107)  \\  & [    1.00]  & [    1.00]  \\ \addlinespace[1ex] 
\bottomrule
\end{tabular}

%% file: tables/spillovers_secondary.tex
\begin{tabular}{l *{2}{S} ScS}
\toprule
\multicolumn{1}{c}{\text{ }} & \multicolumn{1}{c}{\text{\makecell[b]{GD Main\\Village}}} & \multicolumn{1}{c}{\text{\makecell[b]{GD Large\\Village}}} & \multicolumn{1}{c}{\text{\makecell[b]{Control\\Mean}}} & \multicolumn{1}{c}{\text{Observations}} & \multicolumn{1}{c}{\text{$ R^2$}}\\
\midrule
\multicolumn{6}{l}{\emph{A.  Household outcomes}}  \\ 
\addlinespace[1ex] \multirow[t]{ 3 }{0.2\textwidth}{Stock of borrowing$^\dag$ }  &     0.071 &    -0.143 &      6.73 &       475 &      0.18 \\ 
 & (    0.508)  & (    0.538)  \\  & [    1.00]  & [    1.00]  \\ \addlinespace[1ex] 
\multirow[t]{ 3 }{0.2\textwidth}{Stock of saving$^\dag$ }  &    -0.878 &    -1.631\ensuremath{^{***}} &      7.04 &       477 &      0.17 \\ 
 & (    0.536)  & (    0.564)  \\  & [    1.00]  & [    0.08]  \\ \addlinespace[1ex] 
\multirow[t]{ 3 }{0.2\textwidth}{Health knowledge index }  &     0.608 &     0.232 &      2.12 &       477 &      0.07 \\ 
 & (    0.559)  & (    0.733)  \\  & [    1.00]  & [    1.00]  \\ \addlinespace[1ex] 
\multirow[t]{ 3 }{0.2\textwidth}{Sanitation practices index }  &    -0.366 &     0.082 &      0.06 &       477 &      0.12 \\ 
 & (    0.280)  & (    0.410)  \\  & [    1.00]  & [    1.00]  \\ \addlinespace[1ex] 
\multirow[t]{ 3 }{0.2\textwidth}{Productive assets$^\dag$ }  &    -0.325 &    -0.048 &     11.97 &       477 &      0.32 \\ 
 & (    0.218)  & (    0.220)  \\  & [    1.00]  & [    1.00]  \\ \addlinespace[1ex] 
\multirow[t]{ 3 }{0.2\textwidth}{Consumption assets$^\dag$ }  &    -0.404 &     0.053 &     10.08 &       477 &      0.34 \\ 
 & (    0.298)  & (    0.447)  \\  & [    1.00]  & [    1.00]  \\ \addlinespace[1ex] 
\multirow[t]{ 3 }{0.2\textwidth}{House value$^\dag$ }  &     0.087 &     0.079 &     14.13 &       425 &      0.39 \\ 
 & (    0.086)  & (    0.097)  \\  & [    1.00]  & [    1.00]  \\ \addlinespace[1ex] 
\multirow[t]{ 3 }{0.2\textwidth}{Housing quality index }  &    -0.037 &    -0.135 &      0.11 &       477 &      0.25 \\ 
 & (    0.181)  & (    0.250)  \\  & [    1.00]  & [    1.00]  \\ \addlinespace[1ex] 
\multicolumn{6}{l}{\emph{B.  Individual outcomes}}  \\ 
\addlinespace[1ex] \multirow[t]{ 3 }{0.2\textwidth}{Pregnancy }  &    -0.007 &    -0.001 &      0.12 &       745 &      0.13 \\ 
 & (    0.028)  & (    0.032)  \\  & [    1.00]  & [    1.00]  \\ \addlinespace[1ex] 
\multirow[t]{ 3 }{0.2\textwidth}{Live Birth }  &    -0.136 &     0.023 &      0.68 &        86 &      0.48 \\ 
 & (    0.148)  & (    0.158)  \\  & [    1.00]  & [    1.00]  \\ \addlinespace[1ex] 
\multirow[t]{ 3 }{0.2\textwidth}{Birth in Facility }  &    -0.039 &    -0.042 &      0.93 &        59 &      0.45 \\ 
 & (    0.097)  & (    0.130)  \\  & [    1.00]  & [    1.00]  \\ \addlinespace[1ex] 
\multirow[t]{ 3 }{0.2\textwidth}{Any Vaccinations in past year }  &    -0.041 &     0.053 &      0.70 &       290 &      0.42 \\ 
 & (    0.063)  & (    0.070)  \\  & [    1.00]  & [    1.00]  \\ \addlinespace[1ex] 
\multirow[t]{ 3 }{0.2\textwidth}{Completed Vaccinations }  &    -0.035 &     0.084 &      0.44 &       290 &      0.26 \\ 
 & (    0.064)  & (    0.078)  \\  & [    1.00]  & [    1.00]  \\ \addlinespace[1ex] 
\multirow[t]{ 3 }{0.2\textwidth}{Disease Burden }  &    -0.030 &     0.035 &      0.53 &       364 &      0.13 \\ 
 & (    0.073)  & (    0.083)  \\  & [    1.00]  & [    1.00]  \\ \addlinespace[1ex] 
\multirow[t]{ 3 }{0.2\textwidth}{Diarrheal Prevalence }  &     0.015 &     0.035 &      0.08 &       364 &      0.15 \\ 
 & (    0.034)  & (    0.043)  \\  & [    1.00]  & [    1.00]  \\ \addlinespace[1ex] 
\bottomrule
\end{tabular}

%% file: tables/lumpsum_flow_primary.tex
\begin{tabular}{l *{4}{S} ScS}
\toprule
\multicolumn{1}{c}{\text{ }} & \multicolumn{1}{c}{\text{\makecell[b]{Main GD\\Treatment}}} & \multicolumn{1}{c}{\text{\makecell[b]{Main GD\\Lump Sum}}} & \multicolumn{1}{c}{\text{\makecell[b]{Large GD\\Treatment}}} & \multicolumn{1}{c}{\text{\makecell[b]{Large GD\\Lump Sum}}} & \multicolumn{1}{c}{\text{\makecell[b]{Control\\Mean}}} & \multicolumn{1}{c}{\text{Observations}} & \multicolumn{1}{c}{\text{$ R^2$}}\\
\midrule
\multicolumn{8}{l}{\emph{A.  Household outcomes}}  \\ 
\addlinespace[1ex] \multirow[t]{ 3 }{0.2\textwidth}{Consumption$^\dag$ }  &     0.044 &     0.028 &     0.286\ensuremath{^{***}} &     0.027 &     10.69 &      1131 &      0.15 \\ 
 & (    0.097)  & (    0.127)  & (    0.104)  & (    0.118)  \\  & [    1.00]  & [    1.00]  & [    0.04]  & [    1.00]  \\ \addlinespace[1ex] 
\multirow[t]{ 3 }{0.2\textwidth}{Household dietary diversity score }  &     0.068 &     0.014 &     0.614\ensuremath{^{***}} &    -0.268 &      4.77 &      1131 &      0.19 \\ 
 & (    0.159)  & (    0.176)  & (    0.137)  & (    0.237)  \\  & [    1.00]  & [    1.00]  & [    0.00]  & [    1.00]  \\ \addlinespace[1ex] 
\multirow[t]{ 3 }{0.2\textwidth}{Household non-land wealth$^\dag$ }  &     0.012 &    -0.273 &     0.209 &     0.631 &     13.04 &      1131 &      0.25 \\ 
 & (    0.262)  & (    0.502)  & (    0.388)  & (    0.441)  \\  & [    1.00]  & [    1.00]  & [    1.00]  & [    1.00]  \\ \addlinespace[1ex] 
\multicolumn{8}{l}{\emph{B.  Individual outcomes}}  \\ 
\addlinespace[1ex] \multirow[t]{ 3 }{0.2\textwidth}{Height-for-age }  &    -0.002 &    -0.095\ensuremath{^{*}} &     0.074 &     0.047 &     -1.97 &      1380 &      0.74 \\ 
 & (    0.045)  & (    0.057)  & (    0.050)  & (    0.062)  \\  & [    0.93]  & [    0.52]  & [    0.52]  & [    0.56]  \\ \addlinespace[1ex] 
\multirow[t]{ 3 }{0.2\textwidth}{Weight-for-age }  &    -0.036 &     0.125\ensuremath{^{*}} &     0.028 &     0.148\ensuremath{^{**}} &     -1.04 &      1369 &      0.70 \\ 
 & (    0.040)  & (    0.071)  & (    0.041)  & (    0.073)  \\  & [    0.56]  & [    0.52]  & [    0.56]  & [    0.52]  \\ \addlinespace[1ex] 
\multirow[t]{ 3 }{0.2\textwidth}{Mid-upper arm circumference }  &    -0.097 &     0.320\ensuremath{^{***}} &     0.121 &     0.150 &     -0.59 &      1057 &      0.53 \\ 
 & (    0.063)  & (    0.112)  & (    0.082)  & (    0.107)  \\  & [    0.52]  & [    0.11]  & [    0.52]  & [    0.52]  \\ \addlinespace[1ex] 
\multirow[t]{ 3 }{0.2\textwidth}{Child anemia }  &     0.044 &    -0.054 &    -0.007 &    -0.005 &      0.22 &      1544 &      0.08 \\ 
 & (    0.032)  & (    0.038)  & (    0.039)  & (    0.052)  \\  & [    0.52]  & [    0.52]  & [    0.92]  & [    0.93]  \\ \addlinespace[1ex] 
\multirow[t]{ 3 }{0.2\textwidth}{Maternal anemia }  &     0.013 &    -0.033 &    -0.024 &     0.009 &      0.12 &      1025 &      0.12 \\ 
 & (    0.031)  & (    0.035)  & (    0.037)  & (    0.053)  \\  & [    0.72]  & [    0.56]  & [    0.56]  & [    0.92]  \\ \addlinespace[1ex] 
\bottomrule
\end{tabular}

%% file: tables/lumpsum_flow_secondary.tex
\begin{tabular}{l *{4}{S} ScS}
\toprule
\multicolumn{1}{c}{\text{ }} & \multicolumn{1}{c}{\text{\makecell[b]{Main GD\\Treatment}}} & \multicolumn{1}{c}{\text{\makecell[b]{Main GD\\Lump Sum}}} & \multicolumn{1}{c}{\text{\makecell[b]{Large GD\\Treatment}}} & \multicolumn{1}{c}{\text{\makecell[b]{Large GD\\Lump Sum}}} & \multicolumn{1}{c}{\text{\makecell[b]{Control\\Mean}}} & \multicolumn{1}{c}{\text{Observations}} & \multicolumn{1}{c}{\text{$ R^2$}}\\
\midrule
\multicolumn{8}{l}{\emph{A.  Household outcomes}}  \\ 
\addlinespace[1ex] \multirow[t]{ 3 }{0.2\textwidth}{Stock of borrowing$^\dag$ }  &    -1.080\ensuremath{^{***}} &     1.071\ensuremath{^{**}} &     0.024 &    -0.356 &      7.39 &      1131 &      0.16 \\ 
 & (    0.323)  & (    0.469)  & (    0.411)  & (    0.892)  \\  & [    0.02]  & [    0.16]  & [    1.00]  & [    1.00]  \\ \addlinespace[1ex] 
\multirow[t]{ 3 }{0.2\textwidth}{Stock of saving$^\dag$ }  &    -0.584 &     0.888\ensuremath{^{**}} &     0.573 &     0.009 &      5.88 &      1131 &      0.17 \\ 
 & (    0.375)  & (    0.429)  & (    0.410)  & (    0.723)  \\  & [    0.51]  & [    0.22]  & [    0.59]  & [    1.00]  \\ \addlinespace[1ex] 
\multirow[t]{ 3 }{0.2\textwidth}{Health knowledge index }  &     0.229 &     0.115 &     0.186 &     0.176 &      2.89 &      1131 &      0.05 \\ 
 & (    0.360)  & (    0.499)  & (    0.554)  & (    0.591)  \\  & [    1.00]  & [    1.00]  & [    1.00]  & [    1.00]  \\ \addlinespace[1ex] 
\multirow[t]{ 3 }{0.2\textwidth}{Sanitation practices index }  &     0.080 &     0.406 &     0.252 &     0.040 &     -0.68 &      1131 &      0.08 \\ 
 & (    0.302)  & (    0.511)  & (    0.290)  & (    0.544)  \\  & [    1.00]  & [    1.00]  & [    0.91]  & [    1.00]  \\ \addlinespace[1ex] 
\multirow[t]{ 3 }{0.2\textwidth}{Productive assets$^\dag$ }  &     0.199\ensuremath{^{*}} &     0.069 &     0.792\ensuremath{^{***}} &     0.148 &     11.22 &      1131 &      0.29 \\ 
 & (    0.121)  & (    0.295)  & (    0.135)  & (    0.225)  \\  & [    0.47]  & [    1.00]  & [    0.00]  & [    1.00]  \\ \addlinespace[1ex] 
\multirow[t]{ 3 }{0.2\textwidth}{Consumption assets$^\dag$ }  &     0.336 &     0.442 &     0.831\ensuremath{^{***}} &     0.465 &      8.70 &      1131 &      0.38 \\ 
 & (    0.239)  & (    0.352)  & (    0.300)  & (    0.475)  \\  & [    0.59]  & [    0.72]  & [    0.07]  & [    0.90]  \\ \addlinespace[1ex] 
\multirow[t]{ 3 }{0.2\textwidth}{House value$^\dag$ }  &     0.023 &    -0.162\ensuremath{^{**}} &     0.169\ensuremath{^{*}} &     0.012 &     13.81 &      1071 &      0.35 \\ 
 & (    0.065)  & (    0.068)  & (    0.088)  & (    0.111)  \\  & [    1.00]  & [    0.16]  & [    0.27]  & [    1.00]  \\ \addlinespace[1ex] 
\multirow[t]{ 3 }{0.2\textwidth}{Housing quality index }  &    -0.260 &     0.177 &     0.206 &     0.253 &     -0.17 &      1131 &      0.11 \\ 
 & (    0.214)  & (    0.433)  & (    0.206)  & (    0.235)  \\  & [    0.72]  & [    1.00]  & [    0.90]  & [    0.88]  \\ \addlinespace[1ex] 
\multicolumn{8}{l}{\emph{B.  Individual outcomes}}  \\ 
\addlinespace[1ex] \multirow[t]{ 3 }{0.2\textwidth}{Child mortality }  &    -0.001 &    -0.002 &    -0.010\ensuremath{^{**}} &     0.010 &      0.01 &      1751 &      0.02 \\ 
 & (    0.010)  & (    0.014)  & (    0.004)  & (    0.009)  \\  & [    1.00]  & [    1.00]  & [    0.60]  & [    1.00]  \\ \addlinespace[1ex] 
\multirow[t]{ 3 }{0.2\textwidth}{Pregnancy }  &    -0.028 &    -0.031 &    -0.033 &     0.045 &      0.20 &      1646 &      0.09 \\ 
 & (    0.032)  & (    0.034)  & (    0.029)  & (    0.052)  \\  & [    1.00]  & [    1.00]  & [    1.00]  & [    1.00]  \\ \addlinespace[1ex] 
\multirow[t]{ 3 }{0.2\textwidth}{Live birth }  &     0.057 &     0.099 &    -0.087 &    -0.123 &      0.68 &       273 &      0.22 \\ 
 & (    0.088)  & (    0.112)  & (    0.093)  & (    0.148)  \\  & [    1.00]  & [    1.00]  & [    1.00]  & [    1.00]  \\ \addlinespace[1ex] 
\multirow[t]{ 3 }{0.2\textwidth}{Birth in facility }  &     0.077 &    -0.013 &    -0.095 &    -0.033 &      0.84 &       188 &      0.25 \\ 
 & (    0.058)  & (    0.073)  & (    0.100)  & (    0.117)  \\  & [    1.00]  & [    1.00]  & [    1.00]  & [    1.00]  \\ \addlinespace[1ex] 
\multirow[t]{ 3 }{0.2\textwidth}{Any vaccinations in past year }  &    -0.013 &     0.038 &    -0.022 &     0.048 &      0.72 &       838 &      0.27 \\ 
 & (    0.039)  & (    0.059)  & (    0.051)  & (    0.101)  \\  & [    1.00]  & [    1.00]  & [    1.00]  & [    1.00]  \\ \addlinespace[1ex] 
\multirow[t]{ 3 }{0.2\textwidth}{Completed vaccination schedule }  &    -0.006 &     0.013 &    -0.008 &     0.063 &      0.58 &       838 &      0.21 \\ 
 & (    0.039)  & (    0.070)  & (    0.050)  & (    0.103)  \\  & [    1.00]  & [    1.00]  & [    1.00]  & [    1.00]  \\ \addlinespace[1ex] 
\multirow[t]{ 3 }{0.2\textwidth}{Disease burden }  &    -0.016 &    -0.055 &    -0.004 &     0.014 &      0.54 &      1746 &      0.06 \\ 
 & (    0.032)  & (    0.058)  & (    0.044)  & (    0.084)  \\  & [    1.00]  & [    1.00]  & [    1.00]  & [    1.00]  \\ \addlinespace[1ex] 
\multirow[t]{ 3 }{0.2\textwidth}{Diarrheal prevalence }  &    -0.005 &     0.021 &     0.002 &    -0.021 &      0.09 &      1746 &      0.05 \\ 
 & (    0.018)  & (    0.026)  & (    0.019)  & (    0.024)  \\  & [    1.00]  & [    1.00]  & [    1.00]  & [    1.00]  \\ \addlinespace[1ex] 
\bottomrule
\end{tabular}

%% file: tables/choice_primary.tex
\begin{tabular}{l *{3}{S} ScS}
\toprule
\multicolumn{1}{c}{\text{ }} & \multicolumn{1}{c}{\text{\makecell[b]{Got Ones\\Choice}}} & \multicolumn{1}{c}{\text{\makecell[b]{Chose\\Lump Sum}}} & \multicolumn{1}{c}{\text{\makecell[b]{Treated\\Lump Sum}}} & \multicolumn{1}{c}{\text{\makecell[b]{Control\\Mean}}} & \multicolumn{1}{c}{\text{Observations}} & \multicolumn{1}{c}{\text{$ R^2$}}\\
\midrule
\multicolumn{7}{l}{\emph{A.  Household outcomes}}  \\ 
\addlinespace[1ex] \multirow[t]{ 3 }{0.2\textwidth}{Consumption$^\dag$ }  &     0.089 &    -0.018 &    -0.032 &     10.69 &       534 &      0.18 \\ 
 & (    0.115)  & (    0.126)  & (    0.117)  \\  & [    1.00]  & [    1.00]  & [    1.00]  \\ \addlinespace[1ex] 
\multirow[t]{ 3 }{0.2\textwidth}{Household dietary diversity score }  &     0.067 &     0.062 &    -0.204 &      4.77 &       534 &      0.23 \\ 
 & (    0.210)  & (    0.197)  & (    0.178)  \\  & [    1.00]  & [    1.00]  & [    1.00]  \\ \addlinespace[1ex] 
\multirow[t]{ 3 }{0.2\textwidth}{Household non-land wealth$^\dag$ }  &    -0.650 &     0.033 &     0.172 &     13.04 &       534 &      0.29 \\ 
 & (    0.468)  & (    0.471)  & (    0.460)  \\  & [    1.00]  & [    1.00]  & [    1.00]  \\ \addlinespace[1ex] 
\multicolumn{7}{l}{\emph{B.  Individual outcomes}}  \\ 
\addlinespace[1ex] \multirow[t]{ 3 }{0.2\textwidth}{Height-for-age }  &    -0.160\ensuremath{^{**}} &     0.034 &    -0.026 &     -1.97 &       671 &      0.75 \\ 
 & (    0.062)  & (    0.058)  & (    0.063)  \\  & [    0.21]  & [    1.00]  & [    1.00]  \\ \addlinespace[1ex] 
\multirow[t]{ 3 }{0.2\textwidth}{Weight-for-age }  &    -0.042 &     0.013 &     0.093\ensuremath{^{*}} &     -1.04 &       668 &      0.67 \\ 
 & (    0.052)  & (    0.045)  & (    0.049)  \\  & [    1.00]  & [    1.00]  & [    0.36]  \\ \addlinespace[1ex] 
\multirow[t]{ 3 }{0.2\textwidth}{Mid-upper arm circumference }  &     0.070 &     0.025 &     0.153\ensuremath{^{**}} &     -0.59 &       520 &      0.57 \\ 
 & (    0.071)  & (    0.065)  & (    0.067)  \\  & [    1.00]  & [    1.00]  & [    0.21]  \\ \addlinespace[1ex] 
\multirow[t]{ 3 }{0.2\textwidth}{Child anemia }  &     0.015 &    -0.017 &    -0.034 &      0.22 &       750 &      0.08 \\ 
 & (    0.024)  & (    0.030)  & (    0.030)  \\  & [    1.00]  & [    1.00]  & [    1.00]  \\ \addlinespace[1ex] 
\multirow[t]{ 3 }{0.2\textwidth}{Maternal anemia }  &     0.007 &     0.008 &    -0.004 &      0.12 &       496 &      0.14 \\ 
 & (    0.034)  & (    0.043)  & (    0.037)  \\  & [    1.00]  & [    1.00]  & [    1.00]  \\ \addlinespace[1ex] 
\bottomrule
\end{tabular}

%% file: tables/choice_secondary.tex
\begin{tabular}{l *{3}{S} ScS}
\toprule
\multicolumn{1}{c}{\text{ }} & \multicolumn{1}{c}{\text{\makecell[b]{Got Ones\\Choice}}} & \multicolumn{1}{c}{\text{\makecell[b]{Chose\\Lump Sum}}} & \multicolumn{1}{c}{\text{\makecell[b]{Treated\\Lump Sum}}} & \multicolumn{1}{c}{\text{\makecell[b]{Control\\Mean}}} & \multicolumn{1}{c}{\text{Observations}} & \multicolumn{1}{c}{\text{$ R^2$}}\\
\midrule
\multicolumn{7}{l}{\emph{A.  Household outcomes}}  \\ 
\addlinespace[1ex] \multirow[t]{ 3 }{0.2\textwidth}{Stock of borrowing$^\dag$ }  &    -0.558 &    -0.197 &     0.956\ensuremath{^{*}} &      7.39 &       534 &      0.16 \\ 
 & (    0.504)  & (    0.425)  & (    0.494)  \\  & [    1.00]  & [    1.00]  & [    0.67]  \\ \addlinespace[1ex] 
\multirow[t]{ 3 }{0.2\textwidth}{Stock of saving$^\dag$ }  &    -0.222 &     0.074 &     0.132 &      5.88 &       534 &      0.21 \\ 
 & (    0.463)  & (    0.491)  & (    0.450)  \\  & [    1.00]  & [    1.00]  & [    1.00]  \\ \addlinespace[1ex] 
\multirow[t]{ 3 }{0.2\textwidth}{Health knowledge index }  &     0.056 &     0.559 &     0.524 &      2.89 &       534 &      0.09 \\ 
 & (    0.534)  & (    0.479)  & (    0.454)  \\  & [    1.00]  & [    1.00]  & [    1.00]  \\ \addlinespace[1ex] 
\multirow[t]{ 3 }{0.2\textwidth}{Sanitation practices index }  &     0.068 &     0.632\ensuremath{^{*}} &     0.221 &     -0.68 &       534 &      0.15 \\ 
 & (    0.286)  & (    0.349)  & (    0.383)  \\  & [    1.00]  & [    0.67]  & [    1.00]  \\ \addlinespace[1ex] 
\multirow[t]{ 3 }{0.2\textwidth}{Productive assets$^\dag$ }  &    -0.082 &    -0.166 &     0.001 &     11.22 &       534 &      0.29 \\ 
 & (    0.212)  & (    0.207)  & (    0.203)  \\  & [    1.00]  & [    1.00]  & [    1.00]  \\ \addlinespace[1ex] 
\multirow[t]{ 3 }{0.2\textwidth}{Consumption assets$^\dag$ }  &    -0.189 &    -0.917\ensuremath{^{***}} &     0.952\ensuremath{^{***}} &      8.70 &       534 &      0.37 \\ 
 & (    0.362)  & (    0.319)  & (    0.301)  \\  & [    1.00]  & [    0.06]  & [    0.05]  \\ \addlinespace[1ex] 
\multirow[t]{ 3 }{0.2\textwidth}{House value$^\dag$ }  &    -0.065 &    -0.029 &    -0.094 &     13.81 &       508 &      0.40 \\ 
 & (    0.085)  & (    0.090)  & (    0.070)  \\  & [    1.00]  & [    1.00]  & [    1.00]  \\ \addlinespace[1ex] 
\multirow[t]{ 3 }{0.2\textwidth}{Housing quality index }  &    -0.048 &    -0.017 &    -0.114 &     -0.17 &       534 &      0.21 \\ 
 & (    0.175)  & (    0.119)  & (    0.123)  \\  & [    1.00]  & [    1.00]  & [    1.00]  \\ \addlinespace[1ex] 
\multicolumn{7}{l}{\emph{B.  Individual outcomes}}  \\ 
\addlinespace[1ex] \multirow[t]{ 3 }{0.2\textwidth}{Child mortality }  &     0.001 &    -0.011 &     0.000 &      0.01 &       838 &      0.03 \\ 
 & (    0.010)  & (    0.014)  & (    0.015)  \\  & [    1.00]  & [    1.00]  & [    1.00]  \\ \addlinespace[1ex] 
\multirow[t]{ 3 }{0.2\textwidth}{Pregnancy }  &     0.008 &     0.027 &    -0.025 &      0.20 &       757 &      0.11 \\ 
 & (    0.035)  & (    0.032)  & (    0.030)  \\  & [    1.00]  & [    1.00]  & [    1.00]  \\ \addlinespace[1ex] 
\multirow[t]{ 3 }{0.2\textwidth}{Live birth }  &    -0.064 &     0.286\ensuremath{^{*}} &    -0.007 &      0.68 &       129 &      0.32 \\ 
 & (    0.151)  & (    0.162)  & (    0.120)  \\  & [    1.00]  & [    1.00]  & [    1.00]  \\ \addlinespace[1ex] 
\multirow[t]{ 3 }{0.2\textwidth}{Birth in facility }  &     0.137 &     0.077 &    -0.153 &      0.84 &        83 &      0.57 \\ 
 & (    0.127)  & (    0.121)  & (    0.115)  \\  & [    1.00]  & [    1.00]  & [    1.00]  \\ \addlinespace[1ex] 
\multirow[t]{ 3 }{0.2\textwidth}{Any vaccinations in past year }  &     0.012 &     0.102\ensuremath{^{*}} &     0.016 &      0.72 &       434 &      0.25 \\ 
 & (    0.050)  & (    0.062)  & (    0.058)  \\  & [    1.00]  & [    1.00]  & [    1.00]  \\ \addlinespace[1ex] 
\multirow[t]{ 3 }{0.2\textwidth}{Completed vaccination schedule }  &    -0.004 &     0.060 &     0.002 &      0.58 &       434 &      0.19 \\ 
 & (    0.054)  & (    0.065)  & (    0.064)  \\  & [    1.00]  & [    1.00]  & [    1.00]  \\ \addlinespace[1ex] 
\multirow[t]{ 3 }{0.2\textwidth}{Disease burden }  &     0.017 &     0.050 &    -0.026 &      0.54 &       835 &      0.07 \\ 
 & (    0.051)  & (    0.050)  & (    0.049)  \\  & [    1.00]  & [    1.00]  & [    1.00]  \\ \addlinespace[1ex] 
\multirow[t]{ 3 }{0.2\textwidth}{Diarrheal prevalence }  &    -0.001 &    -0.030 &     0.017 &      0.09 &       835 &      0.05 \\ 
 & (    0.024)  & (    0.026)  & (    0.025)  \\  & [    1.00]  & [    1.00]  & [    1.00]  \\ \addlinespace[1ex] 
\bottomrule
\end{tabular}

%% file: 99-Gikuriro-Online_Appendix.tex
\clearpage 

\subsection{Online Supplementary Appendix Tables}

\begin{table}[!hb]
\caption{Balance on Secondary Outcomes}
\label{t:balance_secondary}

\begin{footnotesize}
\begin{center}
\resizebox{0.85\textwidth}{!}{
\input{tables/balance_secondary.tex}

}
\end{center}

\vskip-4ex 

\floatfoot{
	Notes:  See prior table.  Indexes are unweighted sums of z-scores of their underlying components.  Individual secondary outcomes all measured as rates within respective populations.
	
}
\end{footnotesize}
\end{table}

\begin{table}\caption{Balance on Household Covariates}
\label{t:balance_RHS}

\begin{footnotesize}
\begin{center}
\input{tables/balance_RHS.tex}

\end{center}

\vskip-4ex 

\floatfoot{
	Notes:  Columns present coefficients and standard errors from a regression of baseline covariates on treatment indicators, with fixed effects for blocks.   Asterices denote significance at the 10, 5, and 1 percent levels, and are based on clustered standard errors, in parentheses.  \citet{anderson2008multiple} sharpened $q$-values presented in brackets. 
	
}
\end{footnotesize}
\end{table}

\begin{landscape}

\begin{table}
\caption{Determinants of Receiving Gikuriro}
\label{t:GK_Compliance_Determinants_eligibles}

\begin{footnotesize}
\begin{center}
\input{tables/GK_Compliance_Determinants_eligibles.tex}
\end{center}

\vskip-4ex 

\floatfoot{
	Notes:  Outcome is a dummy variable indicating that household received different components of Gikuriro.  First three columns are indicators for receiving training in proper nutrition (1), hygenic cooking habits (2), agricultural extension (3).  Column 4 is an indicator that the household has successfully harvested a farm garden as instructed by FFLS, and Column 5 for the receipt of livestock from Gikuriro.  Regressions are run among all households defined as eligible at baseline based on intended targeting criteria.
	
}
\end{footnotesize}
\end{table}

\end{landscape}

\begin{table}\caption{Anthropometric Impacts using Attrition IPW}
\label{t:ITT_IPW}

	\begin{footnotesize}
	\begin{center}
	\input{tables/ITT_IPW.tex}

	\end{center}
\vskip-4ex 

\floatfoot{
	Notes:  Regressions weighted using the product of standard survey weights and inverse propensity weights calculated from the probability that a child with baseline anthropometrics attrites from the endline.
}
\end{footnotesize}
\end{table}

\clearpage

\begin{table}\caption{Cost Equivalent Total Causal Effects, primary outcomes}
\label{t:ce_tce_primary}

	\begin{footnotesize}
	\begin{center}
	\input{tables/ce_tce_primary.tex}

	\end{center}
\vskip-4ex 

\floatfoot{
	Notes:  Analysis pools eligible and ineligible households and is weighted to be representative of the population in study villages.  First column is a dummy for Gikuriro treatment, giving the differential effect of Gikuriro over cash at equivalent cost.  Second column is a dummy for either treatment, giving the impact of cash at the cost of Gikuriro. Third column is the cost slope, measured as the dollar-value deviation (in hundreds of dollars) of the treatment received from the cost of Gikuriro.   Asterices denote significance at the 10, 5, and 1 percent levels, and are based on clustered standard errors, in parentheses.  \citet{anderson2008multiple} sharpened $q$-values presented in brackets. Variables marked with a $\dag$ are in inverse hyperbolic sines.
}
\end{footnotesize}
\end{table}

\begin{table}\caption{Cost Equivalent Total Causal Effects, secondary outcomes}
\label{t:ce_tce_secondary}
	\begin{footnotesize}
	\begin{center}
	\resizebox{0.85\linewidth}{!}{
	\input{tables/ce_tce_secondary.tex}

	}
	\end{center}
\vskip-4ex 

\floatfoot{
	Notes:  See previous table.
}
\end{footnotesize}
\end{table}

\begin{table}\caption{Heterogeneity by Baseline Malnutrition}
\label{t:het_by_anthro}
	\begin{footnotesize}
	\begin{center}
	\input{tables/anthro_het_baseline.tex}
	\end{center}
\vskip-4ex 

\floatfoot{
	Notes:  Regressions with both baseline and endline outcome measurement are ANCOVA with lagged dependent variables as controls, run on the panel sample.  Regressions include fixed effects for the randomization blocks, and are weighted to be representative of all households in study villages.  Anthropometric outcomes are demeaned prior to interaction so that the uninteracted treatment terms provide impact at average level of baseline anthro measure..
}
\end{footnotesize}
\end{table}

%\begin{table}[htbp]\centering
%\caption{Heterogeneity by Baseline %Malnutrition}\label{t:het_by_anthro}
%	\includegraphics[scale=.9]{tables/het_by_anthro.pdf}
%\end{table}

\begin{landscape}

\begin{table}\caption{Heterogeneity by Baseline Age}
\label{t:het_by_age}
	\begin{footnotesize}
	\begin{center}
	\input{tables/anthro_age_baseline.tex}
	\end{center}
\vskip-4ex 

\floatfoot{
	Notes:  First three columns present an interaction with an indictor for a child in the `First Thousand Days' (<33 months at endline) and the last three columns present interactions with an indicator for `Newborn' (<13 months at endline).  Regressions are endline cross-sections, run on the panel sample, and do not include the lagged outcome variable so as to be able to consider children who are newborns in R2.  Regressions include fixed effects for the randomization blocks, and are weighted to be representative of eligible households in study villages. 
}
\end{footnotesize}
\end{table}

\begin{table}\caption{Cash versus Kind Heterogeneity by Behavioral Attributes}
\label{t:Het_behavior_GK_vs_GD}
	\begin{footnotesize}
	\begin{center}
	\input{tables/Het_behavior_GK_vs_GD.tex}
	\end{center}
\vskip-4ex 

\floatfoot{
	Notes:  Analysis excludes the GD Large arm, and uses interaction with three behavioral parameters to ask whether the effect of GD Main or Gikuriro are heterogeneous by Impatience, Time Inconsistency, or Other Control Problems.  SEs clustered at the village level are in parentheses. 
}
\end{footnotesize}
\end{table}

\begin{table}\caption{Lump Sum vs Flow Heterogeneity by Behavioral Attributes}
\label{t:Het_behavior_LS_Flow}
	\begin{footnotesize}
	\begin{center}
	\input{tables/Het_behavior_LS_Flow.tex}
	\end{center}
\vskip-4ex 

\floatfoot{
	Notes:  Analysis uses only the control and the GD Main arm assigned to Lump Sum or Flow transfers to ask whether the  impact of Lump Sum or FLow transfers is heterogeneous by Impatience, Time Inconsistency, or Other Control Problems.  SEs clustered at the village level are in parentheses. . 
}
\end{footnotesize}
\end{table}

\begin{table}\caption{Getting Flow when one Chose It, Heterogeneity by Behavioral Attributes}
\label{t:Het_behavior_choose_Flow}
	\begin{footnotesize}
	\begin{center}
	\input{tables/Het_behavior_choose_Flow.tex}
	\end{center}
\vskip-4ex 

\floatfoot{
	Notes:  Analysis uses only the GD Main households that chose Flow transfers, and asks whether the impact of actually getting this choice (rather than being assigned to Lump Sum) is heterogeneous by Impatience, Time Inconsistency, or Other Control Problems.  SEs clustered at the village level are in parentheses. 
}
\end{footnotesize}
\end{table}

\end{landscape}

\cleardoublepage

\subsection{Online Supplementary Appendix Figures}

\begin{figure}[htbp]
\begin{center}
\caption{Actual and Assigned Treatment Amounts}\label{f:box_whisker}
	\includegraphics[scale=.8]{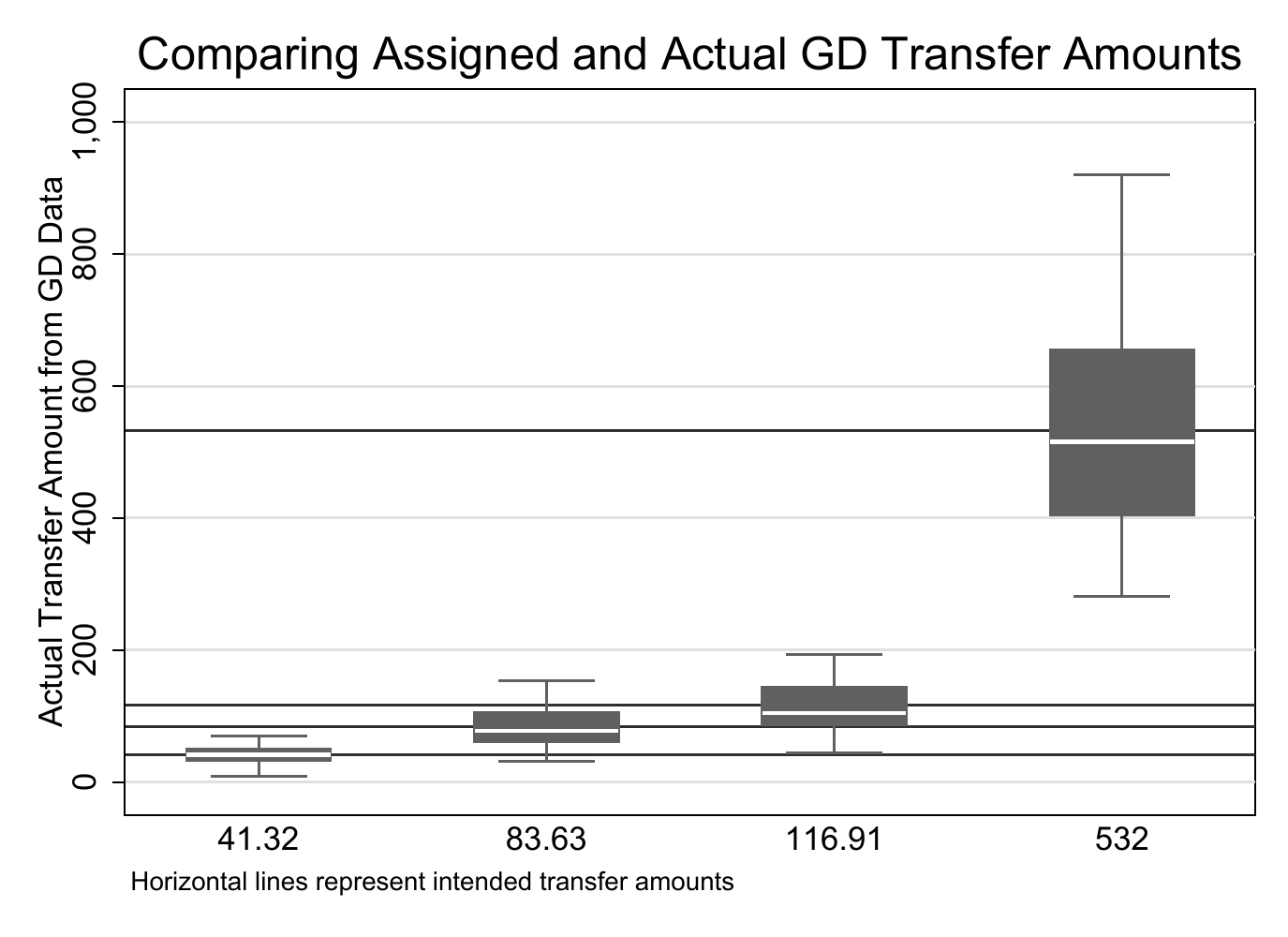}	
	\end{center}
\end{figure}

\begin{figure}[htbp]
\begin{center}
\caption{Fan Regression Impacts by Age}\label{f:HAZ_gd_large_fan}
	\includegraphics[scale=.8]{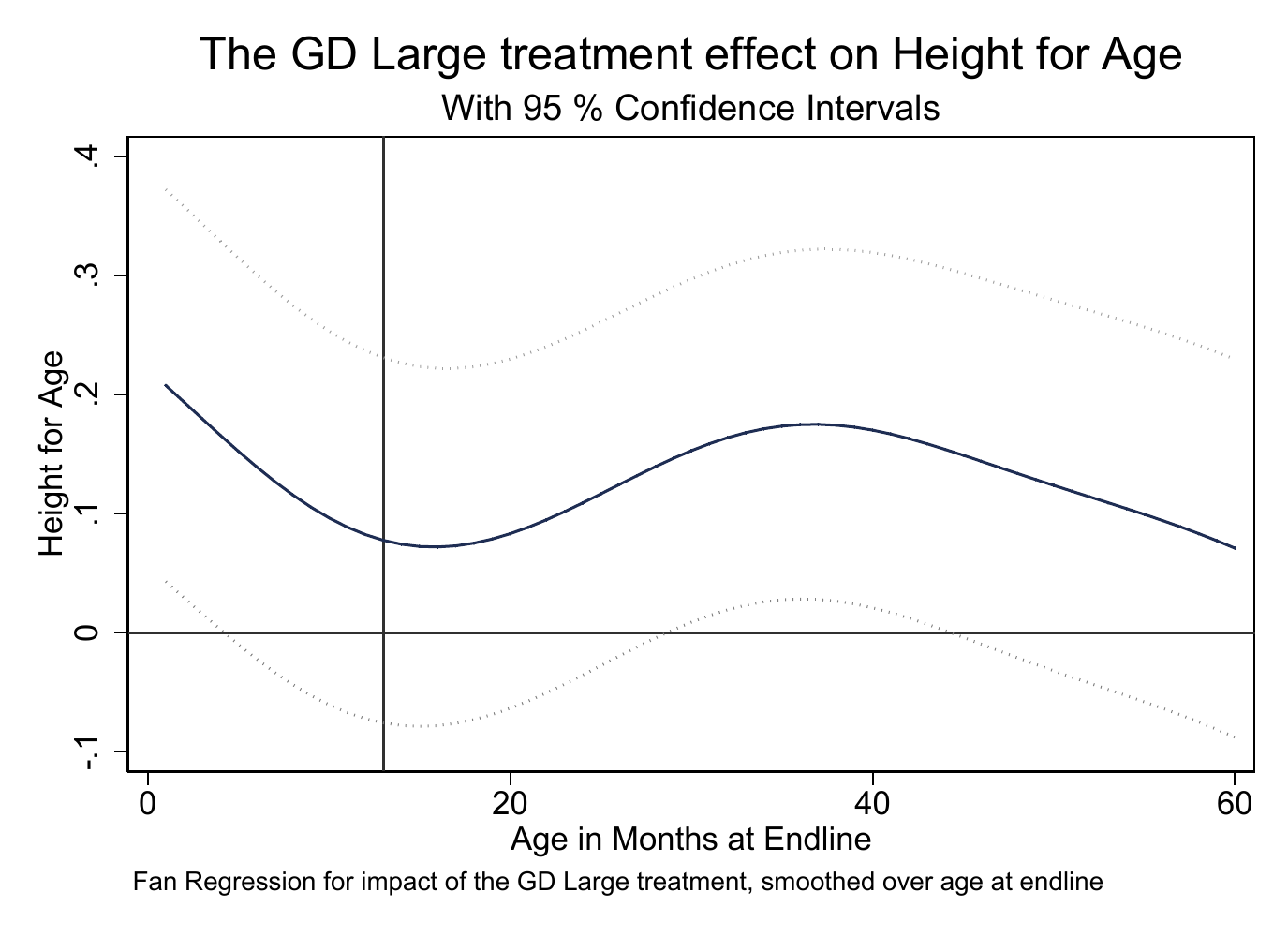}	
	\end{center}
\end{figure}

\begin{figure}[htbp]
\begin{center}
\caption{Fan Regression Impacts by Age}\label{f:anemia_gd_large_fan}
	\includegraphics[scale=.8]{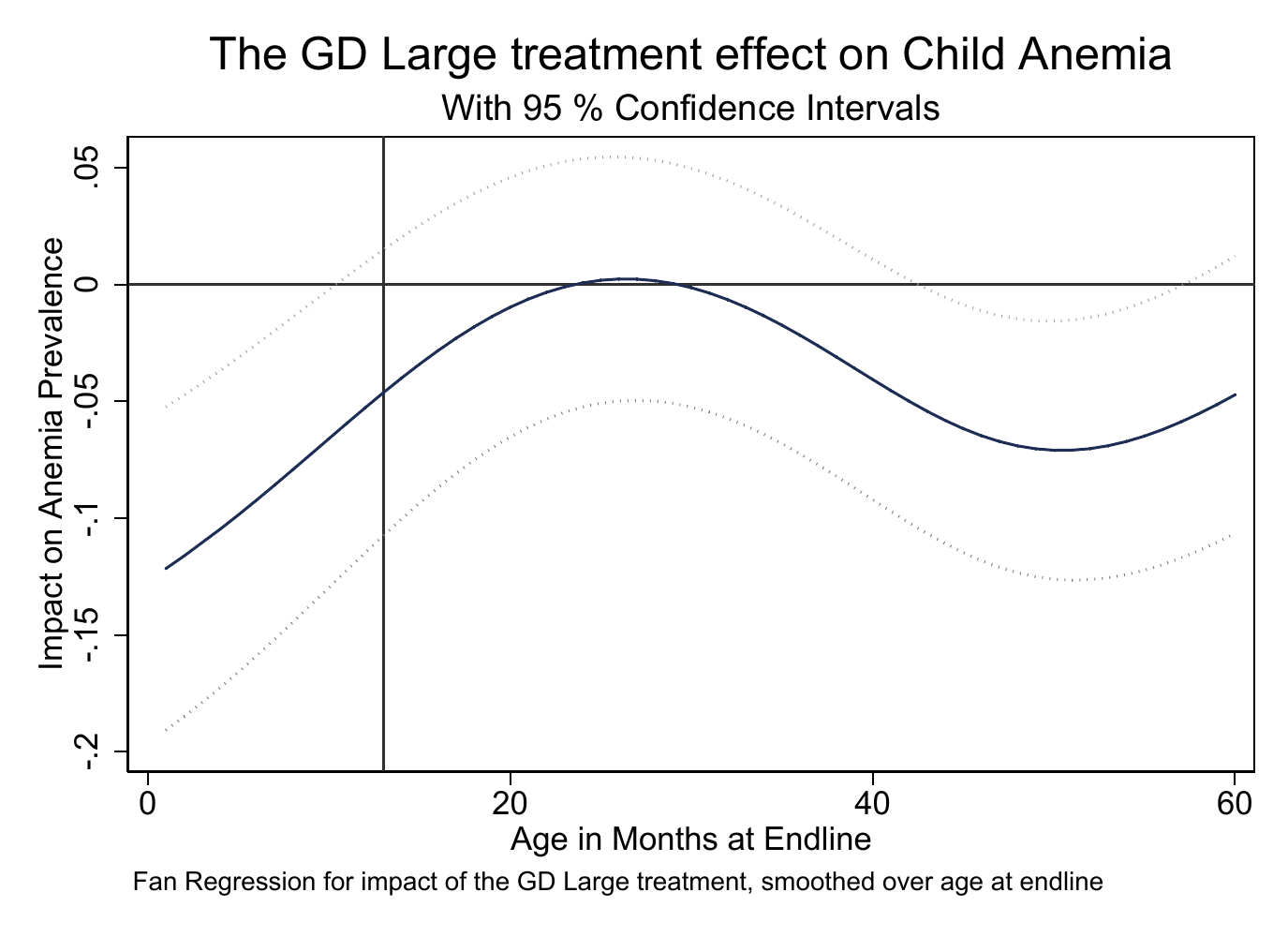}	
	\end{center}
\end{figure}

\begin{figure}
\caption{Predicted impacts on dimensions of child health outcomes are positively associated with one another}
\label{f:heterogeneity_childhealth}

\begin{center}
\subfloat[WAZ and HAZ]{
	\includegraphics[width=0.35\textwidth]{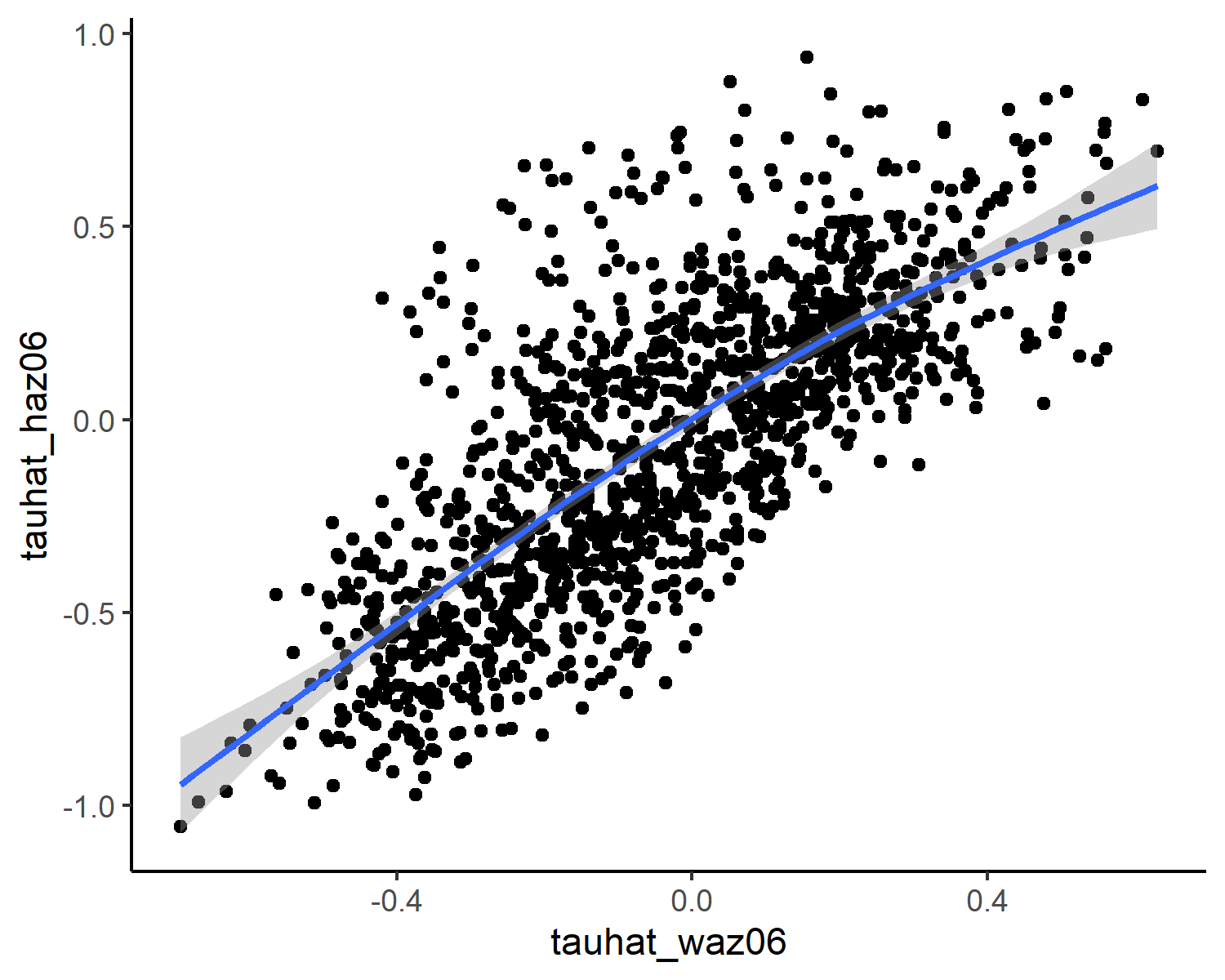}
	\label{f:tauWazHaz}
}
\subfloat[WAZ and MUACZ]{
	\includegraphics[width=0.35\textwidth]{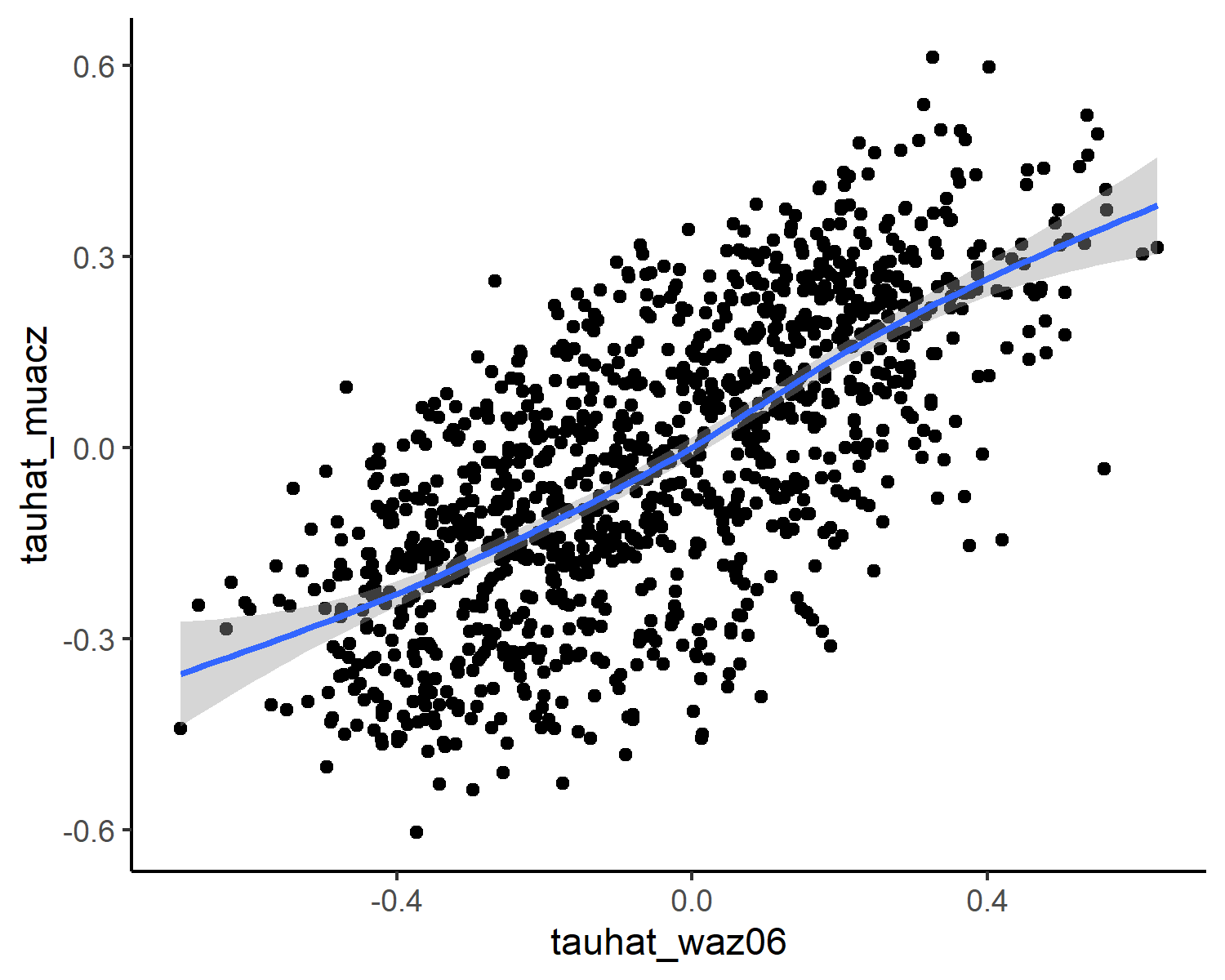}
	\label{f:tauWazMuacz}
}
\\
\subfloat[HAZ and MUACZ]{
	\includegraphics[width=0.35\textwidth]{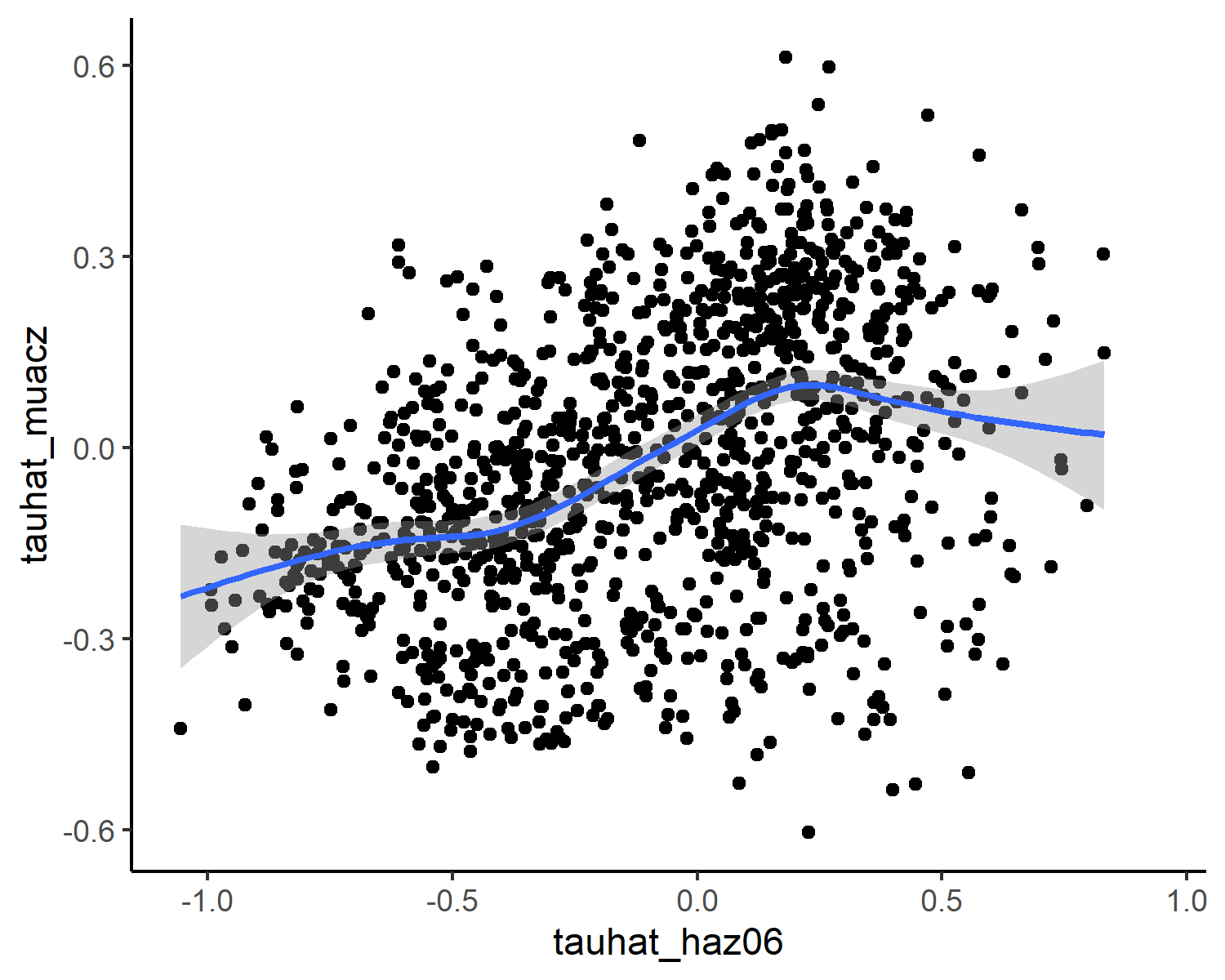}
	\label{f:tauHazMuacz}
}
\end{center}

\vskip-6ex

\floatfoot{
	\begin{footnotesize}
	Notes:  Figure displays associations between predicted impacts of cash relative to kind on weight-for-age z-scores, height-for-age z-scores, and mid-upper-arm-circumference z-scores.  Loess fit and associated 95 percent confidence interval overlaid.
	\end{footnotesize}
}
\end{figure}

\cleardoublepage 

\subsection{Eligibility for the Study}
The study aims to compare nutrition and health gains among poor households with young children across the two programs and a control.  We therefore used a definition of eligibility tailored to Gikuriro's stated target population: namely, households that contained malnourished children, or pregnant and lactating mothers.  A core challenge of the benchmarking endeavor is the need to use a measure of eligibility in a manner that can be defined identically across arms.\footnote{We did not intend the scope of the benchmarking exercise to include the implementers' (potentially different) ability to cost-effectively identify this target population, so as to maintain the interpretation of impacts as being differential impacts on a consistently defined beneficiary group.}  As a result, we established a set of `hard' eligibility criteria on the basis of which beneficiaries would be selected and the survey would be stratified.  Households meeting these criteria would be identified by the survey firm, Innovations for Poverty Action (IPA), prior to sampling for the baseline study, to establish a comparable population of eligible households in all arms---including control---of the study.

CRS and USAID agreed that the following criteria represent the target population for Gikuriro:
\begin{itemize}
	\item  Criteria 1.  All households in a village with a malnourished child (defined by a threshold value of weight/age) were enrolled.
	\begin{itemize}
		\item	Weight/age is used because it is believed that this data is more consistently available than data on middle-upper arm circumference (MUAC) and height/age, and because it is used by CHWs as a basis for referring children to their local Health Centers.
		\item	The threshold weight/age value for inclusion was determined using the Rwandan Ministry of Health standards for malnutrition. The data used to identify eligibles was based on the Community Health Worker data from Growth Monitoring and Promotion visits.
	\end{itemize}
	\item 	Criteria 2. All households in Ubudehe 1 or 2 with children under the age of 5 (Ubudehe is the Rwandan government household-level poverty classification, with 1 being the poorest, 3 being non-poor, and rural areas containing very few of the wealthiest Ubudehe 4 households).
	\item	Criteria 3.  All households in Ubudehe 1 or 2 with a pregnant or lactating mother.
\end{itemize}
Both implementers agreed to attempt to treat all eligible households that were identified as meeting any of these criteria. CRS anticipated an average of 30 eligible households per village, and in principle had established a rationing rule in case that number was exceeded.  As will be described below, the number of households per village that could be identified by the survey firm as meeting these targets turned out to be substantially lower.  We did not try to impose restrictions on how Gikuriro would target outside of the households identified by the survey firm to be eligible.  

We asked IPA to identify the universe of households that they could locate who met these criteria, using three sources.  First, CHW records from the national `Growth Monitoring and Promotion' exercise, which is intended to provide monthly height and weight measurements for all children under two and annual measurements for all children under five;  second, government (census) records of household \emph{Ubudehe} classifications; and finally local health facility information, which provides an alternative data point on children's nutritional status.\footnote{In practice, most children attending local clinics are referred by a CHW and so are also recorded as malnourished in the Growth Monitoring process.} Children were defined as malnourished if they had at least one measurement that met government thresholds for malnourishment definitions in the past year, and households were defined as eligible if they had any individual meeting the criteria above.  In each village we recorded the number of households in each stratum and sampled up to eight eligibles and four ineligibles for inclusion in the study.  Throughout this document we use the words ‘eligible’ and ‘ingeligible’ to refer to the classification made by the survey firm at baseline.

\subsection{Study Outcomes}
\textbf{Primary Outcomes}.  The study focuses on five dimensions. Here we briefly summarize each; details of the construction of these outcomes are included in Appendix A.
\begin{enumerate}
	\item	Household monthly consumption per capita (inverse hyperbolic sine---henceforth IHS---to deal with skewness).
	\item	Household Dietary Diversity, measured using hte WHO standard Household Dietary Diversity Score.
	\item	Anemia:  measured with a biomarker test following DHS protocols at endline only.
	\item	Child growth and development:  measured using in height-for-age, weight-for-age and Mid Upper Arm Circumference at baseline and endline for children under the age of 6 in eligible households.
	\item	Value of household non-land net wealth.  This outcome is the sum of productive and consumption assets; the value of the household’s dwelling, if owned; and the value of the stock of net savings, less the stock of debt (IHS).
\end{enumerate}

\textbf{Secondary Outcomes}. Three types of outcomes are selected to be secondary:  proximate outcomes of one or both interventions that do not have an intrinsic welfare interpretation (such as borrowing and saving stocks); outcomes that have welfare weight but are not within the causal chain of both programs (such as investments in health-seeking behavior, which Gikuriro seeks to impact, or housing quality, which has been identified as a dimension of benefit in prior evaluations of GiveDirectly \citep{haushofer2016short}); or outcomes of common interest on which power is limited (such as disease burden and mortality).

\begin{enumerate}
	\item	Stock of borrowing and stock of savings (IHS).
	\item	Birth outcomes:  the likelihood of pregnancy and likelihood of live birth within 12 months prior to endline.
	\item	Health knowledge and sanitation practices.
	\item	Disease burden and mortality.  Mortality is measured as the likelihood that an individual member of the household from baseline has died prior to endline.  Disease burden is measured as the prevalence of fever, fever with diarrhea or vomiting, or coughing with blood at endline,	
	\item	Health-seeking behavior/preventative care. We focus on the share of pregnancies resulting in births in medical facilities, the share of children under two years of age with at least one vaccination in the prior year, and the share of children under two years of age with a complete dose of vaccines.
	\item	Household productive assets (IHS).
	\item	Housing quality. Two measures are used:  the self-reported replacement cost of the current dwelling (irrespective of ownership status, IHS), and an index of housing construction quality, constructed from measures of wall and roof materials and from the number of rooms in the dwelling.
\end{enumerate}
The inverse hyperbolic sine is commonly used in analysis of outcomes such as consumption, savings, and asset values that tend to be highly right-skewed and also to contain zeros.  The IHS transformation preserves the interpretation of a log (meaning that impacts can be interpreted as percent changes) but does not drop zeros.  Only outcomes that we expected to be skewed were pre-registered to be analyzed using IHS.  All non-binary outcomes are also Winsorized at the 1 percent and 99 percent level (values above the 99th percentile are overwritten with the value at the 99th percentile to reduce skewness and increase statistical power).  Because we restrict the analysis in this paper to the pre-specified primary and secondary outcomes only, we do not correct the results for multiple inference \citep{anderson2008multiple}.

\subsection{Pre-committed Analysis of Heterogeneity}

\subsubsection{Anthropometric effects by baseline malnourishment}

We hypothesized in the Pre-Analysis Plan that the benefits of the treatments in terms of child anthropometrics would be largest for those who began the study most malnourished.  To test this, we run a regression with child anthropometrics (HAZ, WAZ, and MUAC) as the outcomes, using the structure of Equation 1 above and controlling for our battery of baseline covariates, a dummy for all three treatments (GK, GD, and GD large), the baseline biometric outcome, and the interaction between the treatments and baseline biometrics.  The hypothesis is that the interaction terms will be negative, meaning that the programs are most effective for those who had the worst baseline biometric outcomes.  Table \ref{t:het_by_anthro}  the results of this analysis.  The interpretation of the impacts in this table are as follows:  rows 4-6 give the simple impact of the programs when the interacted term is zero (which, in this case, is at the mean).  Rows 1-3 provide a test of the differential impact of the program across baseline anthropometric measures, so the lack of significance in these rows means that the impacts are not heterogeneous by nutrition status at baseline.  The implication is that the  improvement in anthropometrics induced by the GD large treatment were experienced broadly across the baseline distribution of HAZ and WAZ, and were not concentrated among those who began the study most malnourished.

\subsection{Selection of Control Variables.}\label{ss:BCH}
In our pre-analysis plan, we state that control variables for the primary specification  ``will be selected on the basis of their ability to predict the primary outcomes''.  In doing so, we seek to build on recent developments that balance the challenge of using baseline data to select variables that will reduce residual variance  with the danger that researcher freedom in the selection of control variables can lead to $p$-hacking, in which right-hand-side variables are selected specifically on the basis of the statistical significance of the coefficient of interest \citep{CarKrue95aer,CasGlenMig12qje}, thereby invalidating inference. 

To balance these concerns, we follow the \emph{post-double-selection} approach set forth in \citet{BelCheHan14restud}. Those authors advocate a two-step procedure in which, first, Lasso is used to automate the selection of control variables, and second, the post-Lasso estimator \citep{BelCheHan12ecta} is used to estimate the coefficients of primary interest in in the ITT, effectively using Lasso as a model selection device but \emph{not} imposing the shrunken coefficients that results from the Lasso estimates directly. \citet{BelCheHan14restud} demonstrate that this approach not only reduces bias in estimated treatment effects better than alternative approaches---less a concern given the successful randomization in our experiment---but that it may improve power while retaining uniformly valid inference.  

In the first stage, model selection is undertaken by retaining control variables from the union of those chosen either as predictive of the treatment assignment or of the outcome.  This model selection stage can be undertaken after residualizing to account for a set of control variables that the authors have a priori determined belonw in the model, as in \citet{BelCheHan14jep}; in our case, we retain block fixed effects, lagged values of the outcome, and lagged values of (the inverse hyperbolic sine of) household wealth in all specifications, per our pre-analysis plan. We modify the heteroskedasticity-robust Lasso estimator of \citet{BelCheHan12ecta} to incorporate sampling weights consistent with our design, using the Lasso penalty is chosen as a function of the sample size and the number of potential covariates, as in \citet{BelCheHan14jep}.

Resulting covariates selected for each of the primary and secondary outcomes, at household and individual level, are presented in Tables \ref{t:hcontrols} and \ref{t:icontrols}, respectively.

\begin{center}
\begin{footnotesize}
\begin{spacing}{1}
\input{tables/CovariateList_Household.tex}

\end{spacing}
\end{footnotesize}
\end{center}

\clearpage 
\begin{center}
\begin{footnotesize}
\begin{spacing}{1}
\input{tables/CovariateList_Individual.tex}

\end{spacing}
\end{footnotesize}
\end{center}

\subsection{Flyers from GiveDirectly for Cash Beneficiaries.}

	\includegraphics[scale=.75]{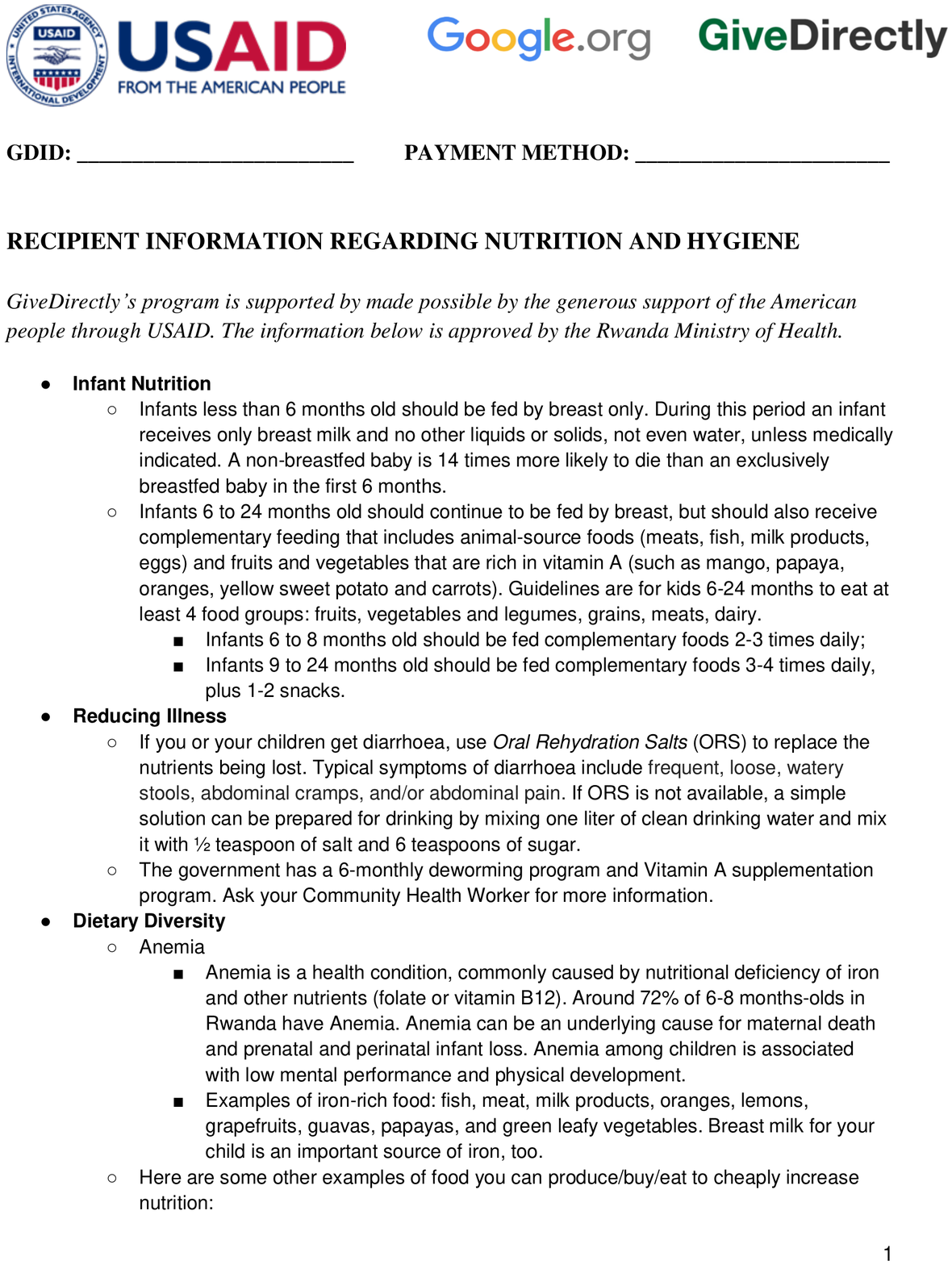}

	\includegraphics[scale=.75]{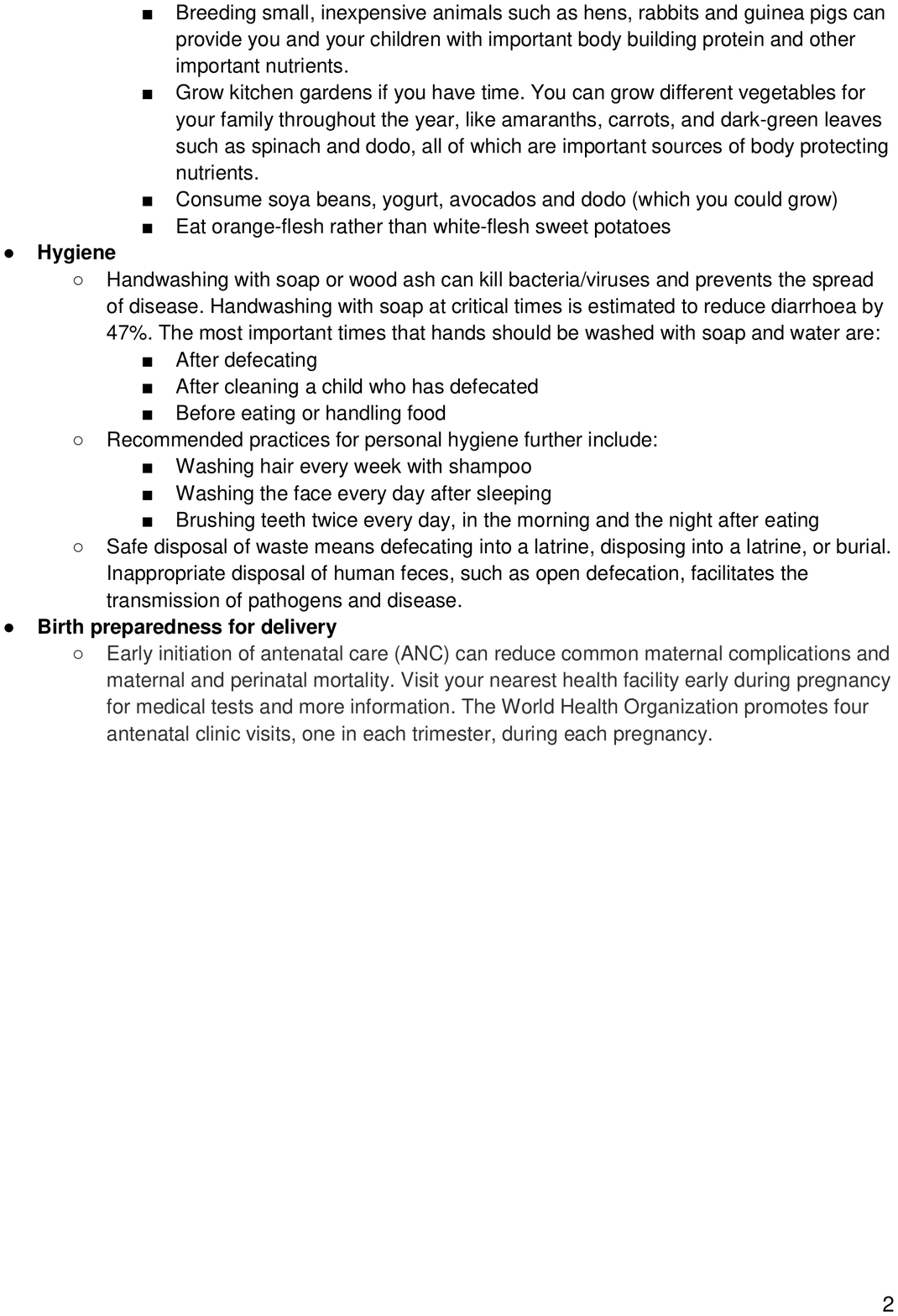}

%% file: tables/balance_secondary.tex
\begin{tabular}{l *{3}{S} ScS}
\toprule
\multicolumn{1}{c}{\text{ }} & \multicolumn{1}{c}{\text{\makecell[b]{Gikuriro\\Village}}} & \multicolumn{1}{c}{\text{\makecell[b]{GD Main\\Village}}} & \multicolumn{1}{c}{\text{\makecell[b]{GD Large\\Village}}} & \multicolumn{1}{c}{\text{\makecell[b]{Control\\Mean}}} & \multicolumn{1}{c}{\text{Observations}} & \multicolumn{1}{c}{\text{$ R^2$}}\\
\midrule
\multicolumn{7}{l}{\emph{A.  Household outcomes}}  \\ 
\addlinespace[1ex] \multirow[t]{ 3 }{0.2\textwidth}{Stock of borrowing$^\dag$ }  &    -0.459 &    -0.007 &    -0.262 &      5.96 &      1751 &      0.04 \\ 
 & (    0.363)  & (    0.409)  & (    0.408)  \\  & [    0.89]  & [    1.00]  & [    1.00]  \\ \addlinespace[1ex] 
\multirow[t]{ 3 }{0.2\textwidth}{Stock of saving$^\dag$ }  &    -0.157 &    -0.665\ensuremath{^{*}} &    -0.269 &      5.18 &      1751 &      0.02 \\ 
 & (    0.378)  & (    0.364)  & (    0.421)  \\  & [    1.00]  & [    0.89]  & [    1.00]  \\ \addlinespace[1ex] 
\multirow[t]{ 3 }{0.2\textwidth}{Health knowledge index }  &    -0.590 &    -0.119 &    -0.225 &      0.19 &      1751 &      0.03 \\ 
 & (    0.366)  & (    0.412)  & (    0.520)  \\  & [    0.89]  & [    1.00]  & [    1.00]  \\ \addlinespace[1ex] 
\multirow[t]{ 3 }{0.2\textwidth}{Sanitation practices index }  &     0.285\ensuremath{^{*}} &    -0.105 &    -0.069 &     -0.23 &      1751 &      0.04 \\ 
 & (    0.169)  & (    0.190)  & (    0.210)  \\  & [    0.89]  & [    1.00]  & [    1.00]  \\ \addlinespace[1ex] 
\multirow[t]{ 3 }{0.2\textwidth}{Productive assets$^\dag$ }  &     0.281\ensuremath{^{**}} &     0.195 &     0.231\ensuremath{^{*}} &     11.41 &      1751 &      0.12 \\ 
 & (    0.125)  & (    0.132)  & (    0.122)  \\  & [    0.89]  & [    0.89]  & [    0.89]  \\ \addlinespace[1ex] 
\multirow[t]{ 3 }{0.2\textwidth}{Consumption assets$^\dag$ }  &     0.158 &    -0.034 &     0.426 &      8.71 &      1751 &      0.08 \\ 
 & (    0.290)  & (    0.316)  & (    0.300)  \\  & [    1.00]  & [    1.00]  & [    0.89]  \\ \addlinespace[1ex] 
\multirow[t]{ 3 }{0.2\textwidth}{House value$^\dag$ }  &    -0.042 &    -0.012 &    -0.067 &     13.59 &      1751 &      0.09 \\ 
 & (    0.059)  & (    0.074)  & (    0.066)  \\  & [    1.00]  & [    1.00]  & [    1.00]  \\ \addlinespace[1ex] 
\multirow[t]{ 3 }{0.2\textwidth}{Housing quality index }  &     0.018 &    -0.195 &    -0.014 &      0.02 &      1751 &      0.04 \\ 
 & (    0.112)  & (    0.132)  & (    0.198)  \\  & [    1.00]  & [    0.89]  & [    1.00]  \\ \addlinespace[1ex] 
\multicolumn{7}{l}{\emph{B.  Individual outcomes}}  \\ 
\addlinespace[1ex] \multirow[t]{ 3 }{0.2\textwidth}{Pregnancy }  &    -0.018 &    -0.031 &    -0.021 &      0.28 &      2358 &      0.03 \\ 
 & (    0.025)  & (    0.022)  & (    0.024)  \\  & [    1.00]  & [    1.00]  & [    1.00]  \\ \addlinespace[1ex] 
\multirow[t]{ 3 }{0.2\textwidth}{Live Birth }  &    -0.017 &    -0.007 &     0.085 &      0.81 &       645 &      0.10 \\ 
 & (    0.050)  & (    0.049)  & (    0.061)  \\  & [    1.00]  & [    1.00]  & [    1.00]  \\ \addlinespace[1ex] 
\multirow[t]{ 3 }{0.2\textwidth}{Birth in Facility }  &     0.011 &    -0.056 &    -0.024 &      0.93 &       544 &      0.11 \\ 
 & (    0.038)  & (    0.043)  & (    0.044)  \\  & [    1.00]  & [    1.00]  & [    1.00]  \\ \addlinespace[1ex] 
\multirow[t]{ 3 }{0.2\textwidth}{Any Vaccinations in past year }  &     0.009 &    -0.006 &     0.001 &      0.93 &      1349 &      0.01 \\ 
 & (    0.019)  & (    0.021)  & (    0.030)  \\  & [    1.00]  & [    1.00]  & [    1.00]  \\ \addlinespace[1ex] 
\multirow[t]{ 3 }{0.2\textwidth}{Completed Vaccinations }  &    -0.015 &    -0.015 &     0.017 &      0.72 &      1347 &      0.02 \\ 
 & (    0.037)  & (    0.045)  & (    0.042)  \\  & [    1.00]  & [    1.00]  & [    1.00]  \\ \addlinespace[1ex] 
\multirow[t]{ 3 }{0.2\textwidth}{Disease Burden }  &     0.030 &     0.004 &     0.007 &      0.42 &      1146 &      0.02 \\ 
 & (    0.040)  & (    0.032)  & (    0.043)  \\  & [    1.00]  & [    1.00]  & [    1.00]  \\ \addlinespace[1ex] 
\bottomrule
\end{tabular}

%% file: tables/balance_RHS.tex
\begin{tabular}{l *{3}{S} ScS}
\toprule
\multicolumn{1}{c}{\text{ }} & \multicolumn{1}{c}{\text{\makecell[b]{Gikuriro\\Village}}} & \multicolumn{1}{c}{\text{\makecell[b]{GD Main\\Village}}} & \multicolumn{1}{c}{\text{\makecell[b]{GD Large\\Village}}} & \multicolumn{1}{c}{\text{\makecell[b]{Control\\Mean}}} & \multicolumn{1}{c}{\text{Observations}} & \multicolumn{1}{c}{\text{$ R^2$}}\\
\midrule
\multirow[t]{ 3 }{0.2\textwidth}{Female Headed }  &     0.036 &     0.043\ensuremath{^{*}} &    -0.018 &      0.16 &      1751 &      0.06 \\ 
 & (    0.025)  & (    0.026)  & (    0.029)  \\  & [    1.00]  & [    0.84]  & [    1.00]  \\ \addlinespace[1ex] 
\multirow[t]{ 3 }{0.2\textwidth}{Agricultural }  &     0.017 &    -0.027 &     0.002 &      0.85 &      1751 &      0.04 \\ 
 & (    0.028)  & (    0.029)  & (    0.035)  \\  & [    1.00]  & [    1.00]  & [    1.00]  \\ \addlinespace[1ex] 
\multirow[t]{ 3 }{0.2\textwidth}{Wage Worker }  &    -0.002 &    -0.063\ensuremath{^{**}} &    -0.084\ensuremath{^{**}} &      0.25 &      1751 &      0.04 \\ 
 & (    0.029)  & (    0.031)  & (    0.035)  \\  & [    1.00]  & [    0.63]  & [    0.46]  \\ \addlinespace[1ex] 
\multirow[t]{ 3 }{0.2\textwidth}{Microenterprise }  &    -0.015 &     0.008 &    -0.024 &      0.13 &      1751 &      0.02 \\ 
 & (    0.025)  & (    0.024)  & (    0.023)  \\  & [    1.00]  & [    1.00]  & [    1.00]  \\ \addlinespace[1ex] 
\multirow[t]{ 3 }{0.2\textwidth}{Savings Group }  &    -0.013 &    -0.022 &     0.026 &      0.25 &      1751 &      0.02 \\ 
 & (    0.038)  & (    0.039)  & (    0.044)  \\  & [    1.00]  & [    1.00]  & [    1.00]  \\ \addlinespace[1ex] 
\multirow[t]{ 3 }{0.2\textwidth}{Village Eligibility Ratio }  &    -0.015 &     0.037 &     0.017 &      0.16 &      1751 &      0.50 \\ 
 & (    0.025)  & (    0.029)  & (    0.033)  \\  & [    1.00]  & [    1.00]  & [    1.00]  \\ \addlinespace[1ex] 
\multirow[t]{ 3 }{0.2\textwidth}{Age of Head }  &     2.186\ensuremath{^{**}} &     2.868\ensuremath{^{**}} &     1.415 &     34.16 &      1751 &      0.07 \\ 
 & (    1.047)  & (    1.200)  & (    1.487)  \\  & [    0.63]  & [    0.46]  & [    1.00]  \\ \addlinespace[1ex] 
\multirow[t]{ 3 }{0.2\textwidth}{Schooling of Head }  &    -0.006 &    -0.002 &    -0.005 &      0.00 &      1751 &      0.02 \\ 
 & (    0.005)  & (    0.006)  & (    0.004)  \\  & [    1.00]  & [    1.00]  & [    1.00]  \\ \addlinespace[1ex] 
\multirow[t]{ 3 }{0.2\textwidth}{Dependency Ratio }  &     0.008 &    -0.007 &     0.003 &      0.59 &      1751 &      0.04 \\ 
 & (    0.012)  & (    0.012)  & (    0.016)  \\  & [    1.00]  & [    1.00]  & [    1.00]  \\ \addlinespace[1ex] 
\multirow[t]{ 3 }{0.2\textwidth}{Household Size }  &    -0.082 &    -0.054 &    -0.183 &      5.18 &      1751 &      0.02 \\ 
 & (    0.134)  & (    0.151)  & (    0.163)  \\  & [    1.00]  & [    1.00]  & [    1.00]  \\ \addlinespace[1ex] 
\multirow[t]{ 3 }{0.2\textwidth}{Poorest Category }  &    -0.040 &    -0.002 &    -0.068\ensuremath{^{*}} &      0.22 &      1751 &      0.05 \\ 
 & (    0.033)  & (    0.045)  & (    0.039)  \\  & [    1.00]  & [    1.00]  & [    0.84]  \\ \addlinespace[1ex] 
\multirow[t]{ 3 }{0.2\textwidth}{Next Poorest Category }  &     0.067\ensuremath{^{*}} &     0.056 &     0.061 &      0.50 &      1751 &      0.12 \\ 
 & (    0.040)  & (    0.046)  & (    0.051)  \\  & [    0.84]  & [    1.00]  & [    1.00]  \\ \addlinespace[1ex] 
\bottomrule
\end{tabular}

%% file: tables/GK_Compliance_Determinants_eligibles.tex
{
\def\sym#1{\ifmmode^{#1}\else\(^{#1}\)\fi}
\begin{tabular}{l*{5}{c}}
\toprule
                    &\multicolumn{1}{c}{Nutrition Training}&\multicolumn{1}{c}{Cooking Training}&\multicolumn{1}{c}{Farmer Training}&\multicolumn{1}{c}{Farmer Harvest}&\multicolumn{1}{c}{Received Livestock}\\
\midrule
Number of Children  &     -0.0639         &     -0.0594         &     -0.0344         &     -0.0496         &      0.0351         \\
                    &    (0.0436)         &    (0.0390)         &    (0.0448)         &    (0.0551)         &    (0.0341)         \\
\addlinespace
Number of Members   &      0.0815\sym{*}  &      0.0894\sym{**} &      0.0679         &      0.0806         &     -0.0271         \\
                    &    (0.0423)         &    (0.0367)         &    (0.0409)         &    (0.0518)         &    (0.0301)         \\
\addlinespace
Female headed HH    &      0.0190         &    -0.00581         &      -0.102         &     -0.0392         &     -0.0472         \\
                    &    (0.0691)         &    (0.0628)         &    (0.0762)         &    (0.0693)         &    (0.0650)         \\
\addlinespace
Age of HH head      &    -0.00543\sym{***}&    -0.00436\sym{**} &    -0.00153         &    -0.00178         &    -0.00377\sym{**} \\
                    &   (0.00204)         &   (0.00190)         &   (0.00182)         &   (0.00182)         &   (0.00143)         \\
\addlinespace
Poorest poverty     &       0.150\sym{**} &       0.190\sym{**} &       0.222\sym{***}&       0.162\sym{**} &       0.238\sym{***}\\
group               &    (0.0746)         &    (0.0766)         &    (0.0740)         &    (0.0756)         &    (0.0745)         \\
\addlinespace
Next poorest group  &       0.165\sym{**} &       0.147\sym{**} &       0.151\sym{**} &      0.0869         &       0.211\sym{***}\\
                    &    (0.0632)         &    (0.0589)         &    (0.0612)         &    (0.0687)         &    (0.0509)         \\
\addlinespace
HH in Agriculture   &      0.0212         &     -0.0215         &     -0.0462         &     -0.0895         &     -0.0458         \\
                    &    (0.0812)         &    (0.0762)         &    (0.0815)         &    (0.0808)         &    (0.0711)         \\
\addlinespace
HH in Wage Work     &      0.0159         &      0.0539         &      0.0795         &      0.0994\sym{*}  &       0.139\sym{***}\\
                    &    (0.0584)         &    (0.0532)         &    (0.0483)         &    (0.0554)         &    (0.0487)         \\
\addlinespace
HH in               &     -0.0460         &      0.0294         &     -0.0501         &      0.0194         &    -0.00755         \\
Microenterprise     &    (0.0688)         &    (0.0802)         &    (0.0604)         &    (0.0645)         &    (0.0662)         \\
\midrule
Mean DV             &       0.628         &       0.511         &       0.568         &       0.480         &       0.335         \\
N                   &         524         &         524         &         524         &         529         &         529         \\
\bottomrule
\end{tabular}
}

%% file: tables/ITT_IPW.tex
\begin{tabular}{l *{3}{S} ScS}
\toprule
\multicolumn{1}{c}{\text{ }} & \multicolumn{1}{c}{\text{Gikuriro}} & \multicolumn{1}{c}{\text{Main}} & \multicolumn{1}{c}{\text{Large}} & \multicolumn{1}{c}{\text{Mean}} & \multicolumn{1}{c}{\text{Obs.}} & \multicolumn{1}{c}{\text{$ R^2$}}\\
\midrule
\multirow[t]{ 3 }{0.2\textwidth}{Height-for-age }  &     0.051 &    -0.021 &     0.091\ensuremath{^{**}} &     -2.06 &      2125 &      0.71 \\ 
 & (    0.045)  & (    0.039)  & (    0.046)  \\  & [    0.62]  & [    1.00]  & [    0.35]  \\ \addlinespace[1ex] 
\multirow[t]{ 3 }{0.2\textwidth}{Weight-for-age }  &     0.038 &     0.010 &     0.067\ensuremath{^{*}} &     -1.06 &      2104 &      0.68 \\ 
 & (    0.040)  & (    0.034)  & (    0.036)  \\  & [    0.69]  & [    1.00]  & [    0.35]  \\ \addlinespace[1ex] 
\multirow[t]{ 3 }{0.2\textwidth}{Mid-upper arm circumference }  &     0.022 &    -0.007 &     0.135\ensuremath{^{*}} &     -0.58 &      1629 &      0.50 \\ 
 & (    0.056)  & (    0.065)  & (    0.078)  \\  & [    1.00]  & [    1.00]  & [    0.35]  \\ \addlinespace[1ex] 
\bottomrule
\end{tabular}

%% file: tables/ce_tce_primary.tex
\begin{tabular}{l *{3}{S} ScS}
\toprule
\multicolumn{1}{c}{\text{ }} & \multicolumn{1}{c}{\text{\makecell[b]{Gikuriro:\\ Differential}}} & \multicolumn{1}{c}{\text{\makecell[b]{Cost-equivalent\\GD impact}}} & \multicolumn{1}{c}{\text{\makecell[b]{Transfer\\Cost}}} & \multicolumn{1}{c}{\text{\makecell[b]{Control\\Mean}}} & \multicolumn{1}{c}{\text{Observations}} & \multicolumn{1}{c}{\text{$ R^2$}}\\
\midrule
\multicolumn{7}{l}{\emph{A.  Household outcomes}}  \\ 
\addlinespace[1ex] \multirow[t]{ 3 }{0.2\textwidth}{Consumption$^\dag$ }  &    -0.016 &    -0.125 &     0.001 &     10.39 &      2717 &      0.15 \\ 
 & (    0.096)  & (    0.097)  & (    0.002)  \\  & [    0.93]  & [    0.81]  & [    0.85]  \\ \addlinespace[1ex] 
\multirow[t]{ 3 }{0.2\textwidth}{Household dietary diversity score }  &     0.184 &    -0.056 &    -0.003 &      4.12 &      2718 &      0.21 \\ 
 & (    0.122)  & (    0.121)  & (    0.002)  \\  & [    0.81]  & [    0.85]  & [    0.81]  \\ \addlinespace[1ex] 
\multirow[t]{ 3 }{0.2\textwidth}{Household non-land wealth$^\dag$ }  &     0.169 &    -0.354 &    -0.001 &     13.28 &      2718 &      0.29 \\ 
 & (    0.248)  & (    0.254)  & (    0.004)  \\  & [    0.85]  & [    0.81]  & [    0.93]  \\ \addlinespace[1ex] 
\multicolumn{7}{l}{\emph{B.  Individual outcomes}}  \\ 
\addlinespace[1ex] \multirow[t]{ 3 }{0.2\textwidth}{Height-for-Age }  &    -0.013 &    -0.001 &     0.001 &     -1.75 &      2618 &      0.74 \\ 
 & (    0.053)  & (    0.051)  & (    0.001)  \\  & [    1.00]  & [    1.00]  & [    1.00]  \\ \addlinespace[1ex] 
\multirow[t]{ 3 }{0.2\textwidth}{Weight-for-Age }  &    -0.057 &    -0.009 &     0.000 &     -0.87 &      2594 &      0.74 \\ 
 & (    0.047)  & (    0.037)  & (    0.001)  \\  & [    1.00]  & [    1.00]  & [    1.00]  \\ \addlinespace[1ex] 
\multirow[t]{ 3 }{0.2\textwidth}{Mid-Upper Arm Circ }  &    -0.065 &    -0.007 &     0.001 &     -0.61 &      1981 &      0.57 \\ 
 & (    0.067)  & (    0.064)  & (    0.001)  \\  & [    1.00]  & [    1.00]  & [    1.00]  \\ \addlinespace[1ex] 
\bottomrule
\end{tabular}

%% file: tables/ce_tce_secondary.tex
\begin{tabular}{l *{3}{S} ScS}
\toprule
\multicolumn{1}{c}{\text{ }} & \multicolumn{1}{c}{\text{\makecell[b]{Gikuriro:\\ Differential}}} & \multicolumn{1}{c}{\text{\makecell[b]{Cost-equivalent\\GD impact}}} & \multicolumn{1}{c}{\text{\makecell[b]{Transfer\\Cost}}} & \multicolumn{1}{c}{\text{\makecell[b]{Control\\Mean}}} & \multicolumn{1}{c}{\text{Observations}} & \multicolumn{1}{c}{\text{$ R^2$}}\\
\midrule
\multicolumn{7}{l}{\emph{A.  Household outcomes}}  \\ 
\addlinespace[1ex] \multirow[t]{ 3 }{0.2\textwidth}{Stock of borrowing$^\dag$ }  &     0.495 &    -0.354 &     0.001 &      5.75 &      2715 &      0.11 \\ 
 & (    0.361)  & (    0.365)  & (    0.005)  \\  & [    0.93]  & [    1.00]  & [    1.00]  \\ \addlinespace[1ex] 
\multirow[t]{ 3 }{0.2\textwidth}{Stock of saving$^\dag$ }  &     0.264 &    -0.482 &    -0.004 &      5.38 &      2718 &      0.15 \\ 
 & (    0.363)  & (    0.354)  & (    0.005)  \\  & [    1.00]  & [    0.93]  & [    1.00]  \\ \addlinespace[1ex] 
\multirow[t]{ 3 }{0.2\textwidth}{Health knowledge index }  &     0.952\ensuremath{^{**}} &     0.497 &    -0.014\ensuremath{^{**}} &     -0.01 &      2718 &      0.06 \\ 
 & (    0.382)  & (    0.373)  & (    0.007)  \\  & [    0.19]  & [    0.93]  & [    0.44]  \\ \addlinespace[1ex] 
\multirow[t]{ 3 }{0.2\textwidth}{Sanitation practices index }  &     0.326 &    -0.162 &     0.007 &      0.03 &      2718 &      0.07 \\ 
 & (    0.206)  & (    0.218)  & (    0.005)  \\  & [    0.93]  & [    1.00]  & [    0.93]  \\ \addlinespace[1ex] 
\multirow[t]{ 3 }{0.2\textwidth}{Productive assets$^\dag$ }  &     0.035 &    -0.116 &     0.005\ensuremath{^{**}} &     11.65 &      2718 &      0.30 \\ 
 & (    0.129)  & (    0.134)  & (    0.002)  \\  & [    1.00]  & [    1.00]  & [    0.19]  \\ \addlinespace[1ex] 
\multirow[t]{ 3 }{0.2\textwidth}{Consumption assets$^\dag$ }  &     0.054 &     0.047 &     0.004 &      9.08 &      2718 &      0.32 \\ 
 & (    0.227)  & (    0.225)  & (    0.004)  \\  & [    1.00]  & [    1.00]  & [    1.00]  \\ \addlinespace[1ex] 
\multirow[t]{ 3 }{0.2\textwidth}{House value$^\dag$ }  &    -0.024 &     0.021 &     0.001 &     13.70 &      2531 &      0.39 \\ 
 & (    0.070)  & (    0.060)  & (    0.001)  \\  & [    1.00]  & [    1.00]  & [    1.00]  \\ \addlinespace[1ex] 
\multirow[t]{ 3 }{0.2\textwidth}{Housing quality index }  &    -0.165 &    -0.007 &    -0.000 &      0.12 &      2718 &      0.16 \\ 
 & (    0.139)  & (    0.148)  & (    0.002)  \\  & [    1.00]  & [    1.00]  & [    1.00]  \\ \addlinespace[1ex] 
\multicolumn{7}{l}{\emph{B.  Individual outcomes}}  \\ 
\addlinespace[1ex] \multirow[t]{ 3 }{0.2\textwidth}{Child Mortality }  &     0.006 &    -0.002 &    -0.000 &      0.01 &      3373 &      0.02 \\ 
 & (    0.006)  & (    0.002)  & (    0.000)  \\  & [    1.00]  & [    1.00]  & [    1.00]  \\ \addlinespace[1ex] 
\multirow[t]{ 3 }{0.2\textwidth}{Pregnancy }  &     0.019 &     0.001 &    -0.000 &      0.12 &      4137 &      0.11 \\ 
 & (    0.020)  & (    0.018)  & (    0.000)  \\  & [    1.00]  & [    1.00]  & [    1.00]  \\ \addlinespace[1ex] 
\multirow[t]{ 3 }{0.2\textwidth}{Live Birth }  &     0.053 &    -0.047 &     0.000 &      0.70 &       594 &      0.13 \\ 
 & (    0.088)  & (    0.082)  & (    0.001)  \\  & [    1.00]  & [    1.00]  & [    1.00]  \\ \addlinespace[1ex] 
\multirow[t]{ 3 }{0.2\textwidth}{Birth in Facility }  &    -0.071 &     0.042 &    -0.001 &      0.90 &       416 &      0.17 \\ 
 & (    0.052)  & (    0.060)  & (    0.001)  \\  & [    0.83]  & [    1.00]  & [    1.00]  \\ \addlinespace[1ex] 
\multirow[t]{ 3 }{0.2\textwidth}{Any Vaccinations in past year }  &     0.023 &     0.092\ensuremath{^{*}} &     0.001 &      0.73 &      1479 &      0.31 \\ 
 & (    0.044)  & (    0.054)  & (    0.001)  \\  & [    1.00]  & [    0.81]  & [    0.81]  \\ \addlinespace[1ex] 
\multirow[t]{ 3 }{0.2\textwidth}{Completed Vaccinations }  &     0.094\ensuremath{^{*}} &     0.104\ensuremath{^{*}} &     0.002\ensuremath{^{**}} &      0.48 &      1479 &      0.17 \\ 
 & (    0.057)  & (    0.060)  & (    0.001)  \\  & [    0.81]  & [    0.81]  & [    0.81]  \\ \addlinespace[1ex] 
\multirow[t]{ 3 }{0.2\textwidth}{Disease Burden }  &     0.039 &    -0.021 &     0.000 &      0.54 &      3366 &      0.06 \\ 
 & (    0.043)  & (    0.045)  & (    0.001)  \\  & [    1.00]  & [    1.00]  & [    1.00]  \\ \addlinespace[1ex] 
\multirow[t]{ 3 }{0.2\textwidth}{Diarrheal Prevalence }  &    -0.026 &     0.033 &    -0.000 &      0.09 &      3366 &      0.05 \\ 
 & (    0.021)  & (    0.021)  & (    0.000)  \\  & [    0.87]  & [    0.81]  & [    1.00]  \\ \addlinespace[1ex] 
\bottomrule
\end{tabular}

%% file: tables/anthro_het_baseline.tex
{
\def\sym#1{\ifmmode^{#1}\else\(^{#1}\)\fi}
\begin{tabular*}{\hsize}{@{\hskip\tabcolsep\extracolsep\fill}l*{3}{c}}
\hline\hline
                &\multicolumn{1}{c}{(1)}&\multicolumn{1}{c}{(2)}&\multicolumn{1}{c}{(3)}\\
                &\multicolumn{1}{c}{Height-for-Age}&\multicolumn{1}{c}{Weight-for-Age}&\multicolumn{1}{c}{Mid-Upper Arm Circ}\\
\hline
Baseline outcome x Gikuriro&  -0.0416         &  -0.0349         &   0.0852         \\
                & (0.0444)         & (0.0619)         & (0.0564)         \\
[1em]
Baseline outcome x GD Main&  -0.0247         &  -0.0654         &   0.0776         \\
                & (0.0457)         & (0.0445)         & (0.0653)         \\
[1em]
Baseline outcome x GD Large&   0.0220         &  0.00599         &   0.0804         \\
                & (0.0433)         & (0.0461)         & (0.0603)         \\
[1em]
Gikuriro        &   0.0434         &   0.0323         &   0.0253         \\
                & (0.0428)         & (0.0362)         & (0.0557)         \\
[1em]
GD Main         &  -0.0252         &  0.00182         & -0.00498         \\
                & (0.0398)         & (0.0357)         & (0.0647)         \\
[1em]
GD Large        &   0.0940\sym{*}  &   0.0641         &    0.135\sym{*}  \\
                & (0.0517)         & (0.0392)         & (0.0795)         \\
[1em]
Baseline Outcome&    0.768\sym{***}&    0.748\sym{***}&    0.600\sym{***}\\
                & (0.0336)         & (0.0355)         & (0.0425)         \\
\hline
Observations    &     2125         &     2104         &     1629         \\
Mean of DV      &   -2.031         &   -1.043         &   -0.572         \\
R squared       &    0.696         &    0.673         &    0.507         \\
\hline\hline
\multicolumn{4}{l}{\footnotesize Standard errors in parentheses}\\
\multicolumn{4}{l}{\footnotesize \sym{*} \(p<0.10\), \sym{**} \(p<0.05\), \sym{***} \(p<0.01\)}\\
\end{tabular*}
}

%% file: tables/anthro_age_baseline.tex
{
\def\sym#1{\ifmmode^{#1}\else\(^{#1}\)\fi}
\begin{tabular*}{\hsize}{@{\hskip\tabcolsep\extracolsep\fill}l*{6}{c}}
\hline\hline
                &First Thousand Days         &                  &                  &  Newborn         &                  &                  \\
                &\multicolumn{1}{c}{(1)}&\multicolumn{1}{c}{(2)}&\multicolumn{1}{c}{(3)}&\multicolumn{1}{c}{(4)}&\multicolumn{1}{c}{(5)}&\multicolumn{1}{c}{(6)}\\
                &\multicolumn{1}{c}{Height-for-Age}&\multicolumn{1}{c}{Weight-for-Age}&\multicolumn{1}{c}{Mid-Upper Arm Circ}&\multicolumn{1}{c}{Height-for-Age}&\multicolumn{1}{c}{Weight-for-Age}&\multicolumn{1}{c}{Mid-Upper Arm Circ}\\
\hline
Indicator x Gikuriro& -0.00731         &  -0.0206         &    0.115         &    0.599         &    0.251         &    0.282         \\
                &  (0.138)         &  (0.113)         &  (0.109)         &  (0.645)         &  (0.505)         &  (0.491)         \\
[1em]
Indicator x GD Main&   -0.300\sym{**} &   -0.152         &    0.159         &    0.382         &    0.594         &    0.666         \\
                &  (0.138)         &  (0.115)         &  (0.104)         &  (0.522)         &  (0.495)         &  (0.506)         \\
[1em]
Indicator x GD Large&   -0.115         &  -0.0159         &    0.160         &    0.407         &    0.729         &    0.304         \\
                &  (0.139)         &  (0.122)         &  (0.144)         &  (0.396)         &  (0.469)         &  (0.281)         \\
[1em]
Gikuriro        &   0.0105         &   0.0114         &  -0.0813         &  0.00489         &  0.00325         &  -0.0263         \\
                &  (0.106)         & (0.0801)         & (0.0822)         & (0.0833)         & (0.0629)         & (0.0690)         \\
[1em]
GD Main         &    0.115         &   0.0779         &   -0.108         &  -0.0171         &  0.00905         &  -0.0346         \\
                &  (0.119)         & (0.0829)         & (0.0905)         & (0.0988)         & (0.0678)         & (0.0723)         \\
[1em]
GD Large        &    0.246\sym{**} &    0.191\sym{**} &   0.0773         &    0.196\sym{**} &    0.185\sym{***}&    0.159\sym{**} \\
                &  (0.105)         & (0.0817)         &  (0.111)         & (0.0848)         & (0.0668)         & (0.0783)         \\
[1em]
Indicator       &    0.141         &    0.123         &  -0.0101         &  -0.0102         &   0.0247         &    0.177         \\
                &  (0.148)         &  (0.117)         &  (0.142)         &  (0.254)         &  (0.295)         &  (0.270)         \\
\hline
Observations    &     2360         &     2347         &     2020         &     2360         &     2347         &     2020         \\
Mean of DV      &   -2.031         &   -1.043         &   -0.572         &   -2.031         &   -1.043         &   -0.572         \\
R squared       &   0.0722         &   0.0356         &   0.0726         &   0.0699         &   0.0358         &   0.0740         \\
\hline\hline
\multicolumn{7}{l}{\footnotesize Standard errors in parentheses}\\
\multicolumn{7}{l}{\footnotesize \sym{*} \(p<0.10\), \sym{**} \(p<0.05\), \sym{***} \(p<0.01\)}\\
\end{tabular*}
}

%% file: tables/Het_behavior_GK_vs_GD.tex
{
\def\sym#1{\ifmmode^{#1}\else\(^{#1}\)\fi}
\begin{tabular}{l*{9}{c}}
\hline\hline
                    &\multicolumn{1}{c}{(1)}&\multicolumn{1}{c}{(2)}&\multicolumn{1}{c}{(3)}&\multicolumn{1}{c}{(4)}&\multicolumn{1}{c}{(5)}&\multicolumn{1}{c}{(6)}&\multicolumn{1}{c}{(7)}&\multicolumn{1}{c}{(8)}&\multicolumn{1}{c}{(9)}\\
                    &\multicolumn{1}{c}{Consumption$^\dag$}&\multicolumn{1}{c}{Diet Diversity}&\multicolumn{1}{c}{Wealth$^\dag$}&\multicolumn{1}{c}{Borrowing$^\dag$ }&\multicolumn{1}{c}{Saving$^\dag$}&\multicolumn{1}{c}{Health}&\multicolumn{1}{c}{Sanitation}&\multicolumn{1}{c}{Prod assets$^\dag$}&\multicolumn{1}{c}{Cons assets$^\dag$}\\
                    &        b/se   &        b/se   &        b/se   &        b/se   &        b/se   &        b/se   &        b/se   &        b/se   &        b/se   \\
\hline
Inconsistent x GK   &        0.22   &        0.37   &       -0.34   &        0.85   &        1.04   &       -0.34   &        0.54   &        0.56** &        0.23   \\
                    &      (0.20)   &      (0.25)   &      (0.39)   &      (0.77)   &      (0.82)   &      (0.68)   &      (0.42)   &      (0.27)   &      (0.42)   \\
Inconsistent x GD   &        0.42*  &        0.31   &     0.00021   &        0.25   &        0.69   &        0.40   &        0.68   &        0.56*  &        0.66   \\
                    &      (0.23)   &      (0.27)   &      (0.50)   &      (0.77)   &      (0.87)   &      (0.70)   &      (0.47)   &      (0.30)   &      (0.52)   \\
Impatient x GK      &      -0.044   &       -0.49   &       0.025   &       -0.86   &       -0.57   &        0.78   &       -0.90*  &       -0.59** &       -1.06*  \\
                    &      (0.20)   &      (0.32)   &      (0.52)   &      (0.85)   &      (0.87)   &      (0.78)   &      (0.53)   &      (0.26)   &      (0.57)   \\
Impatient x GD      &        0.30   &      -0.037   &       -0.19   &       -0.29   &      -0.072   &      -0.020   &        0.21   &        0.11   &       -0.86*  \\
                    &      (0.22)   &      (0.32)   &      (0.52)   &      (1.02)   &      (0.83)   &      (0.90)   &      (0.47)   &      (0.33)   &      (0.48)   \\
Lack Other Cont x GK&        0.16   &       -0.33   &       -0.67*  &        0.89   &       -0.54   &       -1.51***&        0.28   &       -0.32   &      -0.060   \\
                    &      (0.22)   &      (0.26)   &      (0.40)   &      (0.64)   &      (0.54)   &      (0.57)   &      (0.39)   &      (0.24)   &      (0.48)   \\
Lack Other Cont x GD&        0.16   &       -0.27   &       -0.63   &       -0.34   &       -0.44   &       -0.60   &       -0.11   &       -0.34   &       -0.23   \\
                    &      (0.22)   &      (0.28)   &      (0.45)   &      (0.68)   &      (0.65)   &      (0.73)   &      (0.43)   &      (0.29)   &      (0.52)   \\
Time Inconsistent   &       -0.17   &       -0.22   &      -0.041   &       -0.41   &       -0.98   &       -0.19   &       -0.67** &       -0.61***&       -0.67** \\
                    &      (0.17)   &      (0.20)   &      (0.25)   &      (0.61)   &      (0.70)   &      (0.42)   &      (0.27)   &      (0.17)   &      (0.29)   \\
Impatient           &       -0.18   &        0.34   &       -0.10   &       0.088   &        0.62   &       -0.27   &       0.094   &        0.14   &        1.07***\\
                    &      (0.16)   &      (0.27)   &      (0.33)   &      (0.72)   &      (0.62)   &      (0.60)   &      (0.38)   &      (0.20)   &      (0.36)   \\
Lack Other Control  &       -0.16   &       0.050   &        0.25   &        0.20   &        1.01** &        0.77*  &      -0.045   &        0.20   &       -0.16   \\
                    &      (0.16)   &      (0.19)   &      (0.21)   &      (0.47)   &      (0.41)   &      (0.39)   &      (0.27)   &      (0.19)   &      (0.27)   \\
Gikuriro            &       -0.29*  &        0.16   &        0.43   &       -0.54   &        0.69   &        0.43   &       -0.51   &       -0.12   &       -0.31   \\
                    &      (0.16)   &      (0.21)   &      (0.33)   &      (0.63)   &      (0.60)   &      (0.68)   &      (0.32)   &      (0.19)   &      (0.32)   \\
GiveDirectly        &       -0.33** &       0.011   &        0.23   &       -0.81   &       -0.47   &        0.14   &       -0.39   &      -0.080   &        0.10   \\
                    &      (0.16)   &      (0.23)   &      (0.49)   &      (0.60)   &      (0.63)   &      (0.69)   &      (0.39)   &      (0.22)   &      (0.42)   \\
\hline
Control Mean        &        10.4   &        4.16   &        12.9   &        5.96   &        5.18   &        0.19   &       -0.23   &        11.4   &        8.71   \\
Observations        &        1508   &        1509   &        1509   &        1509   &        1509   &        1509   &        1509   &        1509   &        1509   \\
$R^2$               &        0.15   &        0.18   &        0.22   &        0.12   &        0.17   &       0.049   &       0.079   &        0.32   &        0.38   \\
Self\_control\_GD=GK  &        0.29   &        0.80   &        0.52   &        0.38   &        0.61   &        0.35   &        0.77   &        0.98   &        0.42   \\
Other\_control\_GD=GK &        0.97   &        0.85   &        0.95   &       0.066   &        0.88   &        0.22   &        0.38   &        0.93   &        0.77   \\
\hline\hline
\end{tabular}
}

%% file: tables/Het_behavior_LS_Flow.tex
{
\def\sym#1{\ifmmode^{#1}\else\(^{#1}\)\fi}
\begin{tabular}{l*{9}{c}}
\hline\hline
                    &\multicolumn{1}{c}{(1)}&\multicolumn{1}{c}{(2)}&\multicolumn{1}{c}{(3)}&\multicolumn{1}{c}{(4)}&\multicolumn{1}{c}{(5)}&\multicolumn{1}{c}{(6)}&\multicolumn{1}{c}{(7)}&\multicolumn{1}{c}{(8)}&\multicolumn{1}{c}{(9)}\\
                    &\multicolumn{1}{c}{Consumption$^\dag$}&\multicolumn{1}{c}{Diet Diversity}&\multicolumn{1}{c}{Wealth$^\dag$}&\multicolumn{1}{c}{Borrowing$^\dag$ }&\multicolumn{1}{c}{Saving$^\dag$}&\multicolumn{1}{c}{Health}&\multicolumn{1}{c}{Sanitation}&\multicolumn{1}{c}{Prod assets$^\dag$}&\multicolumn{1}{c}{Cons assets$^\dag$}\\
                    &        b/se   &        b/se   &        b/se   &        b/se   &        b/se   &        b/se   &        b/se   &        b/se   &        b/se   \\
\hline
Inconsistent x Lump Sum&        0.37   &        0.34   &        0.74   &       0.097   &        0.48   &       0.061   &        1.10** &        0.77** &        0.96   \\
                    &      (0.25)   &      (0.30)   &      (0.70)   &      (1.05)   &      (0.91)   &      (0.97)   &      (0.48)   &      (0.33)   &      (0.69)   \\
Inconsistent x Flow &        0.35*  &        0.13   &        0.72   &        0.17   &        0.51   &       -0.48   &        0.58   &        1.13***&        1.09*  \\
                    &      (0.19)   &      (0.30)   &      (0.61)   &      (0.89)   &      (0.80)   &      (0.86)   &      (0.38)   &      (0.27)   &      (0.61)   \\
Impatient x Lump Sum&        0.31   &       -0.27   &        0.27   &       -1.61   &       -0.58   &       0.047   &        0.47   &        0.49   &       -0.42   \\
                    &      (0.26)   &      (0.32)   &      (0.76)   &      (1.10)   &      (1.12)   &      (0.80)   &      (0.48)   &      (0.56)   &      (0.77)   \\
Impatient x Flow    &        0.18   &       -0.40   &       -1.17*  &        1.23   &      -0.059   &        0.21   &       0.086   &       -0.41   &       -0.95   \\
                    &      (0.23)   &      (0.34)   &      (0.68)   &      (1.06)   &      (0.83)   &      (0.90)   &      (0.48)   &      (0.36)   &      (0.73)   \\
Lack Other Cont x Lump Sum&        0.28   &        0.11   &       -0.75   &       0.087   &       -0.27   &       -0.78   &       -0.68   &       0.093   &        0.41   \\
                    &      (0.21)   &      (0.33)   &      (0.88)   &      (1.08)   &      (0.71)   &      (0.75)   &      (0.71)   &      (0.34)   &      (0.80)   \\
Lack Other Cont x Flow&        0.19   &       -0.25   &      -0.028   &       0.076   &        0.15   &        0.33   &       -0.54   &       -0.21   &       -0.64   \\
                    &      (0.23)   &      (0.30)   &      (0.55)   &      (0.81)   &      (0.70)   &      (0.84)   &      (0.49)   &      (0.29)   &      (0.48)   \\
Time Inconsistent   &       -0.13   &       -0.19   &       -0.25   &       -0.32   &       -0.72   &        0.22   &       -0.73***&       -0.81***&       -0.61*  \\
                    &      (0.15)   &      (0.18)   &      (0.23)   &      (0.55)   &      (0.64)   &      (0.42)   &      (0.23)   &      (0.18)   &      (0.36)   \\
Impatient           &       -0.15   &        0.40*  &      -0.029   &      -0.064   &        0.53   &       -0.48   &        0.10   &       0.090   &        0.56   \\
                    &      (0.17)   &      (0.23)   &      (0.28)   &      (0.61)   &      (0.58)   &      (0.59)   &      (0.31)   &      (0.19)   &      (0.43)   \\
Lack Other Control  &       -0.12   &      -0.067   &       0.082   &       0.054   &        0.86** &        0.48   &       0.019   &        0.14   &       -0.13   \\
                    &      (0.16)   &      (0.18)   &      (0.21)   &      (0.43)   &      (0.43)   &      (0.36)   &      (0.27)   &      (0.18)   &      (0.26)   \\
GD Lump Sum         &       -0.27   &     -0.0054   &       -0.35   &        0.35   &        0.54   &        0.34   &       -0.26   &       -0.20   &        0.15   \\
                    &      (0.20)   &      (0.29)   &      (0.69)   &      (0.82)   &      (0.72)   &      (0.76)   &      (0.51)   &      (0.29)   &      (0.61)   \\
GD Flow             &       -0.21   &        0.47*  &       -0.28   &       -1.08*  &       0.050   &      -0.029   &       -0.14   &      -0.094   &     -0.0084   \\
                    &      (0.16)   &      (0.25)   &      (0.54)   &      (0.64)   &      (0.62)   &      (0.65)   &      (0.31)   &      (0.22)   &      (0.44)   \\
\hline
Control Mean        &        10.4   &        4.16   &        12.9   &        5.96   &        5.18   &        0.19   &       -0.23   &        11.4   &        8.71   \\
Observations        &        1131   &        1131   &        1131   &        1131   &        1131   &        1131   &        1131   &        1131   &        1131   \\
$R^2$               &        0.15   &        0.19   &        0.25   &        0.16   &        0.18   &       0.060   &       0.091   &        0.31   &        0.38   \\
Self\_control\_LS=Flow&        0.93   &        0.60   &        0.97   &        0.95   &        0.97   &        0.61   &        0.31   &        0.27   &        0.85   \\
Other\_control\_LS=Flow&        0.65   &        0.31   &        0.49   &        0.99   &        0.58   &        0.26   &        0.85   &        0.40   &        0.26   \\
\hline\hline
\end{tabular}
}

%% file: tables/Het_behavior_choose_Flow.tex
{
\def\sym#1{\ifmmode^{#1}\else\(^{#1}\)\fi}
\begin{tabular}{l*{9}{c}}
\hline\hline
                    &\multicolumn{1}{c}{(1)}&\multicolumn{1}{c}{(2)}&\multicolumn{1}{c}{(3)}&\multicolumn{1}{c}{(4)}&\multicolumn{1}{c}{(5)}&\multicolumn{1}{c}{(6)}&\multicolumn{1}{c}{(7)}&\multicolumn{1}{c}{(8)}&\multicolumn{1}{c}{(9)}\\
                    &\multicolumn{1}{c}{Consumption$^\dag$}&\multicolumn{1}{c}{Diet Diversity}&\multicolumn{1}{c}{Wealth$^\dag$}&\multicolumn{1}{c}{Borrowing$^\dag$ }&\multicolumn{1}{c}{Saving$^\dag$}&\multicolumn{1}{c}{Health}&\multicolumn{1}{c}{Sanitation}&\multicolumn{1}{c}{Prod assets$^\dag$}&\multicolumn{1}{c}{Cons assets$^\dag$}\\
                    &        b/se   &        b/se   &        b/se   &        b/se   &        b/se   &        b/se   &        b/se   &        b/se   &        b/se   \\
\hline
Inconsistent x Got It&        0.61   &       0.075   &        1.93   &       -0.28   &        0.52   &       -2.14   &       -1.07   &        0.77   &        1.46   \\
                    &      (0.41)   &      (0.84)   &      (1.41)   &      (1.86)   &      (1.45)   &      (1.73)   &      (1.06)   &      (0.69)   &      (1.20)   \\
Impatient x Got It  &       -0.32   &       -0.79   &       -3.60*  &        3.60   &        0.78   &        1.94   &      -0.056   &       -1.03   &       -2.85** \\
                    &      (0.46)   &      (0.71)   &      (1.85)   &      (2.74)   &      (2.58)   &      (1.80)   &      (1.35)   &      (0.81)   &      (1.28)   \\
Lack Other Cont x Got It&       0.084   &        0.29   &        1.54   &       -0.86   &        0.53   &       -1.57   &       -1.46   &        0.12   &      -0.095   \\
                    &      (0.48)   &      (0.74)   &      (2.13)   &      (2.01)   &      (1.77)   &      (1.76)   &      (0.99)   &      (0.78)   &      (1.24)   \\
Time Inconsistent   &       -0.38   &        0.18   &       -0.24   &        0.49   &      -0.039   &        1.76   &        0.80   &       -0.40   &       -0.43   \\
                    &      (0.36)   &      (0.79)   &      (0.71)   &      (1.33)   &      (1.30)   &      (1.30)   &      (0.91)   &      (0.61)   &      (0.91)   \\
Impatient           &        0.42   &        0.30   &        2.78** &       -2.92   &       -0.34   &       -2.75*  &        0.59   &        0.71   &        1.84*  \\
                    &      (0.41)   &      (0.62)   &      (1.37)   &      (2.21)   &      (2.40)   &      (1.50)   &      (1.14)   &      (0.71)   &      (1.02)   \\
Lack Other Control  &       -0.22   &       -0.86   &       -2.81   &       -0.38   &       -0.95   &        2.40*  &        0.34   &       -0.53   &       -1.97** \\
                    &      (0.37)   &      (0.59)   &      (2.03)   &      (1.66)   &      (1.63)   &      (1.42)   &      (0.87)   &      (0.72)   &      (0.97)   \\
Got Choice in Choice Experiment&       -0.41   &     -0.0070   &       -1.50   &       -1.37   &       -1.40   &        0.40   &        1.18   &       -0.75   &       -1.39   \\
                    &      (0.40)   &      (0.86)   &      (1.15)   &      (1.49)   &      (1.29)   &      (1.31)   &      (0.81)   &      (0.66)   &      (0.96)   \\
\hline
Control Mean        &        10.4   &        4.16   &        12.9   &        5.96   &        5.18   &        0.19   &       -0.23   &        11.4   &        8.71   \\
Observations        &         200   &         200   &         200   &         200   &         200   &         200   &         200   &         200   &         200   \\
$R^2$               &        0.27   &        0.36   &        0.33   &        0.21   &        0.31   &        0.25   &        0.27   &        0.33   &        0.37   \\
\hline\hline
\end{tabular}
}

%% file: tables/CovariateList_Household.tex
\begin{longtable}{p{.25\textwidth}p{.75\textwidth}}% p{0.1\textwidth}}
\caption{Covariates selected in Belloni et al. (2014) post-double-lasso selection procedure for household outcomes} \label{t:hcontrols} \\ 

\hline \hline
Outcome & Control set \\% & Share missing \\ 
\hline
\endfirsthead

\multicolumn{2}{r}{Table \ref{t:hcontrols} (continued)} \\ 
\hline \hline
Outcome & Control set \\ 
\hline
\endhead

\hline \multicolumn{2}{r}{{Continued on next page}}
\endfoot

\hline
\multicolumn{2}{p{\textwidth}}{Note:  block fixed effects and lag of the relevant outcome included in all specifications. Specifications that include both eligible and ineligible households include an indicator for eligibility status.} \\ 
\hline \hline
\endlastfoot

consumption\_asinh & Baseline value of consumption\_asinh, present in both rounds \\ 
 & L.Lhh\_wealth\_asinh \\%  \\ 
 & L.Fraction of village defined eligible by IPA \\% \\ 
Household dietary diversity score & Baseline value of dietarydiversity, present in both rounds \\ 
 & L.Lhh\_wealth\_asinh \\ 
 & L.Fraction of village defined eligible by IPA \\ 
 & Lsavingsstock\_asinh3 \\ 
 & Lconsumpti\_x\_Ldietarydi \\ 
 & Lconsumpti\_x\_Lproductiv \\ 
 & Ldietarydi\_x\_Lassetscon \\ 
wealth\_asinh & Baseline value of wealth\_asinh, present in both rounds \\ 
 & L.Lhh\_wealth\_asinh \\% \\ 
 & L.Fraction of village defined eligible by IPA \\% \\ 
 & L.Own dwelling\\%  \\ 
borrowingstock\_asinh & Baseline value of borrowingstock\_asinh, present in both rounds \\ 
 & L.Lhh\_wealth\_asinh\\%  \\ 
 & L.Fraction of village defined eligible by IPA\\%  \\ 
savingsstock\_asinh & Baseline value of savingsstock\_asinh, present in both rounds \\ 
 & L.Lhh\_wealth\_asinh \\ 
 & L.Fraction of village defined eligible by IPA \\ 
 & Lconsumpti\_x\_Lproductiv \\ 
 & Lconsumpti\_x\_Lassetscon \\ 
Health Knowledge Index & Baseline value of health\_knowledge, present in both rounds \\ 
 & L.Lhh\_wealth\_asinh \\ 
 & L.Fraction of village defined eligible by IPA \\ 
Sanitation Practices Index & Baseline value of sanitation\_practices, present in both rounds \\ 
 & L.Lhh\_wealth\_asinh \\ 
 & L.Fraction of village defined eligible by IPA \\ 
 & Lproductiv\_x\_Lassetscon \\ 
productiveassets\_asinh & Baseline value of productiveassets\_asinh, present in both rounds \\ 
 & L.Lhh\_wealth\_asinh \\ 
 & L.Fraction of village defined eligible by IPA \\ 
 & Lconsumpti\_x\_Lassetscon \\ 
assetsconsumption\_asinh & Baseline value of assetsconsumption\_asinh, present in both rounds \\ 
 & L.Lhh\_wealth\_asinh \\ 
 & L.Fraction of village defined eligible by IPA \\ 
 & L.Number of rooms \\ 
 & L.Durables expenditure (12-month recall) \\ 
 & Ldietarydi\_x\_Lassetscon \\ 
 & Lproductiv\_x\_Lassetscon \\ 
selfcostdwell\_asinh & Baseline value of selfcostdwell\_asinh, present in both rounds \\ 
 & L.Lhh\_wealth\_asinh \\ 
 & L.Fraction of village defined eligible by IPA \\ 
 & L.Number of rooms \\ 
 & L.Durables expenditure (12-month recall) \\ 
Housing Quality Index & Baseline value of housing\_quality, present in both rounds \\ 
 & L.Lhh\_wealth\_asinh \\ 
 & L.Fraction of village defined eligible by IPA \\ 
 & L.Number of rooms \\ 
\end{longtable}

%% file: tables/CovariateList_Individual.tex
\begin{longtable}{p{.25 \textwidth}p{.2\textwidth}p{.55\textwidth}}
\caption{Covariates selected in Belloni et al. (2014) post-double-lasso selection procedure for individual outcomes} \label{t:icontrols} \\ 

\hline \hline
Outcome & Sample & Control set \\ 
\hline
\endfirsthead

Table \ref{t:icontrols} (continued) \\ 
\hline \hline
Outcome & Sample & Control set \\ 
\hline
\endhead

\hline \multicolumn{3}{r}{{Continued on next page}}
\endfoot

\hline
\multicolumn{3}{p{\textwidth}}{Note:  block fixed effects and lag of the relevant outcome included in all specifications. Specifications that include both eligible and ineligible households include an indicator for eligibility status.} \\ 
\hline \hline
\endlastfoot

haz06, Winsorized fraction .005, high only  & Under 5s  & L.haz06, Winsorized fraction .005, high only \\ 
 &  & female \\ 
 &  & agemonths \\ 
 &  & agemonths\_sq \\ 
 &  & agemonths\_cu \\ 
 &  & L.Lhh\_wealth\_asinh \\ 
 &  & L.Food expenditure (weekly recall) \\ 
 &  & L.Food consumption-value own production (weekly recall) \\ 
 &  & L.waz06, Winsorized fraction .005, high only \\ 
 &  & Lconsumpti\_x\_Lselfcostd \\ 
waz06, Winsorized fraction .005, high only  & Under 5s  & L.waz06, Winsorized fraction .005, high only \\ 
 &  & female \\ 
 &  & agemonths \\ 
 &  & agemonths\_sq \\ 
 &  & agemonths\_cu \\ 
 &  & L.Lhh\_wealth\_asinh \\ 
 &  & L.Food expenditure (weekly recall) \\ 
 &  & L.Food consumption-value own production (weekly recall) \\ 
 &  & Lconsumpti\_x\_Lproductiv \\ 
muacz, Winsorized fraction .01  & Under 5s  & L.muacz, Winsorized fraction .01 \\ 
 &  & female \\ 
 &  & agemonths \\ 
 &  & agemonths\_sq \\ 
 &  & agemonths\_cu \\ 
 &  & L.Lhh\_wealth\_asinh \\ 
 &  & L.waz06, Winsorized fraction .005, high only \\ 
 &  & Lconsumpti\_x\_Lproductiv \\ 
anemia\_dummy  & Under 5s  & female \\ 
 &  & agemonths \\ 
 &  & agemonths\_sq \\ 
 &  & agemonths\_cu \\ 
 &  & L.Lhh\_wealth\_asinh \\ 
anemia\_dummy  & Pregnant/lactating women  & agemonths \\ 
 &  & agemonths\_sq \\ 
 &  & agemonths\_cu \\ 
 &  & L.Lhh\_wealth\_asinh \\ 
mortality  & All  & female \\ 
 &  & agemonths \\ 
 &  & agemonths\_sq \\ 
 &  & agemonths\_cu \\ 
 &  & L.Lhh\_wealth\_asinh \\ 
Was this women pregnant at any point in the past 12 months  & Pregnant/lactating women  & agemonths \\ 
 &  & agemonths\_sq \\ 
 &  & agemonths\_cu \\ 
 &  & L.Lhh\_wealth\_asinh \\ 
 &  & L.Lwealth\_asinh \\ 
Did pregnancy conclude in live birth  & Pregnant/lactating women  & agemonths \\ 
 &  & agemonths\_sq \\ 
 &  & agemonths\_cu \\ 
 &  & L.Lhh\_wealth\_asinh \\ 
 &  & L.Food expenditure (weekly recall) \\ 
 &  & L.Food consumption-value own production (weekly recall) \\ 
 &  & Lconsumpti\_x\_Lwealth\_as \\ 
facility\_birth  & Pregnant/lactating women  & agemonths \\ 
 &  & agemonths\_sq \\ 
 &  & agemonths\_cu \\ 
 &  & L.Lhh\_wealth\_asinh \\ 
anthro\_vacc\_year  & Under 3s  & female \\ 
 &  & agemonths \\ 
 &  & agemonths\_sq \\ 
 &  & agemonths\_cu \\ 
 &  & L.Lhh\_wealth\_asinh \\ 
 &  & Lconsumpti\_x\_Lproductiv \\ 
anthro\_vacc\_complete  & Under 3s  & female \\ 
 &  & agemonths \\ 
 &  & agemonths\_sq \\ 
 &  & agemonths\_cu \\ 
 &  & L.Lhh\_wealth\_asinh \\ 
Any fever, diarrhea, or coughing blood at individual/round level  & Under 5s  & female \\ 
 &  & agemonths \\ 
 &  & agemonths\_sq \\ 
 &  & agemonths\_cu \\ 
 &  & L.Lhh\_wealth\_asinh \\ 
 &  & L.Food consumption-value own production (weekly recall) \\ 
Individual reported with diarrhea/vomiting/fever now  & Under 5s  & female \\ 
 &  & agemonths \\ 
 &  & agemonths\_sq \\ 
 &  & agemonths\_cu \\ 
 &  & L.Lhh\_wealth\_asinh \\ 
\end{longtable}